\documentclass[12pt]{article}
\usepackage[utf8]{inputenc}
\usepackage[T1]{fontenc}
 \usepackage{relsize}
\usepackage{epsfig}
\usepackage{amsfonts} 
\usepackage{amsmath} 
\usepackage{mathrsfs} 
\usepackage{color}

\usepackage{indentfirst}
\usepackage{hyperref}
  
  \usepackage{comment}
\usepackage{graphicx}
\newcommand{\cev}[1]{\reflectbox{\ensuremath{\vec{\reflectbox{\ensuremath{#1}}}}}}

\def\half{{1 \over 2}}

\def\m{\mu}

\def\a{\alpha}

\def\Or[#1]{{\text{O}}\left({#1}\right)}
\def\dotl[#1,#2]{\left\langle #1, #2 \right\rangle}
\def\dotlb[#1,#2]{[ #1, #2 ]}
\def\dotp[#1,#2]{(#1) \cdot (#2)}
\def\aff[#1,#2]{\hat{#1}(#2)}
\def\n4sym{{\cal N}=4 SYM}
\def\>{\rangle}
\def\<{\langle}
\def\weight[#1,#2,#3]{\{(#1),#2,#3\}}
\def\ads[#1]{$\text{AdS}_{#1}$}

\newcommand{\ba}{\begin{eqnarray}}
\newcommand{\ea}{\end{eqnarray}}

\newcommand{\be}{\begin{equation}}
\newcommand{\ee}{\end{equation}}  
\newcommand{\bi}{\begin{itemize}}
\newcommand{\ei}{\end{itemize}}

\newcommand{\Ncal}{{\mathcal N}}
\newcommand{\Ocal}{{\mathcal O}}

\newcommand{\Tcal}{{\mathcal T}}

\newcommand{\aslash}[1]{\,\,{\raise.15ex\hbox{/}\mkern-12mu #1}}
\newcommand{\bslash}[1]{\,\,{\raise.15ex\hbox{/}\mkern-9mu #1}}

\renewcommand{\bar}{\overline}
\renewcommand{\tilde}{\widetilde}
\renewcommand{\hat}{\widehat}

\newcommand\myatop[2]{\genfrac{}{}{0pt}{}{#1}{#2}}

\newcommand\lrpar{\raise .8ex\hbox{$^\leftrightarrow$} \hspace{-9pt}
\partial}
\newcommand\lpar{\raise .8ex\hbox{$^\leftarrow$} \hspace{-9pt}
\partial}
\newcommand\rpar{\raise .8ex\hbox{$^\rightarrow$} \hspace{-9pt}
\partial}
\newcommand\lrd{\raise .8ex\hbox{$^\leftrightarrow$} \hspace{-9pt}
\nabla}

\newcommand{\gsim}{\lower.7ex\hbox{$\;\stackrel{\textstyle>}{\sim}\;$}}
\newcommand{\lsim}{\lower.7ex\hbox{$\;\stackrel{\textstyle<}{\sim}\;$}}

\let\a=\alpha 
\let\b=\beta 
\let\g=\gamma 
\let\d=\delta

\let\k=\kappa
\let\l=\lambda 
\let\m=\mu 
\let\n=\nu 
 
\let\r=\rho
\let\s=\sigma 
\let\t=\tau

\let\D=\Delta 
 
\let\L=\Lambda

\let\G=\Gamma

\renewcommand{\ba}{\begin{eqnarray}}
\renewcommand{\ea}{\end{eqnarray}}
\newcommand{\bea}{\begin{eqnarray}}
\newcommand{\eea}{\end{eqnarray}}

\begin{document}

\begin{titlepage}

\begin{center}
\vspace{1cm}

{\Large \bf Factorization of Mellin amplitudes}

\vspace{0.8cm}

{\bf Vasco Gon\c calves,  João Penedones, Emilio Trevisani}

\vspace{.5cm}

{\it 
Centro de Física do Porto \\
Departamento de Física e Astronomia \\
Faculdade de Ciências da Universidade do Porto \\
Rua do Campo Alegre 687, 4169-007 Porto, Portugal
 }

\end{center}
\vspace{1cm}

\begin{abstract}
We introduce Mellin amplitudes for correlation functions of $k$ scalar operators and one  operator with  spin 
in conformal field theories (CFT) in general dimension.
We show that Mellin amplitudes for scalar operators have simple poles with residues that factorize in terms of lower point Mellin amplitudes, similarly to what happens for scattering amplitudes in flat space.
Finally, we study the flat space limit of Anti-de Sitter (AdS) space, in the context of the AdS/CFT correspondence, and generalize a formula relating CFT Mellin amplitudes to scattering amplitudes of the bulk theory, including particles with spin.
\end{abstract}

\end{titlepage}

\tableofcontents

\newpage

\section{Introduction}

Mellin amplitudes are an alternative representation of conformal correlation functions that are analogous to scattering amplitudes. 
In particular, we shall show that the Operator Product Expansion (OPE) leads to the factorization  of the residues of the poles of Mellin amplitudes. 
In the future, we hope these factorization properties can be used to compute Mellin amplitudes more efficiently with BCFW-type recursion relations.

\begin{figure}
\begin{centering}
\includegraphics[scale=0.4]{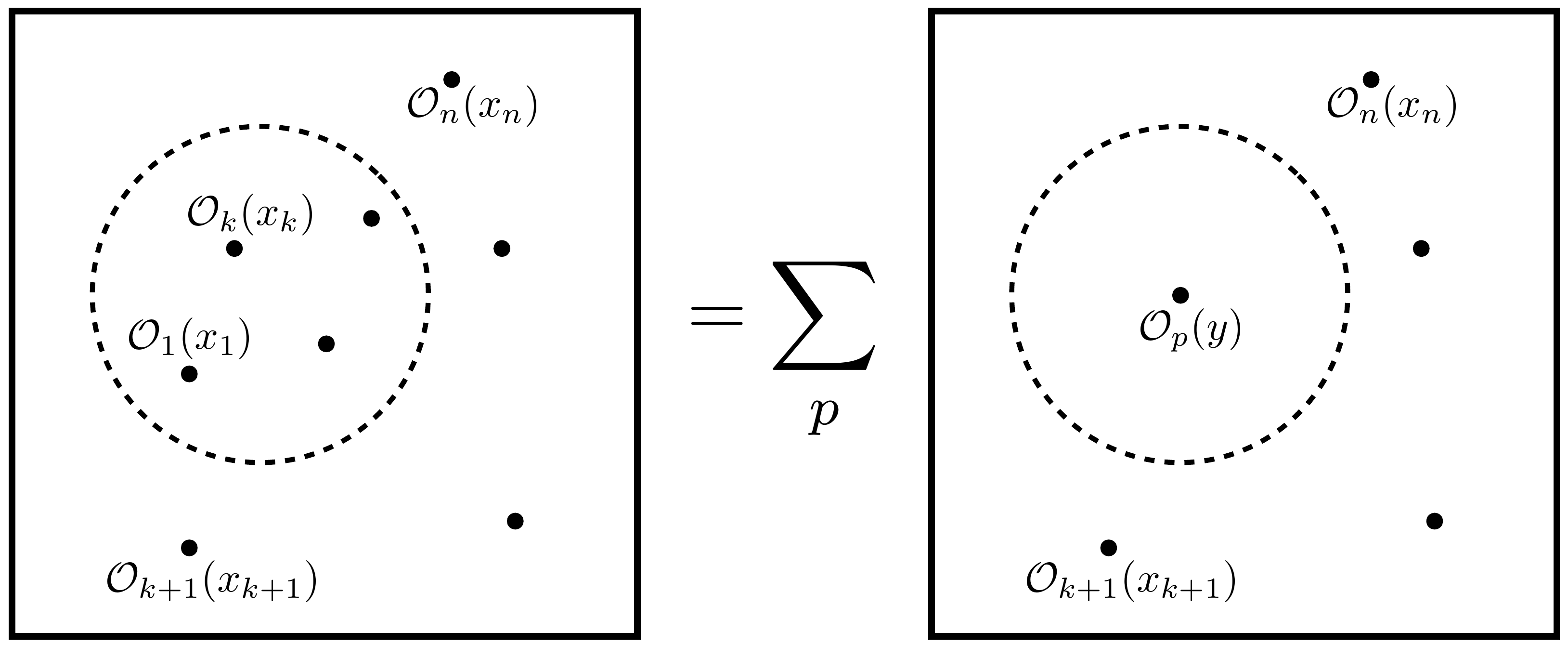}
\par\end{centering}
\caption{\label{fig:OPEconvergence}
In a CFT correlation function, one can replace multiple operators inside a sphere by a  (infinite) sum of local operators inserted at the center of the sphere.
}
\end{figure}

The existence of a convergent OPE is a basic  property of a Conformal Field Theory (CFT). This means that we can replace the product of $k$ local operators (inside a correlation function) by an infinite sum of local operators
\begin{align}
\Ocal_1 (x_1 )\dots\Ocal_k (x_k )=\sum_{p} C_{\mu_1\dots\mu_J}^{(1\dots k,p)}
\left(x_1,\dots,x_k,y,\partial_y\right) \Ocal_p ^{\mu_1\dots\mu_J}\left(y\right),
\label{OPE}
\end{align} 
where $p$ runs over all primary local operators.
This sum converges inside a $n$-point correlation function if there is a sphere centred at $y$ that contains all points $x_1,\dots,x_k$ and does not contain any of the other $n-k$ points,
as depicted in figure \ref{fig:OPEconvergence}.
Therefore, we can write
\begin{align}
\!\!\!\langle\Ocal_1 (x_1 )\dots\Ocal_n (x_n )\rangle=\sum_{p} C_{\mu_1\dots\mu_J}^{(1\dots k,p)}
\left(x_1,\dots,x_k,y,\partial_y\right) 
\langle \Ocal_p ^{\mu_1\dots\mu_J} (y )
\Ocal_{k+1} (x_{k+1} )\dots\Ocal_n (x_n )\rangle \ .
\label{npointCB}
\end{align} 
Notice that the OPE coefficient function is entirely determined by the $(k+1)$-point function,
\begin{align}
\langle\Ocal_1\left(x_1\right)\dots\Ocal_k\left(x_k\right)\Ocal_p ^{\nu_1\dots\nu_J}\left(z\right)\rangle=  C_{\mu_1\dots\mu_J}^{(1\dots k,p)}
\left(x_1,\dots,x_k,y,\partial_y\right) 
\langle\Ocal_p ^{\mu_1\dots\mu_J}\left(y\right)\Ocal_p ^{\nu_1\dots\nu_J}\left(z\right)\rangle\ ,
\end{align} 
where we have chosen a basis of operators that diagonalizes the two-point functions.
This suggests that using the OPE one should be able to \emph{factorize} $n$-point functions in products of lower point functions (in this case, $k+1$ times $n-k+1$).
Following Mack \cite{Mack, MackSummary}, we shall argue that this factorization is best formulated in Mellin
space.

Consider the Mellin amplitude associated with a $n$-point function of scalar primary operators $\Ocal_i$ of dimension $\D_i$,
\footnote{We use  $\g_{ij}$ for Mellin variables, instead of the standard notation $\d_{ij}$ to avoid confusion with the Kronecker-deltas that proliferate in this work.
Throughout this paper, $M(\g_{ij})$ denotes a function $M(\g_{12},\g_{13},\dots)$ of all Mellin variables.
}
\begin{align}
\left\langle \Ocal_1\left(x_1\right)...\Ocal_n\left(x_n\right) \right\rangle=\int 
\left[d\g\right]M\left(\g_{ij}\right)\prod_{1\leq i<j\leq n} \Gamma\left(\g_{ij}\right)\left(x_{ij}\right)^{-\g_{ij}}\ ,
\label{Mellinscalar}
\end{align}
where the integration $\left[d\g\right]$ is subject to the constraints
\footnote{The notation $\left[d\g\right]$ includes the a factor of $\frac{1}{2\pi i}$ for each one of the $ n(n-3)/2$ independent $\g_{ij}$ variables.}
\be
\sum_{i=1}^n \g_{ij}=0\ , \qquad
\g_{ij}=\g_{ji}\ , \qquad
\g_{ii}=-\D_i\ , 
\label{constraintsdelta}
\ee
ensuring that the correlation function transforms appropriately under conformal transformations.
The integration contours for the independent $\g_{ij}$ variables run parallel to the imaginary axis. The Mellin amplitude $M$ depends on  the  variables $\g_{ij}$ subject to the constraints (\ref{constraintsdelta}) but we shall often keep this dependence implicit to simplify our formulas.

It is convenient to introduce a set of auxiliary vectors $\{p_1,\dots,p_n\}$ such that $\g_{ij}=p_i\cdot p_j$. Then, the constraints (\ref{constraintsdelta}) follow from 
momentum conservation $\sum_{i=1}^n p_i=0$ and 
on-shellness $p_i^2 = -\D_i$.

\begin{figure}
\begin{centering}
\includegraphics[scale=0.4]{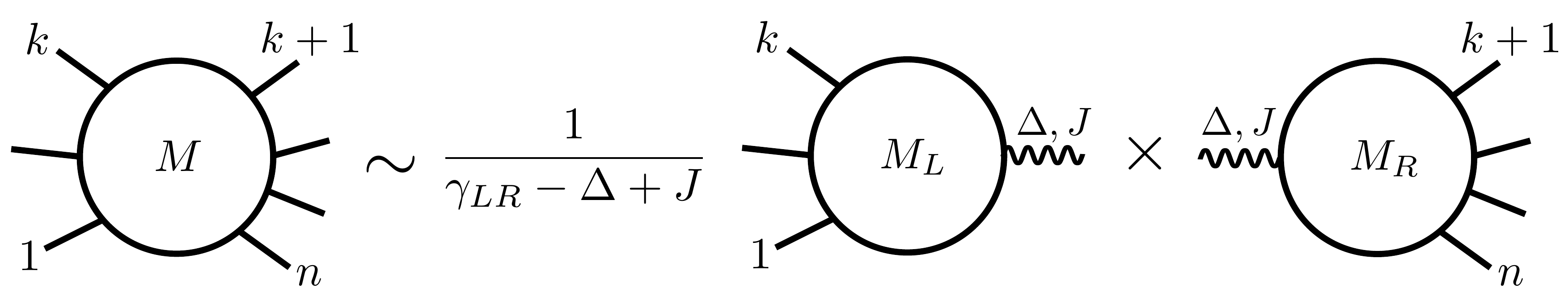}
\par\end{centering}
\caption{\label{fig:Factorization}
The multiple OPE of figure \ref{fig:OPEconvergence} leads to the factorization of the residues of the poles of the Mellin amplitude in terms of lower-point Mellin amplitudes.
}
\end{figure}

As explained in \cite{NaturalMellin} (section 2.1),
for each primary operator $\Ocal_p$, with dimension $\D$ and spin $J$, that appears in the OPE (\ref{OPE}), the Mellin amplitude has an infinite sequence of poles 
\begin{align}
M \approx 
\frac{\mathcal{Q}_m  }{\g_{LR}-\left(\Delta-J+2m\right)}\ ,\ \ \ \ \ \ \ 
m=0,1,2,\dots \ ,
\label{MellinFacPoles}
\end{align} 
in the variable 
\be \label{def:gammaLR}
\g_{LR}=-\left(\sum_{a=1}^{k}p_a\right)^2=\sum_{a=1}^{k}\sum_{i=k+1}^n\g_{ai}\ .
\ee
Notice that the position of the poles can be thought as the on-shellness condition for the total momentum injected into the first $k$ operators.
Moreover,  the residues $\mathcal{Q}_m$ factorize in terms of the Mellin amplitudes for a $(k+1)$-point function (Left) and a  $(n-k+1)$-point function (Right), as depicted in figure \ref{fig:Factorization}.
When the exchanged operator has spin zero, the formulas are particularly simple 
\footnote{These formulas were first derived in \cite{NaturalMellin} for correlation functions given by Witten diagrams in AdS. 
This expression differs by a factor of 
\be
\mathcal{C}_\D=\frac{\G(\D)}{2\pi^{\frac{d}{2}}
\G\left(1+\D-\frac{d}{2}\right)}
\ee
from the result of \cite{NaturalMellin}.
This mismatch follows from a different choice of normalization for the operators $\Ocal$. In \cite{NaturalMellin}, the following choice was made
$\langle \Ocal(x) \Ocal(0)\rangle=\mathcal{C}_\D |x|^{-2\D}$.}
\begin{align} 
\mathcal{Q}_0   =-2\G(\D) \, M_L 
M_R \ , \qquad
\mathcal{Q}_m    =
\frac{-2\G(\D) m!}{ 
 \left(1+\D-\frac{d}{2}  \right)_m }
 L_m R_m \ , \label{Eq:FacScalar} 
\end{align}
where $(x)_m=\G(x+m)/\G(x)$ is the Pochhammer symbol and $d$ is the spacetime dimension. $M_L\ (M_R)$ is the Mellin amplitude associated with the left (right) sub-diagram in figure \ref{fig:Factorization} and 
\be
 L_m =\sum_{ \myatop{n_{ab}\ge 0}{\sum n_{ab}=m} } 
M_L(\g_{ab}+n_{ab})  
\prod_{1\le a<b\le k} \frac{ 
\left(\g_{ab}\right)_{n_{ab}}}{n_{ab}!}
\label{LmSum}
\ee
and similarly for $R_m $.
In sections \ref{sec:FacShadow} and \ref{sec:FacCasimir} we will derive these and  similar factorization formulas for the residues associated with primary operators with non-zero spin.
However, before that we must discuss the generalization of the Mellin representation (\ref{Mellinscalar}) for correlation functions involving tensor operators (section \ref{sec:MellinSpin}).
In order to make the analogy with scattering amplitudes more explicit, we start by reviewing their factorization properties in section \ref{sec:reviewFS}.
In section \ref{sec:FSL}, we generalize a formula proposed in \cite{JPMellin}, relating the Mellin amplitude of a CFT correlator to the scattering amplitude of the dual bulk theory through the flat space limit of Anti-de Sitter (AdS) spacetime.
In addition in appendix \ref{app:FSLfactorization} we check that, in the flat space limit, our factorization formulas for Mellin amplitudes reduce to the standard factorization formulas for scattering amplitudes.
Finally, we conclude in section \ref{sec:Conclusion} with some   ideas for the future.

\section{Factorization of scattering amplitudes}
\label{sec:reviewFS}

\begin{figure}
\begin{centering}
\includegraphics[scale=0.4]{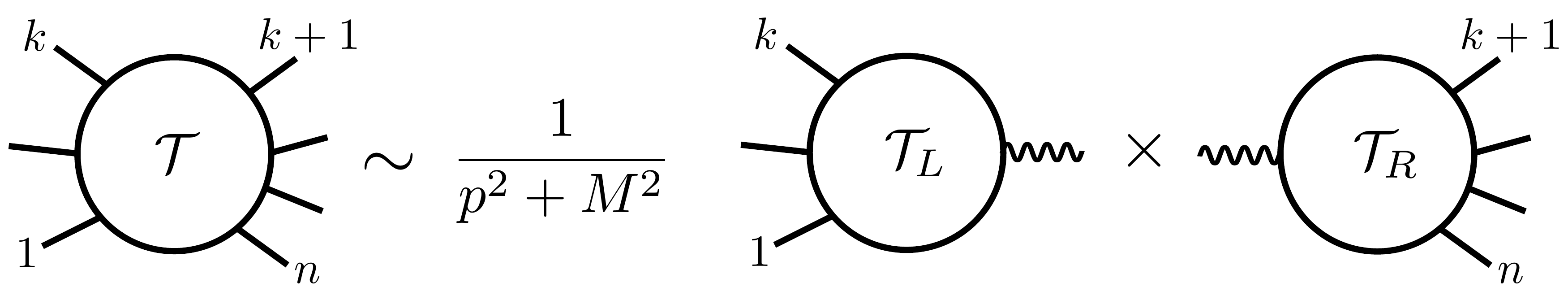}
\par\end{centering}
\caption{\label{fig:FacScat}
Scattering amplitudes have poles when the momentum $p=\sum_{a=1}^kp_a$ approaches the mass shell, $p^2+M^2=0$, of a  particle in the theory.
The residue of this pole factorizes in terms of lower point scattering amplitudes.
}
\end{figure}

In this section, we 
review the factorization properties of scattering amplitudes of $n$ scalar particles.
In particular, we study their poles associated to the exchange of  particle with mass $M$ and spin $J$. The residues of these poles factorize in terms of left and right scattering amplitudes (see figure \ref{fig:FacScat}) involving the exchanged particle as an external state. The case of spin $J=0$ is particularly simple,
\begin{align}
\underset{p^2=-M^2}{\rm Res}\left(\mathcal{T}\right)&=       \mathcal{T}_L \mathcal{T}_R  \label{FSLfacJ0}
\end{align}
where $p=\sum_{a=1}^k p_a$ is the total momentum injected on the left part of the diagram  and $\mathcal{T}_L$ and $\mathcal{T}_R$ are the scattering amplitudes associated to the left and right scattering amplitudes (see figure \ref{fig:FacScat}). Before generalizing this formula for general spin $J$ we must introduce some notation for scattering amplitudes involving one particle with spin $J$.

\subsection{Scattering amplitudes for spinning particles}
\label{subsec:SAspin}

\begin{figure}
\begin{centering}
\includegraphics[scale=0.35]{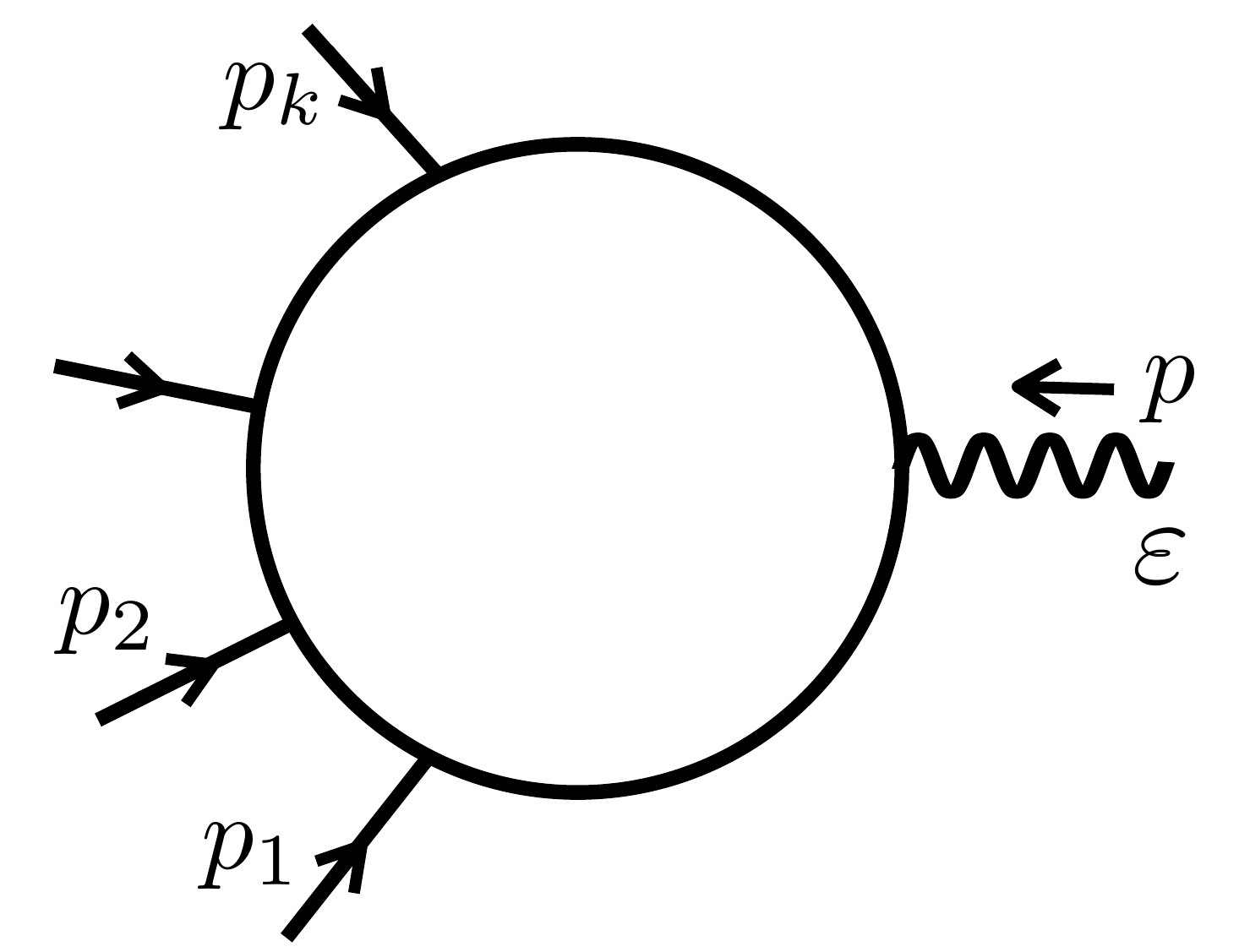}
\par\end{centering}
\caption{\label{fig:ScatVector}
Scattering amplitude of $k$ scalar particles and one particle with spin. The polarization of the spinning particle is encoded in the null vector $\varepsilon$. 
}
\end{figure}

A scattering amplitude of a massive vector boson and $k$ scalars is a function of the polarization vector $\varepsilon$ and the momenta $p$ and $p_i$ (see figure \ref{fig:ScatVector}). It can be written as
\be
\mathcal{T}(\varepsilon,p,p_i)=
\sum_{a=1}^k \varepsilon \cdot p_a
\, \mathcal{T}^a(p_i\cdot p_j)
\label{ScatVector}
\ee
where the $\mathcal{T}^a$ are functions of the Lorentz invariants $p_i\cdot p_j$. Notice that momentum conservation 
\be
p+\sum_{a=1}^k p_a=0
\ee
can be used to write $p\cdot p_i$ in terms of $p_i\cdot p_j$.
Moreover, the condition $\varepsilon \cdot p=0$ leads to the redundancy 
\be
\mathcal{T}^a(p_i\cdot p_j) \to
\mathcal{T}^a(p_i\cdot p_j)+
\Lambda(p_i\cdot p_j)
\ee
for any function $\L$ of the Lorentz invariants $p_i\cdot p_j$. If the vector particle is massless (photon) then we have gauge invariance
\be
\mathcal{T}(\varepsilon+\l p,p,p_i)=
\mathcal{T}(\varepsilon,p,p_i)\ ,\ \ \ \ 
 \forall \l \in \mathbb{R} \ ,
\ee
which leads to the constraint
\be
\sum_{a=1}^k p \cdot p_a
\, \mathcal{T}^a(p_i\cdot p_j)=0\ .
\label{transversalitySA}
\ee

Consider now  the scattering amplitude of a massive spin $J$ boson and $k$ scalars.  In this case the  polarization vector is a symmetric and traceless tensor with $J$ indices, which we shall encode   with a null polarization vector with just one index $\varepsilon^\m$.
In this way we can write the scattering amplitude as a function of $\varepsilon$ and the momenta $p$ and $p_i$ as follows
\be
\mathcal{T}(\varepsilon,p,p_i)=
\sum_{a_1,\dots,a_J=1}^k \left( \prod_{\ell=1}^J \varepsilon \cdot p_{a_\ell}\right) \, \mathcal{T}^{a_1 \dots a_J}(p_i\cdot p_j) \label{TRepres}
\ee
where   $ \mathcal{T}^{a_1 \dots a_J}(p_i\cdot p_j)$ are functions of the Lorentz invariants $p_i\cdot p_j$, totally symmetric under permutations of the indices $a_1 \dots a_J$. The condition $\varepsilon \cdot p=0$ leads to the redundancy
\be \label{RedundancySpinJT}
\mathcal{T}^{a_1 \dots a_J}(p_i\cdot p_j) \to
\mathcal{T}^{a_1 \dots a_J}(p_i\cdot p_j)+
\sum_{m=1}^J \Lambda^{a_1 \dots a_{m-1}a_{m+1}\dots a_J}(p_i\cdot p_j)
\ee
where $\Lambda^{a_1 \dots a_{m-1}a_{m+1}\dots a_J}(p_i\cdot p_j)$ is any function of $p_i\cdot p_j$ that depends on one less index and it is symmetric under permutations of its indices.

When the spin $J$ field is massless, the gauge  invariance $
\mathcal{T}(\varepsilon+\l p,p,p_i)=
\mathcal{T}(\varepsilon,p,p_i)
$
leads to the constraint
\be
\sum_{a_1=1}^k p \cdot p_{a_1}
\, \mathcal{T}^{a_1 \dots a_J}(p_i\cdot p_j)=0.
\ee

\subsection{Factorization  on a vector particle}

We start from the spin $J=1$ case. We consider a vector field $A^\m$  with mass $M$ and the following Proca propagator
\be
\frac{\Pi_{\m \n}(p)}{p^2+M^2} =
\frac{1}{p^2+M^2} \left(g_{\m \n}+\frac{1}{M^2} p_{\m} p_\n \right) \,
\ee
where $g_{\mu\nu}$ is the Euclidean space metric.
If such a particle exists then we expect a pole in the scattering amplitude $\Tcal$ of $n$ scalar particles when the sum of momenta of the first $k$ particles,
\be
p=\sum_{a=1}^k p_a \ ,
\ee
satisfies the on-shell condition
$p^2+M^2=0$. Moreover, the residue of this pole is given by
\begin{align}
\underset{p^2=-M^2}{\rm Res}\left(\mathcal{T}\right)&=  \sum_{a=1}^k 
 \sum_{i=k+1}^n \mathcal{T}_L^a  \mathcal{T}_R^i  \,p_a^\m \left(g_{\m \n}+\frac{1}{M^2} p_{\m} p_\n\right) p_i^\n  =  \sum_{a=1}^k 
 \sum_{i=k+1}^n  \mathcal{T}^a_L \mathcal{T}^i_R\, \Omega_{a i}  \label{FSLfacJ1}
\end{align}
where $\mathcal{T}_L^a$ is the scattering amplitude for the first $k$ scalars and one vector particle in the notation of equation (\ref{ScatVector}) (similarly for $\mathcal{T}_R^i$).
For later convenience we introduced
 \be \label{Omegaij}
\Omega_{i j}  \equiv p_i^\m p_j^\n 
\Pi_{\m \n}(p)= (p_i \cdot p_j) +\frac{1}{M^2} (p \cdot p_i) (p \cdot p_j).
\ee
We point out that in (\ref{FSLfacJ1}) there is no divergence when $M=0$ due to the  transversality condition (\ref{transversalitySA}).

\subsection{Factorization on a spin 2 particle}

A particle with mass $M$ and spin $J=2$ can be represented as a symmetric and traceless tensor $h^{\m \n}$ such that $\partial_\m h^{\m \n}=0$. These conditions can be used to fix its propagator in momentum space $D_{\m \n \s \r}(p)= \frac{P_{\m \n \s \r}(p)}{p^2+M^2}$  where $P_{\m \n \s \r}(p)$ must be
\begin{itemize}
\item transverse with respect to the momentum $p$ of the particle, \emph{i.e.} $p^\m P_{\m \n \s \r}(p)=0$,
\item symmetric in the exchange of $\m \leftrightarrow \n$ and $\r \leftrightarrow \s$,
\item traceless, \emph{i.e.} $g^{\m \n} P_{\m \n \s \r}(p)=0$,
\item a projector, \emph{i.e.} $P_{\m \n}^{\ \  \a \b}(p) P_{\a \b \s \r}(p)=P_{\m \n \s \r}(p)$.
\end{itemize}
The result is 
\be
D_{\m \n \s \r}(p)=\frac{1}{p^2+M^2}\left[ \frac{1}{2}\Pi_{\m \s}\Pi_{\n \r}+\frac{1}{2}\Pi_{\m \r}\Pi_{\n \s} -\frac{1}{d} \Pi_{\m \n}\Pi_{\s \r} \right]
\ee
where $\Pi_{\m \n} =g_{\m \n} +\frac{1}{M^2} p_\m p_\n$ is the on-shell projector onto the space orthogonal to $p$ and $g_{\m \n}$ is the flat metric for $\mathbb{R}^{d+1}$. \footnote{We work in $d+1$ dimensions for later convenience when considering the flat space limit of $AdS_{d+1}$ dual to a CFT in $d$ dimensions.}

The residue of a scattering amplitude of $n$ scalars at the pole associated to an exchanged spin 2 particle, factorizes as follows 
\begin{align} \label{FS:J2Factoriz}
\underset{p^2=-M^2}{\rm Res}(\mathcal{T})
&=   \sum_{a,b=1}^k 
 \sum_{i,j=k+1}^n \mathcal{T}^{a b}_L \mathcal{T}^{i j}_R\, \left[ \Omega_{a i}\Omega_{b j} 
-\frac{1}{d} \Omega_{a b}\Omega_{i j} \right] 
\end{align}
where $\mathcal{T}_L^{a b}$ is   the scattering amplitude of the first $k$ scalar particles and one spin 2 particle in the notation of equation
(\ref{TRepres}). 

\subsection{Factorization on a spin $J$ particle}
A particle with mass $M$ and spin $J$ can be represented as a symmetric, traceless and divergenceless tensor with $J$ indices. Once again we can fix its propagator in momentum space $D_{\m_1 \dots\m_J \m'_1 \dots \m'_J}(p)= \frac{1}{p^2+M^2} P_{\m_1 \dots\m_J \m'_1 \dots \m'_J}(p)$  providing $P(p)$ to be a  traceless projector transverse to the  momentum $p$ and totally symmetric in the exchange of the $\m_i$ and the $\m'_i$ separately.
The result can be written as \cite{Ingraham}
\begin{align}
P(p)&=\sum_{r=0}^{[J/2]} b_r \overbrace{\Pi \dots \Pi}^r \overbrace{\Pi' \dots \Pi'}^{J-2r}\overbrace{\Pi'' \dots \Pi''}^{r}\\
b_r&=\frac{(-1)^r}{ 2^{2 r}r!} \frac{(1+J-2r)_{2 r}}{\left(\frac{d}{2}-1+J-r\right)_{r}}
\end{align}
where we have suppressed the indices in the expression and $\Pi$, $\Pi'$ and $\Pi''$ are the already defined projector $\Pi_{\m \n}$ with respectively zero, one or two primed indices. The  $\Pi'$ are defined with the unprimed index first and each term in the sum is meant to be symmetrized in the primed and unprimed indices separately.

The factorization formula for the residue associated with the exchange of a spin $J$ particle is then given by 
\begin{align} 
\underset{p^2=-M^2}{\rm Res}(\mathcal{T})
&=   \sum_{a_\ell=1}^k 
 \sum_{i_\ell=k+1}^n \mathcal{T}_L^{a_1 \dots a_J}\mathcal{T}_R^{i_1 \dots i_J} \sum_{r=0}^{[J/2]} b_r \left(\prod_{t=1}^r \Omega_{a_{2 t-1} a_{2t}}\Omega_{i_{2t-1} i_{2t}}\right) \left(\prod_{q=2r+1}^J\Omega_{a_{q} i_{q}} \right)   \label{FlatJFactoriz}
\end{align}
where $\mathcal{T}^{a_1 \dots a_J}$ is defined in  equation (\ref{TRepres}). 

%

To better understand the physical meaning of $\Omega_{ij} $ we  consider the scattering of $k$ particles going into $n-k$ with energy-momenta   $p_i^\m=(E_i, \vec{p}_i)$  in the center of mass frame. Calling $E_{c m}=\sum_{a=1}^k E_a$ the total energy in the center of mass we have $p^\m \equiv \sum_{a=1}^k p_a^\m = (E_{cm}, \vec{0})$. Noticing that at the factorization pole $E_{cm}=M$, we find
\be \label{omegacm}
\Omega_{ij} = -E_i E_j +\vec{p}_i \cdot \vec{p}_j + \frac{(-E_i E_{cm})(-E_j E_{cm})}{M^2}=\vec{p}_i \cdot \vec{p}_j \ .
\ee 

In the case of 2 $\to$ 2 scattering ($k=2$ and $n=4$), we have $\vec{p}_2=-\vec{p}_1$, $\vec{p}_4=-\vec{p}_3$ and $\vec{p}_1\cdot \vec{p}_3 =
|\vec{p}_1||\vec{p}_3|\cos \theta$, with $\theta$ the scattering angle. This gives
\begin{align}
\underset{p^2=-M^2}{\rm Res}(\mathcal{T}) 
&=\frac{J! }{2^J\left(\frac{d}{2}-1\right)_J}   \mathcal{T}_L \; \mathcal{T}_R\; |\vec{p}_1|^J|\vec{p}_3|^J  C^{d/2-1}_J(\cos \theta)    
\end{align}
where $C^{d/2-1}_J(\cos \theta)$ is the Gegenbauer polynomial and 
\begin{align}
\mathcal{T}_L= \sum_{a_\ell= 1, 2} (-1)^{\sum_{\ell=1}^J \d_{a_\ell}^1}  \mathcal{T}_L^{a_1 \dots a_J}  , \qquad
\mathcal{T}_R= \sum_{i_\ell =3, 4}(-1)^{\sum_{\ell=1}^J\d_{i_\ell}^3}\mathcal{T}_R^{i_1 \dots i_J}   . \label{TLTR}
\end{align}
Notice that the scattering amplitude between two scalars and a spin $J$ particle is characterized by a single number. This is not obvious in the representation (\ref{TRepres}) which involves $J+1$ terms ($\mathcal{T}_L^{11\dots 1}$, $\mathcal{T}_L^{21\dots1}$, ... ,$\mathcal{T}_L^{22\dots2}$), but it follows from the redundancy (\ref{RedundancySpinJT}) which can be used to eliminate $J$ terms.
As a consistency check one can verify that $\mathcal{T}_L$  and $\mathcal{T}_R$  are invariant under the redundancy (\ref{RedundancySpinJT}).

\section{Mellin representations for tensor operators}
\label{sec:MellinSpin}

The goal of this section is to generalize the Mellin representation (\ref{Mellinscalar}) for correlation functions of scalar operators, to correlation functions  involving tensor operators. 
In particular, we shall focus our attention on correlators of $k$ scalar operators and one tensor operator because this is what we need to write factorization formulas for $n$-point functions of scalar operators.
 
Throughout  this paper, we will make significant use of the embedding formalism for CFTs \cite{Dirac, Mack:1969rr, Boulware:1970ty, Ferrara:1973yt, ourDIS, Weinberg:2010fx}.
In this formalism, points in $\mathbb{R}^d$ are mapped to null rays through the origin of $\mathbb{M}^{d+2}$ and the conformal group acts as the Lorentz group $SO(d+1,1)$.
Primary operators of dimension $\D$ and spin $J$ are encoded into homogeneous functions of two null vectors $P,Z \in {M}^{d+2}$, 
\be
\Ocal(\l P,\a Z   )=\l^{-\D} \a^J \Ocal(P,Z)\ ,\ \ 
\ \ \ \ \forall \a  ,\l \in \mathbb{R}\ .
\label{homogeneousO}
\ee
The vector $P$ parametrizes the light cone of the embedding space $\mathbb{M}^{d+2}$ and $Z$ is an auxiliary polarization vector that encodes the tensor nature of the operator. Finally, we impose $Z\cdot P=0$ and
\be
\Ocal(  P,  Z  +\b P )= \Ocal(P,Z)\ ,\ \ 
\ \ \ \ \forall \b \in \mathbb{R}\ .
\label{shiftO}
\ee
for each operator to avoid over-counting of degrees of freedom \cite{SpinningCC}.
 
\subsection{Vector operator}

Let us start with the case of a vector primary  operator and $k$ scalars,
\be
\langle
\Ocal(P,Z)\Ocal_1(P_1)\dots \Ocal_k(P_k) 
\rangle\ .
\label{corrVector}
\ee
A possible Mellin representation for this correlator is 
\be 
\sum_{a=1}^k ( Z \cdot P_{a})
  \int [d\g]\,
 M^{a}
  \prod_{i,j=1\atop i<j}^k \frac{\G(\g_{ij})}{(-2P_i\cdot P_j)^{\g_{ij}}}
\prod_{i=1}^k \frac{\G(\g_{i}+\d_i^{a})}{(-2P_i\cdot P)^{\g_{i}+\d_i^{a} }}\ ,
  \label{Mrep}
\ee 
where $\d_i^{a}$ is the Kronecker-delta and 
\be
\g_{i}= -\sum_{j=1}^k \g_{i j}\ ,\qquad
\g_{ij}=\g_{ji}\ , \qquad
\g_{ii}=-\D_i\ ,
\label{gammai}
\ee
as required by (\ref{homogeneousO}) applied to each scalar operator.
Imposing (\ref{homogeneousO}) for the vector operator, we obtain the final constraint
\be
\sum_{i,j=1}^k \g_{i j}=1-\D\ .
\label{constraintM1}
\ee
In this case, it is convenient to think of $\g_{ij}$ for $1\le i<j\le k$ as the independent Mellin variables
subject to the single constraint (\ref{constraintM1}) (recall that $\g_{ii}=-\D_i$).
From (\ref{shiftO}), we conclude that the Mellin amplitudes $M^a$ are constrained by 
\be
\sum_{a=1}^k\g_aM^a=0\ .
\label{constraintMa}
\ee

Another possible Mellin representation for the correlator (\ref{corrVector}) is 
\be 
\sum_{a=1}^k D_{a} 
  \int [d\g]\,
 \check{M}^a
 \prod_{i,j=1\atop i<j}^k \frac{\G(\g_{ij})}{(-2P_i\cdot P_j)^{\g_{ij}}}
\prod_{i=1}^k \frac{\G(\g_{i}+\d_i^{a})}{(-2P_i\cdot P)^{\g_{i}+\d_i^{a} }}\ ,
  \label{Mcheckrep}
\ee 
where the  $D_{a}$ is the following differential operator, 
\begin{align} 
D_{a}= (P \cdot P_a )(Z\cdot \partial_P )-  (Z\cdot P_a)(P\cdot \partial_P -Z\cdot\partial_Z)  \ . \label{eq:McheckSpinJ}
\end{align}
This was suggested in \cite{PaulosMellin} from the study of Witten diagrams.
In this representation, the Mellin variables obey the same constraints (\ref{gammai}) and (\ref{constraintM1}).
Acting with the differential operator, it is not hard to see that the two representations are related through
\be \label{ChangeRepres}
  M^a
 =  \sum_{b=1}^k 
 \g_b\left( \check{M}^a -   \check{M}^b\right)\ .
\ee
Notice that the constraint (\ref{constraintMa}) on  $M^a$ is automatic in terms of $\check{M}^a$. 
On the other hand, the second description $\check{M}^a$ is redundant because the shift
\be
\check{M}^a \to \check{M}^a + \Lambda 
\ee
leaves $M^a$ invariant for any function $\L$ of the Mellin variables $\g_{ij}$.
Since these two representations are equivalent, we shall use them according to convenience. For example, $M^a$ seems to be more useful to formulate factorization and impose conservation, while $\check{M}^a$ leads to a simpler formula for the flat space limit.

\subsection{Tensor operator}

Let us now generalize the Mellin representation for the correlator (\ref{corrVector}) involving one primary operator $\Ocal(P,Z)$ with spin $J$.
The first representation is   
\begin{align} 
\sum_{a_1,\dots,a_J=1}^k 
\left(\prod_{\ell=1}^J ( Z \cdot P_{a_\ell})\right) 
  \int [d\g]\,
 M^{\{a\}}  \prod_{i,j=1\atop i<j}^k \frac{\G(\g_{ij})}{(-2P_i\cdot P_j)^{\g_{ij}}}
\prod_{i=1}^k \frac{\G(\g_{i}+\{a\}_i)}{( -2P_i\cdot P)^{\g_{i}+\{a\}_i}}  \label{SpinJRep1}
\end{align}
where $\{ a\}$ stands for the set $a_1,\dots,a_J$ and $\{ a\}_i$ counts the number of occurrences of $i$ in the list $a_1,\dots,a_J$, \emph{i.e.}
\be
\{ a\}_i=\d_i^{a_1}+\dots+\d_i^{a_J} \ . 
\ee
The constraints on the Mellin variables are
\be
\g_{i}= -\sum_{j=1}^k \g_{i j}\ ,
\qquad
\g_{ij}=\g_{ji}\ , \qquad
\g_{ii}=-\D_i\ ,
\qquad
\sum_{i,j=1}^k \g_{i j}=J-\D\ ,
\label{constraintMJ}
\ee
and the Mellin amplitudes are symmetric under permutations of the indices $a_1, \dots, a_J$ and obey
\be \label{TransverseJ} 
\sum_{a_1=1}^k (\g_{a_1}+\d_{a_1}^{a_2}+\d_{a_1}^{a_3}+\dots+\d_{a_1}^{a_J} ) M^{a_1 a_2 \dots a_J}=0\ .
\ee

The generalization of the second representation is
\begin{align} 
\sum_{a_1,\dots,a_J=1}^k 
\left(\prod_{\ell=1}^J D_{a_\ell} \right) 
  \int [d\g]\,
 \check{M}^{\{a\}}  \prod_{1\le i<j\le k} \frac{\G(\g_{ij})}{(-2P_i\cdot P_j)^{\g_{ij}}}
\prod_{1\le i\le k} \frac{\G(\g_{i}+\{a\}_i)}{(-2P_i\cdot P)^{\g_{i}+\{a\}_i}} \ .
 \label{SpinJRep2}
\end{align}
Since $[D_{a},D_{b}]=0$, we can choose $\check{M}^{a_1 \dots a_J}$   invariant under permutation  of the indices $a_l$. 
Moreover, from the identity
\begin{align}  
\sum_{a_1=1}^k   D_{a_1}  
\prod_{1\le i\le k} \frac{\G(\g_{i}+\{a\}_i)}{(-2P_i\cdot P)^{\g_{i}+\{a\}_i}} =0
\end{align}
we conclude that the correlator is invariant under
\be
\check{M}^{a_1\dots a_J} \to
\check{M}^{a_1\dots a_J}+\sum_{m=1}^J
\L^{a_1\dots a_{m-1}a_{m+1}\dots a_J}\ ,
\label{redundancyMcheck}
\ee
where $\L^{a_2\dots a_J}$ is any function of the Mellin variables that depends on one less index.
Notice that this is the direct analogue of the redundancy (\ref{RedundancySpinJT}) of scattering amplitudes.

To see how is the relation between the two representations we first give the example of $J=2$ \begin{align}
M^{a_1 a_2}&=
\sum_{b_1,b_2=1}^k (\g_{b_1}+\d_{b_1}^{a_2}+\d_{b_1}^{b_2})(\g_{b_2}+\d_{b_2}^{a_1}) \left[\check{M}^{a_1 a_2}-\check{M}^{a_1 b_2}-\check{M}^{b_1 a_2}+\check{M}^{b_1 b_2} \right]\nonumber \\
&=\sum_{b_1, b_2=1}^k (\g_{b_1}+\d_{b_1}^{a_2})(\g_{b_2}+\d_{b_2}^{a_1}+\d_{b_2}^{b_1}) \left[\check{M}^{a_1 a_2}-\check{M}^{a_1 b_2}-\check{M}^{b_1 a_2}+\check{M}^{b_1 b_2} \right]  \label{MtoMcheckJ2}
\end{align}
From (\ref{MtoMcheckJ2}) it is clear that 
the map
\be
\check{M}^{a b}(\g_{ij}) \rightarrow \check{M}^{a b}(\g_{ij}) + \left[\Lambda^{a}(\g_{ij})+\Lambda^{b}(\g_{ij}) \right]
\ee
leaves $M^{a b}$ invariant.
Moreover, since $\check{M}^{a_1 a_2}-\check{M}^{b_1 a_2}-\check{M}^{a_1 b_2}+\check{M}^{b_1 b_2} $ is antisymmetric in the exchange of $a_1 \leftrightarrow b_1$, it is immediate to see that the constraint
\be
\sum_{a_1} (\g_{a_1}+\d_{a_1}^{a_2}) M^{a_1 a_2}=0
\ee
is automatically satisfied by the $\check{M}$ representation.

For a generic $J$ we conjecture that the relation between the two representations is
\be
  M^{\{ a\}}= \sum_{\{b\}=1}^k \sum_{q=0}^J (-1)^q \check{M}^{\{b\}\{a\}}_{q} \prod_{\ell=1}^{J}\biggl{(} \g_{b_\ell}+\sum_{\myatop{\ell_1=1}{\ell_1 \neq \ell}}^J \d_{b_\ell}^{a_{\ell_1}}+\sum_{\ell_2=\ell+1}^J \d_{b_\ell}^{b_{\ell_2}}\biggr{)}    ,
  \label{MofMcheck}
\ee
where, given the permutation group $S_J$ of $J$ elements, we defined 
\be \label{Mpermutations}
\check{M}^{\{b\}\{a\}}_{q}=\frac{1}{q! (J-q)!}\sum_{\s \in S_J} \check{M}^{b_{\s(1)}\dots b_{\s(q)}a_{\s(q+1)} \dots a_{\s(J)}}
\ee
as the sum over all possible terms with $q$ indices from the set $\{a\}$ and $J-q$   indices from the set $\{b\}$. This formula was built as a generalization of (\ref{ChangeRepres}) and (\ref{MtoMcheckJ2}) in a way such that the following two properties hold: the constraint (\ref{TransverseJ}) is automatically satisfied for any $\check{M}^{a_1 \dots a_J}$ and (\ref{MtoMcheckJ2}) is invariant under the redundancy (\ref{redundancyMcheck}).

\subsection{Conserved currents}

When the vector operator is a conserved current (thus $\D=d-1$), its correlation functions must satisfy
\be
\frac{\partial}{\partial P} \cdot 
\frac{\partial}{\partial Z}
\langle\Ocal(P,Z)\Ocal_1(P_1)\dots \Ocal_k(P_k) 
\rangle=0\ .
\ee
Using the representation (\ref{Mrep}), this is equivalent to 
\be \label{ConservedCurrentJ1}
\sum_{ a, b=1 \atop a\neq b}^k \g_{ab} \,[M^a ]^{ab}=0\ ,
\ee
where, given a function $f $ of the variables $\g_{i j}$, we define
\begin{equation} 
\label{def:SquareBrackets}
[f(\g_{i j })]^{a b}=f(\g_{i j } +\d_{i}^a\d_{j}^b+\d_{j}^a\d_{i}^b)\ , 
\qquad [f(\g_{i j })]_{a b}=f(\g_{ij }-\d_{i}^a\d_{j}^b-\d_{j}^a\d_{i}^b)\ .
\end{equation} 
The variables $\g_{ab}$ in equation (\ref{ConservedCurrentJ1}) are subject to the constraints
\be
\g_{ab}=\g_{ba}\ , \qquad
\g_{aa}=-\D_a\ , \qquad
\sum_{a,b=1}^k\g_{ab}=-d\ ,
\label{shiftedgammas}
\ee
so that the arguments of the Mellin amplitude 
$M^a$ satisfy the constraint (\ref{constraintM1}).

A conserved tensor (with $\D=J+d-2$) satisfies a similar equation \cite{SpinningCC},
\be
\frac{\partial}{\partial P_{M}} D^{M}_Z
\langle\Ocal(P,Z)\Ocal_1(P_1)\dots \Ocal_k(P_k) 
\rangle=0\ ,
\ee
where $D^{M}_Z$ is the  operator defined by
\begin{align}
D_Z^{M}=\left(\frac{d}{2}-1+Z\cdot\frac{\partial}{\partial Z}\right)\frac{\partial}{\partial Z_{M}}-\frac{1}{2} Z^{M} \frac{\partial}{\partial Z}\cdot\frac{\partial}{\partial Z}\ .
\label{TodorovOper}
\end{align}
This  imposes an additional constraint on the Mellin amplitudes,   
\begin{align} \label{ConservedCurrentRelation}
(2J+d-4)\sum_{a, b=1 \atop a\neq b}^k\g_{a b}[M^{a c_2\dots c_J}]^{a b}=
(J-1)
\sum_{a, b=1 \atop a\neq b}^k\g_{a b}[M^{a b c_3\dots c_J}]^{a b}\ ,
\end{align}
where the variables $\g_{ab}$ satisfy the constraints (\ref{shiftedgammas}).

\section{Factorization from the shadow operator formalism}
\label{sec:FacShadow}

Using the multiple OPE (\ref{OPE}), one can write a CFT $n$-point function  as a sum over the contribution of each primary operator (and its descendants), as written in equation (\ref{npointCB}). 
As explained in \cite{NaturalMellin}, each term in this sum gives rise to a series of poles in the  Mellin amplitude.
In this section, our strategy to obtain these poles is to use the \emph{projector} 
\cite{Ferrara:1972uq, SimmonsDuffin:2012uy}
\be
|\Ocal| = \frac{1}{\Ncal_{\D}} \int
d^d y d^d z|\Ocal(y) \rangle 
\frac{\G(d-\D)}{(y-z)^{2(d-\D)}}
\langle \Ocal(z)|
\label{ConformalProjector}
\ee
inside the correlation function.
The conformal integral
\be
\frac{1}{\Ncal_\D}\int d^dyd^dz
\langle \Ocal_1(x_1)\dots \Ocal_k(x_k) \Ocal(y) \rangle \frac{\G(d-\D)}{(y-z)^{2(d-\D)}} \langle \Ocal(z) \Ocal_{k+1}(x_{k+1})\dots \Ocal_n(x_n)\rangle\ ,
\label{CBplusShadow}
\ee
gives  the contribution of the operator $\Ocal$ in the multiple OPE of $\Ocal_1 \dots \Ocal_k $, to the $n$-point function $\langle \Ocal_1(x_1)\dots \Ocal_n(x_n)\rangle$. 
In fact, this integral  includes an extra shadow contribution, which  can be removed by doing an appropriate monodromy projection \cite{SimmonsDuffin:2012uy}.
Fortunately, if we are only interested in the poles of the Mellin amplitude, this monodromy projection is very simple to perform in Mellin space.
The reason is that  the Mellin amplitude of (\ref{CBplusShadow}) has poles 
associated with the operator $\Ocal$ and other poles associated with its shadow. 
Therefore, the monodromy projection amounts to focusing on the first set of poles.
This follows from the fact that each series of poles in Mellin space gives rise to a power series expansion in position space with different monodromies.


Let us start by considering the case where the exchanged operator $\Ocal$ is a scalar.
 If we normalize the operator to have unit two point function $\langle \Ocal(x) \Ocal(0)\rangle=|x|^{-2\D}$, the normalization constant
 $\Ncal_\D$  is given by 
\be
\Ncal_\D=\frac{\pi^d \,
\G\left(\D-\frac{d}{2}\right)
\G\left(\frac{d}{2}-\D \right)}{\G(\D)}\ ,
\ee
as we show in appendix \ref{ap:projector}.
We start by writing the correlation functions that appear in (\ref{CBplusShadow}) in the Mellin representation,
\be
\langle \Ocal_1(x_1)\dots \Ocal_k(x_k) \Ocal(y) \rangle = \int [d\l ]\,M_L  \prod_{1\le a<b\le k} \frac{\G(\l_{ab})}{(x_{ab}^2)^{\l_{ab}}}
\prod_{1\le a\le k}  
\frac{\G(\l_{a})}{(x_{a}-y)^{2\l_{a}}}
\label{MellinLeft}
\ee
where
\be
\l_{a}= -\sum_{b=1}^k \l_{ab}\ ,
\qquad
\l_{ab}=\l_{ba}\ ,
\qquad
\l_{aa}=-\D_a\ ,
\qquad
\sum_{a,b=1}^k \l_{ab}=-\D\ .
\label{tildeconstraint}
\ee
The integration measure $[d\l]$ denotes $(k-2)(k+1)/2$ integrals running parallel to the imaginary axis, over the independent variables $\l_{ab}$ that remain after solving the constraints (\ref{tildeconstraint}).
Similarly, the  second correlator in (\ref{CBplusShadow}) reads
\be
\langle \Ocal(z) \Ocal_{k+1}(x_{k+1})\dots \Ocal_n(x_n)  \rangle = \int [d\r]\,M_R \prod_{k< i<j\le n} \frac{\G(\r_{ij})}{(x_{ij}^2)^{\r_{ij}}}
\prod_{k< i\le n}  
\frac{\G(\r_{i})}{(x_{i}-z)^{2\r_{i}}}
\label{MellinRight}
\ee
where
\be
\r_{i}= -\sum_{j=k+1}^n \r_{ij}\ ,
\qquad
\r_{ij}=\r_{ji}\ ,
\qquad
\r_{ii}=-\D_i\ ,
\qquad
\sum_{i,j=k+1}^n \r_{ij}
=-\D\ .
\ee
We use $a,b$ to label the first $k$ points of the $n$-point function and $i,j$ to denote the other $n-k$ points.
In appendix \ref{ap:projectorScalar}, we
insert (\ref{MellinLeft}) and (\ref{MellinRight}) in (\ref{CBplusShadow}) and simplify  the resulting integral in Mellin space,
until we arrive at
\begin{align}
M_\Ocal =&\frac{\pi^d}{\Ncal_\D}
\frac{\G\left(\frac{2\D-d}{2}\right)\G\left(\frac{d-\D-\g_{LR}}{2}\right)}{\G\left(\frac{\D-\g_{LR}}{2}\right)}
\ F_L\times F_R
\label{MellinBlockPlusShadow}
\end{align}
where $\g_{LR}=\sum_{a=1}^k\sum_{i=k+1}^n \g_{ai}$ and 
\begin{align}
&F_L=
\int  [d\l]  \,
M_L(\l_{ab})  
\prod_{1\le a<b\le k} 
\frac{\G( \l_{ab}) 
\G( \g_{ab}-\l_{ab} )}{\G( \g_{ab})} 
\label{LeftPartB+S}
\\
&F_R=\int    [d\r]\,
M_R(\r_{ij}) 
\prod_{k< i<j\le n} 
\frac{\G(\r_{ij})
\G(\g_{ij}-\r_{ij})}{\G(\g_{ij})}
\end{align}
Expression (\ref{MellinBlockPlusShadow}) is the final result for the Mellin amplitude $M_\Ocal$ of the conformal integral (\ref{CBplusShadow}). 
As expected, the Mellin amplitude $M_\Ocal$ has poles at
$\g_{LR}=d-\D+2m$ for $m=0,1,2,\dots$, which are associated with the shadow of $\Ocal$ . We are not interested in these poles because they are not poles of the Mellin amplitude $M$ of the   physical $n$-point function.
On the other hand,  $M_\Ocal$ has poles at
$\g_{LR}=\D+2m$ for $m=0,1,2,\dots$, which are also present in the $n$-point Mellin amplitude $M$
with exactly the same residues.
Our goal is to compute these residues.
From (\ref{MellinBlockPlusShadow}), we conclude that both $F_L$ and $F_R$ must have simple poles at $\g_{LR}=\D+2m$.
In appendix \ref{ap:projectorScalar}, we   deform the integration contours in (\ref{LeftPartB+S}) and arrive at the following formula for the residues of $F_L$,
 \be
 F_L\approx \frac{-2(-1)^m}{\g_{LR}-\D-2m} \sum_{n_{ab}\ge 0 \atop\sum n_{ab}=m} 
M_L(\g_{ab}+n_{ab})  
\prod_{1\le a<b\le k} \frac{ 
\left(\g_{ab}\right)_{n_{ab}}}{n_{ab}!}\ .
\ee
We can now return to (\ref{MellinBlockPlusShadow}) and  conclude that the poles of the Mellin amplitude associated with the exchange of a scalar operator $\Ocal$ (of dimension $\D$) between the first $k$ operators and the other $(n-k)$ operators of a $n$-point function, is given by
equations (\ref{MellinFacPoles}--\ref{LmSum}).


With this method, one can find similar factorization formulas for the residues of poles associated with tensor operators.
One just needs to generalize the projector 
(\ref{ConformalProjector}) for tensor operators and perform several conformal integrals using (a generalized) Symanzik's formula.
We describe this calculation in appendix \ref{app:FacVector} for the case of vector operators.
The result is that the residues $\mathcal{Q}_m$ in equation (\ref{MellinFacPoles}) are given by
\begin{align}
\label{spin1final} 
\mathcal{Q}_m=&
\frac{\D \G(\D-1)m!}{ \left(1+\D-\frac{d}{2}\right)_m} 
\sum_{a=1}^k \sum_{i=k+1}^n 
  L_m^a\,R_m^i
  \left[\g_{ai}
  +\frac{d-2\D}{2m(\D-d+1)} \sum_{b=1}^k
  \g_{ab}
  \sum_{j=k+1}^n\g_{ij}\right]
 \end{align}
where
\be \label{def:Lma}
L_m^a = \sum_{n_{ab}\ge 0 \atop \sum n_{ab}=m} 
M^a_L(\g_{ab}+n_{ab})  
\prod_{1\le a<b\le k} \frac{ 
\left(\g_{ab}\right)_{n_{ab}}}{n_{ab}!}
\ee
is constructed in terms of the Mellin amplitude $M_L^a$  of the correlator of the first $k$ scalar operators and the exchanged vector operator, as defined in (\ref{Mrep}) and similarly for the right factor   $R_m^i$.
Notice that the second term vanishes for $m=0$ due to the transversality constraint (\ref{constraintMa}). This leads to a particularly simple formula for the residue of the first pole
\begin{align} 
\mathcal{Q}_0=&
\D\G(\D-1)   
\sum_{a=1}^k \sum_{i=k+1}^n \g_{ai}
  M_L^a\,M_R^i \ . 
 \end{align}
In appendix \ref{ap:Spin2FacProj} we extend this method to compute factorization formulas for operators with spin 2. However, the calculations quickly become very lengthy as spin increases. 
In the next section, we shall describe an alternative  method to achieve the same goal.
\section{Factorization from the conformal Casimir equation}
\label{sec:FacCasimir}
Given a $n$-point function, we can perform a multiple OPE expansion of the first $k$ operators as described in (\ref{npointCB}) to obtain a sum over the contributions of the  exchanged primary operators $\Ocal_p$ and their descendants, 
\be
\langle \Ocal_1(P_1) \dots \Ocal_n(P_n) \rangle=
\sum_p  G_p(P_1, \dots, P_n)\ .
\ee
Let us define the conformal Casimir for the firsts $k$ operators as
\begin{equation}
 \mathscr{C}=\frac{1}{2} \left[\sum_{i=1}^k \mathscr{J} _i\,\right]^2,
\end{equation}
where 
\begin{equation}
\mathscr{J}_{A B}= P_A \frac{\partial}{\partial P^B}-P_B \frac{\partial}{\partial P^A}, \qquad P\in \mathbb{M}^{d+2}\ ,
\end{equation}
are the generators of the Lorentz group acting on the embedding space   $\mathbb{M}^{d+2}$. Then each $G_p(P_1, \dots, P_n)$ is an eigenfunction of the Casimir $\mathscr{C}$ with eigenvalue 
$$c_{\Delta J}=\Delta (\Delta - d) +J (J + d - 2)\ ,$$
where $J$ and $\D$ are the spin and the conformal dimension of the exchanged operator $\Ocal_p$, \emph{i.e.}
\begin{equation}
\mathscr{C} G_p(P_1, \dots, P_n)=- c_{\Delta J} G_p(P_1, \dots, P_n) \ .
\end{equation}
This equation takes a simpler form in Mellin space. In fact, the  Mellin transform $M_p(\g_{ij})$ of $G_p(P_1, \dots, P_n)$ has to satisfy the following shifting relation  \cite{NaturalMellin}
\begin{align} \label{shiftrelJ}
&[(\g_{LR} -\Delta)(d -\Delta-\g_{LR})+J(J+d-2)] M_p +\\
&\qquad\qquad+ \sum_{a, b=1 \atop a\neq b}^k
\sum_{i ,j=k+1  \atop i\neq j}^n \left[ \g_{a i}\g_{bj} \left(M_p- [M_p]^{a i, b j}_{a j, b i}\right)+\g_{ab}\g_{ij} [M_p]^{a b, i j}_{a i, b j}\right]=0\ ,
\nonumber
\end{align} 
where we recall that $\g_{LR}=\sum_{a=1}^k\sum_{i=k+1}^n \g_{ai}$ 
and that the definition for the square brackets was given in (\ref{def:SquareBrackets}).
The Mellin amplitude $M_p(\g_{\m \n})$ has the following pole structure 
\begin{align}
M_p \approx 
\frac{\mathcal{Q}_m  }{\g_{LR}-\left(\Delta+2m-J\right)}\ ,\ \ \ \ \ \ \ 
m=0,1,2,\dots \ .
\label{polesMp}
\end{align} 
where the residues $\mathcal{Q}_m$ are functions of the Mellin variables $\g_{\m \n}$ which satisfies the \emph{on shell} condition $\g_{LR}=\D+2m-J$. Therefore, the full Mellin amplitude $M=\sum_p M_p$ will also have these poles with the same residues. \footnote{
In an   interacting CFT, we do not expect that different primary operators give rise to coincident poles.
}
Studying equation (\ref{shiftrelJ}) close to the poles (\ref{polesMp}), we obtain an  equation for the residues
$\mathcal{Q}_m$,
which can be written as
\be 
\label{eq:CasimirEquationQm}
\hat{C}\left( \mathcal{Q}_m \right)=0 \ ,
\qquad m=0,1,2,\dots\ ,
\ee
where $\hat{C}$ is the  operator 
\begin{align} \label{CasimirOperatorJm}
\hat{C}(\mathcal{Q}_m)\equiv\eta \mathcal{Q}_m + \sum_{a, b=1 \atop a\neq b}^k
\sum_{i ,j=k+1  \atop i\neq j}^n \left[ \g_{a i} \g_{bj} \left(\mathcal{Q}_m- [\mathcal{Q}_m]^{a i, b j}_{a j, b i}\right)+\g_{ab}\g_{ij} [\mathcal{Q}_{m-1}]^{a b, i j}_{a i, b j}\right]
\end{align}
and $\eta=(2 m-J) (d - 2 \Delta - 2 m+J)+J(J+d-2)$.
In particular, we notice that for $m>0$ (\ref{eq:CasimirEquationQm}) is a recurrence equation for $\mathcal{Q}_m$ in terms of $\mathcal{Q}_{m-1}$, while for  $m=0$ (\ref{eq:CasimirEquationQm}) reduces to a
constraint on $\mathcal{Q}_0$ (since $\mathcal{Q}_{-1}=0$). 
In the rest of this section, we present a way to find $\mathcal{Q}_m$ using  
(\ref{eq:CasimirEquationQm}).
\subsection{Factorization for   scalar exchange}

In the scalar case  it is natural to guess a factorization formula of the kind
\be
\mathcal{Q}_{m}=  \k_{\D 0}\frac{ m!}{(1-\frac{d}{2}+\D)_m} L_m R_m\ ,
\ee
where $L_m$ and  $R_m$ are respectively functions of the Mellin variables on the left ($\g_{ab}$ with $a,b=1,\dots k$) and on the right  ($\g_{ij}$ with $i,j=k+1,\dots n$) such that $L_0=M_L$ and $R_0=M_R$.
The  overall constant $\k_{\D\, 0}$   will be fixed later and $ \frac{m!}{(1-\frac{d}{2}+\D)_m} $ is a function of $m$ that we introduced for convenience and that could in principle  be absorbed in the definition of $L_m$ and  $R_m$.
Since $\mathcal{Q}_{m}$ does not depend on the mixed variables $\g_{a i}$ (with $a=1,\dots k$ and   $i=k+1,\dots n$), then $[\mathcal{Q}_{m}]^{a i, b j}_{a j, b i}=\mathcal{Q}_{m}$ trivially.
Therefore, equation (\ref{eq:CasimirEquationQm}) reduces to
 \begin{equation} \label{resrel1}
 2 m  (d - 2 \Delta - 2 m) \mathcal{Q}_m + \sum_{a, b=1 \atop a\neq b}^k
\sum_{i ,j=k+1  \atop i\neq j}^n   \g_{ab}\g_{ij} [\mathcal{Q}_{m-1}]^{a b, i j} =0\ .
\end{equation}
This equation is automatically satisfied for $m=0$.
Notice that the ansatz $\mathcal{Q}_{m}$ is consistent with equation  (\ref{resrel1}).
In fact, given a $\mathcal{Q}_{m-1}$ factorized in functions of left and right Mellin variables, (\ref{resrel1}) implies that $\mathcal{Q}_{m}$ 
is also factorized in the same way. 
Replacing $\mathcal{Q}_{m}$ in (\ref{resrel1}) we get a recurrence equation for $L_m$ and $R_m$
\begin{equation} \label{resrel2}
L_m R_m =\Big(\frac{1}{2m} \sum_{a, b=1 \atop a\neq b}^k
 \g_{ab} [L_{m-1}]^{a b}\Big) \Big(\frac{1}{2m}\sum_{i ,j=k+1  \atop i\neq j}^n\g_{ij} [R_{m-1}]^{i j}\Big)\ ,
\end{equation}
which can be solved separately for  $L_m$ and $R_m$ in terms of $L_0=M_L$ and $R_0=M_R$.
In appendix \ref{App:DemRec1}, we show that  
\begin{align} \label{LmRecScalar}
L_m&=\frac{1}{2m} \sum_{a, b=1 \atop a\neq b}^k \g_{ab} [L_{m-1}]^{a b}& \Longleftrightarrow& \qquad L_m=
 \sum_{n_{ab}\ge 0 \atop \sum  n_{ab}=m} 
M_L(\g_{ab}+n_{ab})  
\prod_{a,b=1 \atop a<b}^k \frac{ 
\left(\g_{ab}\right)_{n_{ab}}}{n_{ab}!}
\end{align}
and similarly for $R_m$. The final result exactly matches (\ref{Eq:FacScalar}) up to an overall factor that cannot be fixed by the Casimir equation, since it is a homogeneous equation. We will discuss how to fix the normalization in section \ref{sec:FourPtFactorization}.
\subsection{Factorization for vector exchange}
For spin $J=1$ the left and right Mellin amplitudes can be represented as functions $M_L^a$ and  $M^i_R$ that satisfy the transersality condition $\sum_{a=1}^k \g_{a} M_L^a=0$ as discussed in (\ref{constraintMa}), where we recall that $\g_{a}=-\sum_{b=1}^k \g_{ab}=\sum_{i>k}^n \g_{a i}$.

The solution of (\ref{eq:CasimirEquationQm}) should depend on the left and the right Mellin amplitudes $M_L^a$ and $M_R^i$ in a
form invariant under permutations of the left points $P_a$ with $a=1, \dots, k$ and of the right points $P_i$ with $i=k+1,\dots,n$.
Considering that the scalar solution takes the form (\ref{sec:FourPtFactorization}), the first natural ansatz for the vector case is
\be
\mathcal{Q}^{(1)}_{m}=  \sum_{a=1}^k\sum_{i= k+1}^n\g_{a i} L^a_m R^i_m \ ,
\ee
with $L_m^a$ and  $R_m^i$  defined in (\ref{def:Lma}). This ansatz is actually the complete solution in the case $m=0$, but it fails to solve the Casimir equation (\ref{eq:CasimirEquationQm}) for higher $m$. 
In fact,
acting with the Casimir operator (\ref{CasimirOperatorJm}) on the ansatz  $ \mathcal{Q}^{(1)}_{m}$ times a function $f^{(1)}_m$ that does not depend on the Mellin variables, we find
(see appendix \ref{App:J=1CasimirFac})
\begin{align}
 \label{C(Q1)J=1}
\hat{C}(f^{(1)}_m \mathcal{Q}^{(1)}_{m})=&2m \left( (d-2\D-2m)
 f^{(1)}_m+2m f^{(1)}_{m-1}\right)\mathcal{Q}^{(1)}_{m} +2\left(f^{(1)}_m-f^{(1)}_{m-1}\right)\mathcal{Q}^{(2)}_{m}
\end{align}
where
\be
\mathcal{Q}^{(2)}_{m}=\dot{L}_m \dot{R}_m
\ ,\qquad{\rm with}\qquad
\dot{L}_m=- \sum_{a=1}^k \g_{a} L^{a}_m 
\ee
and similarly for $\dot{R}_m$. Notice that  $L_0^a= M_L^a$ so that $\dot{L}_0 =- \sum_{a=1}^k \g_{a}M_L^{a}=0$ due to the transversality condition
(\ref{constraintMa}).
Therefore $\mathcal{Q}^{(2)}_{0}=0$ and    $\mathcal{Q}^{(1)}_{0}$ automatically solves (\ref{C(Q1)J=1}) for $m=0$. 
Acting with the Casimir operator (\ref{CasimirOperatorJm}) on $f^{(2)}_{m} \mathcal{Q}^{(2)}_m$ we find (see appendix \ref{App:J=1CasimirFac})
\be
 \label{C(Q2)J=1}
\hat{C}(f^{(2)}_{m} \mathcal{Q}^{(2)}_{m})=\left( \eta f^{(2)}_{m}  + 4 (m-1)^2 f^{(2)}_{m-1} \right) \mathcal{Q}^{(2)}_{m}.
\ee 
Notice that the action of the Casimir operator $\hat{C}$ on the structures $\mathcal{Q}^{(1)}_{m}$ and $\mathcal{Q}^{(2)}_{m}$ closes because it does not produce any new structure.
Thus, we can   find the solution of the problem fixing the functions $f^{(1)}_m$ and  $f^{(2)}_{m}$ such that 
\be \label{C(Q1+Q2)}
\hat{C}(f^{(1)}_m \mathcal{Q}^{(1)}_{m}+f^{(2)}_m \mathcal{Q}^{(2)}_{m})=0. 
\ee
Since $\mathcal{Q}^{(1)}_{m}$ and $\mathcal{Q}^{(2)}_{m}$ are linearly independent,  we   need to set to zero the coefficients multiplying each structure in (\ref{C(Q1+Q2)}).
Setting to zero the coefficient of $\mathcal{Q}^{(1)}_{m}$ we get a recurrence relation for $f^{(1)}_{m}$,
\be \label{coeffRes1}
(d-2\D-2m)f^{(1)}_m+2m f^{(1)}_{m-1}=0\ ,
\qquad m\ge0\ ,
\ee
that can be solved up to an overall constant $f^{(1)}_0$ that we will  call $\k_{\D 1}$,
\be
\label{f1mJ1}
f^{(1)}_m=\k_{\D 1} \frac{m!}{(1-d/2+\D)_m}\ .
\ee
Setting to zero the terms multiplying $\mathcal{Q}^{(2)}_{m}$, we find a recurrence relation for $f^{(2)}_{m}$,
\be \label{coeffRes2}
  \eta f^{(2)}_{m}  + 4 (m-1)^2 f^{(2)}_{m-1} +2\left(f^{(1)}_m-f^{(1)}_{m-1}\right)=0\ ,
\qquad m\ge1\ ,
\ee
that, once we substitute (\ref{f1mJ1}), can be solved as 
\be
f^{(2)}_{m}=\k_{\D 1} \frac{m!}{(1-d/2+\D)_m} \frac{d-2 \Delta }{2 m (\Delta-d+1 )} .
\ee
Therefore the final result is
\be \label{facJ1}
\mathcal{Q}_m=\k_{\D 1}  \frac{m!}{(1-d/2+\D)_m}  \sum_{a=1}^k\sum_{i=k+1}^n L^a_m R^i_m \left( \g_{ai} + \frac{d-2 \Delta } {2 m (\Delta-d+1 )} \g_{a}\g_{i} \right) \ ,
\ee
which matches the result (\ref{spin1final}) that we found in the previous section using the shadow method.  
\subsubsection{Conserved currents}
The conformal dimension of a conserved current is $\D=d-1$. For such a value of $\D$ equation (\ref{facJ1}) naively looks divergent. 
However,  the conserved current relation (\ref{ConservedCurrentJ1})    implies that $\mathcal{Q}^{(2)}_{m}=0$. To see this, we use the fact  that $\dot{L}_m$ can be defined by (formula (\ref{Lhatm2})  as explained in appendix \ref{ap:CasimirFormulasJ1})
\be \label{Lhats}
\dot{L}_m= \sum_{a, b=1 \atop a\neq b}^k \g_{a b} \left[ L^a_{m-1} \right]^{a b} \ , \qquad
m\ge 1 \ . 
\ee
Using   (\ref{Lhats}) it is easy to see that 
\be
\dot{L}_1=
\sum_{a, b=1 \atop a\neq b}^k \g_{a b}
 \left[ M^a_L \right]^{a b}
=0\ ,
\ee
using the conservation constraint (\ref{ConservedCurrentJ1}).
Since $\dot{L}_m$ can also be defined recursively (see (\ref{Lhatm1}) in appendix \ref{ap:CasimirFormulasJ1})
\be \label{Lhats2}
  \dot{L}_m=\frac{1}{2 (m-1)}  \sum_{a, b=1 \atop a\neq b}^k \g_{a b} \left[ \dot{L}_{m-1} \right]^{a b} \ , \qquad m\ge2\ ,
\ee
we conclude that  $\dot{L}_m=0$ for any $m$.
So we can simplify (\ref{facJ1}) as follows
\be \label{facJ1current}
\mathcal{Q}_m=\k_{d-1, 1}  \frac{m!}{
\left(\frac{d}{2}\right)_m} \, \sum_{a=1}^k\sum_{i=k+1}^n  \g_{ai} L^a_m R^i_m.
\ee
\subsection{Factorization for spin $J=2$ exchange}

As in the vector case,   a natural way to construct the ansatz for $J=2$ is to take the left and right Mellin amplitudes $M^{a b}_L$ and $M^{i j}_R$ and ``contract'' them with two mixed variables $\g_{a i}$ (with $a=1,\dots k$ and   $i=k+1,\dots n$), 
as follows
\be
 \sum_{a,b=1}^k\sum_{i,j=k+1}^n  \g_{ai} \g_{bj}L_m^{a b} R_m^{i j} \ ,
\ee
where $L^{a b}_m$ is defined by 
\be
L^{a b}_m=\sum_{{n_{ef}\ge 0 \atop \sum n_{ef}=m}}M^{a b}_L(\g_{e f}+n_{e f})\prod_{e,f=1 \atop e<f}^k \frac{(\g_{e f})_{n_{e f}}}{n_{e f}!}
 \ee
 and similarly for $R_{m}^{i j}$.
This time our guess does not work even for $m=0$
but this is easily fixed including the structure
$\sum \g_{ai}M_L^{aa}M_R^{ii}$, as explained in appendix \ref{App. SpinJ=2 Fact Formula}.
The final result can be written in a concise way as
\be \label{ResJ2m0}
\mathcal{Q}_{0} =\k_{\D 2} \sum_{a,b=1}^k\sum_{i,j=k+1}^n    \g_{ai}(\g_{bj}+\d^a_b \d^i_j) M^{a b}_L M^{i j}_R \ ,
\ee
where $\k_{\D 2}$ is an overall constant not yet fixed. 

To find the result for general $m$ we use the same idea as in the vector case. We promote $\mathcal{Q}_{0} $ to be a function of $m$   replacing $M_{L}^{a b}$ by $L^{a b}_m$ and $M_{R}^{ij}$ by $R^{ij}_m$. Then we act on this ansatz with the Casimir operator $\hat{C}$  and we get new structures until the action of $\hat{C}$ closes.  The structures that we find are (see appendix \ref{App. SpinJ=2 Fact Formula})
\begin{align} 
\label{QsJ=2}
\begin{split}
& \mathcal{Q}^{(1)}_{m} =  \sum_{a,b=1}^k\sum_{i,j=k+1}^n    \g_{ai}(\g_{bj}+\d^a_b \d^i_j) L^{a b}_m R^{i j}_m \ ,
 \qquad \qquad\mathcal{Q}^{(2)}_{m}= \sum_{a=1}^k\sum_{i=k+1}^n  \g_{ai} \dot{L}_m^a \dot{R}_m^i \ , \\
& \mathcal{Q}^{(3)}_{m}=\ddot{L}_m \ddot{R}_m \ ,
\qquad \qquad \, \, \, \,
\mathcal{Q}^{(4)}_{m}= \tilde{L}_m \ddot{R}_m + \ddot{L}_m  \tilde{R}_m \ ,
\qquad \qquad \, \, \, \,
\mathcal{Q}^{(5)}_{m} = \tilde{L}_m \tilde{R}_m\ , 
\end{split}
\end{align}
where
 \begin{align} 
 \dot{L}_m^a
=\sum_{b=1}^k (\g_{b}+\d^a_b )L^{a b}_m \ , \qquad 
\ddot{L}_m=\sum_{a=1}^k \g_{a} \dot{L}_m^{a}  \ , \qquad
 \tilde{L}_m=\sum_{a ,b=1 \atop a\neq b}^k  \g_{a b} \left[L^{a b}_{m-1}\right]^{a b}  \ .
 \end{align}
Finally, we take a linear combination of all the structures and  fix all the coefficients imposing  (\ref{eq:CasimirEquationQm}), as detailed in appendix \ref{App. SpinJ=2 Fact Formula}.
The final result is  
\be \label{finalresultJ2}
\mathcal{Q}_m=\k_{\D 2} \frac{ m!}{\left(1-\frac{d}{2}+\Delta \right)_m}\sum_{s=1}^5 h^{(s)}_m \mathcal{Q}^{(s)}_m\ ,
\ee
with \footnote{Notice that $\mathcal{Q}^{(2)}_{0}=0$ which means that the divergence of $h^{(2)}_{0}$ is immaterial and $\mathcal{Q}_{0}$ is finite. Moreover, since $2\mathcal{Q}^{(3)}_{1}=\mathcal{Q}^{(4)}_{1}=2\mathcal{Q}^{(5)}_{1}$, also the residue $\mathcal{Q}_{1}$ is finite because the combination $h^{(3)}_{m}+2h^{(4)}_{m}+h^{(5)}_{m}$ does not diverge at $m=1$.}
\begin{align}
\begin{split}
&h^{(1)}_m=1 \ , \qquad \qquad \,
h^{(2)}_{m}=\frac{d-2  \Delta}{ m  (\Delta-d )}\ , 
\qquad \qquad \,
 h^{(3)}_m=\frac{  
 d-2\Delta-2   }{4  (m-1) (\Delta-d +1)}
 h^{(2)}_{m} \ , \\ 
&h^{(4)}_{m}=- 
\frac{ 2\Delta-d +2m}{2\Delta-d +2} 
h^{(3)}_m \ ,  \qquad 
h^{(5)}_{m}=-\left[1+2(m-1)\left( 1-\frac{  \Delta(\Delta-1 ) }{d \left(2\Delta-d \right)}\right)\right] h^{(4)}_{m}\ .
\end{split}
\end{align}

%

\subsubsection{Stress-energy tensor}
For the stress-energy tensor  ($\D=d$) we can once again simplify the factorization formula (\ref{finalresultJ2}) using the conserved tensor relation (\ref{ConservedCurrentRelation}). As we show  in appendix \ref{LmJ2Formulas}, one can define   $\dot{L}^c_m$ as
\be 
\dot{L}^c_m=\sum_{a,b=1 \atop a \neq b}^k \g_{a b} \left[L^{a c}_{m-1}\right]^{a b}   \ .
\ee
Then, relation (\ref{ConservedCurrentRelation}) can be rewritten as
\be
\dot{L}^c_1= \frac{\tilde{L}_1}{d} \qquad \forall \; c=1,\dots,n\ .
\ee
Since $\dot{L}^c_m$ and $ \tilde{L}_m  $ satisfy exactly the same recurrence relation
(equations  (\ref{App:RecRelLHatc}) and (\ref{App:RecRelLtilde}) in appendix \ref{LmJ2Formulas}), we conclude that $\dot{L}^c_m=\frac{\tilde{L}_m}{d} $.
Using this fact we can rewrite the structures $ \mathcal{Q}^{(2)}_m, \mathcal{Q}^{(3)}_m, \mathcal{Q}^{(4)}_m$ in terms of $ \mathcal{Q}^{(5)}_m$ as follows
\begin{align}
\mathcal{Q}^{(2)}_m=\frac{\bar\g_{LR}}{d^2}\mathcal{Q}^{(5)}_m\ , \qquad
\mathcal{Q}^{(3)}_m=\frac{\bar\g_{LR}^2}{d^2}\mathcal{Q}^{(5)}_m\ , \qquad
\mathcal{Q}^{(4)}_m=\frac{2 \bar\g_{LR}}{d}\mathcal{Q}^{(5)}_m \ ,
\end{align}
where $\bar\g_{LR}=d+2m-2$.
The final form of the residues is then
 \be \label{finalresultJ2conserved}
\mathcal{Q}_m=\k_{d\, 2} \frac{ m!}{\left( \frac{d}{2}+1 \right)_m} \left[\mathcal{Q}^{(1)}_m-\left(\frac{1}{2m}+\frac{1}{d}\right) \mathcal{Q}^{(5)}_m \right].
\ee

\subsection{Factorization for  general spin $J$ exchange}
The hope of having a closed formula for general spin $J$ and general $m$ was lost after we found the $J=2$ result. The proliferation of different structures and the complicated functions that multiply them do not seem easy to generalize. However, we conjecture the following factorization formula for the residue of the first pole,
\footnote{In the appendix \ref{App:Q0DownIndex} we write (\ref{genericm0}) in a more compact way at the expense of introducing more notation.}
\be
\mathcal{Q}_0=\k_{\D J} \sum_{\{a\} =1}^k \sum_{\{ i\} =k+1}^n 
M_L^{\{a\}} M_R^{\{ i\}} \;
\prod_{\ell=1}^J (\g_{a_\ell i_\ell}+\d_{a_\ell}^{a_{\ell+1}} \d_{i_\ell}^{i_{\ell+1}}+\cdots+\d_{a_\ell}^{a_{J}} \d_{i_\ell}^{i_{J}})
 \ ,
\label{genericm0}
\ee
where $\k_{\D J}$ is a coefficient that we will fix in section \ref{sec:FourPtFactorization} and $\{a\}=a_1 \dots a_J$ and $\{ i\}=i_1 \dots i_J$.
We were not able to prove that (\ref{genericm0}) solves the Casimir equation (\ref{eq:CasimirEquationQm})  in full generality, but we proved it for $J$ up to 7 using \emph{Mathematica} (and also for higher spin $J$ fixing $k$ and $n$). 

The knowledge of (\ref{genericm0}) is itself a fairly interesting result. 
In fact, using (\ref{genericm0}) 
as  the seed of the recurrence relation (\ref{eq:CasimirEquationQm})
and imposing that $\mathcal{Q}_m$ is a polynomial of degree $J$ in the mixed variables $\g_{ai}$,
one can  recursively compute $\mathcal{Q}_m$ for any $m$. 
Moreover, (\ref{genericm0}) encodes the contribution of the exchanged primary operator and all its descendants with minimal twist (dimension minus spin), which dominate in the Lorentzian OPE limit.

\subsection{Factorization of the four point function} \label{sec:FourPtFactorization}
In the four-point function case ($n=4$ and $k=2$),  the residues of the Mellin amplitudes  for any $J$  are known \cite{Mack, Costa:2012cb}.
We want to match this result with
our conjecture (\ref{genericm0}) for the first residue $\mathcal{Q}_0$.
To do so we need to replace the actual expressions for the left and right Mellin amplitudes in (\ref{genericm0}). 
In this case, the left and right Mellin amplitudes are just constants because they correspond to   three-point functions.
We first consider   the left three point function of two scalar operators $\mathcal{O}_1$, $\mathcal{O}_2$ and the exchanged  operator $\mathcal{O}$ with spin $J$ and conformal dimension $\D$ 
\be
\left\langle \mathcal{O}_1(P_1)\mathcal{O}_2(P_2) \mathcal{O}(Z_3,P_3) \right\rangle 
=c_{1 2 \Ocal}\frac{ \big((Z_3\cdot P_1)(-2 P_2\cdot P_3)-(Z_3\cdot P_2)(- 2 P_1\cdot P_3)\big)^{J}}
{(-2 P_1 \cdot P_{2})^{\g_{12}}
(-2 P_{1} \cdot P_{3})^{\g_1+J}
(-2 P_{2} \cdot P_{3})^{\g_2 +J}}\ ,
\ee
where $c_{1 2 \Ocal}$ is the usual structure constant   and 
\begin{align}
\!\!\!\g_{12}=\frac{\Delta_1+\Delta_2-\Delta+J}{2}\ , \qquad \g_{1}=\frac{\Delta_1+\Delta-\Delta_2-J}{2}\ , \qquad \g_{2}=\frac{\Delta_2+\Delta-\Delta_1-J}{2} \ . \label{gammas3pt}
\end{align}
Comparing with the Mellin representation (\ref{SpinJRep1}), we find  (as shown in detail in appendix \ref{Qm=0anyJ})
\begin{align} \label{eq:3ptMellinMaintext}
M_L^{\overset{j}{\scriptsize \overbrace{1\dots 1}} \, {\overset{J-j}{ \scriptsize \overbrace{2\dots 2}}}} =c_{1 2 \Ocal} \frac{ (-1)^{J-j}}{ \Gamma(\g_{12}) \Gamma(\g_1+j) \Gamma(\g_2+J-j)}
\end{align}
and all other components of $M_L^{a_1\dots a_J}$ are related to this by permutations of the indices, which in this case can only take the values 1 or 2.
Replacing this expression for  $M_L$ (and similarly  for $M_R$ with $1 \rightarrow 3$ and $2 \rightarrow 4$)  in   (\ref{genericm0}) we obtain 
 \footnote{We derive this result in appendix \ref{Qm=0anyJ}.
 Actually we had to use a combinatorial identity that we were not able to prove in general but which we verified extensively.}
\begin{align} \label{M0Simple1}
\mathcal{Q}_0
&=\k_{\D J}\frac{c_{1 2 \Ocal} c_{3 4 \Ocal} \,  (-1)^J(\D-1)_J }{\G(\g_{12}) \G(\g_{34}) \prod_{\ell=1}^{4}\G(\g_{\ell}+J )} \; \sum _{j=0}^J (-1)^j   \frac{(\g_{14})_{J-j}   (\g_{23})_{J-j}}{(J-j)!}   \frac{(\g_{13})_{j}  (\g_{24})_{j} }{j!} \ . 
\end{align}
Matching this result with the one computed   in \cite{Costa:2012cb} we find agreement if
\be
\k_{\D J} = (-2)^{1-J} (\D+J-1) \G(\D-1) \ . \label{kappaDeltaJ}
\ee
Even though $\k_{\D J}$ was determined for the case of $k=2$ and $n=4$, we conjecture that $\k_{\D J}$  does not depend on the parameters $k$ and $n$.
This conjecture is supported by the results obtained using the shadow method for $J=0,1,2$. 
Moreover, in appendix \ref{App: FSL general spin J factorization} we find an independent hint that (\ref{kappaDeltaJ}) holds. 
In particular we matched  the leading  behavior at large $\D$ of (\ref{kappaDeltaJ}) asking that, in the flat space limit, formula (\ref{genericm0}) reproduces  a piece of the amplitude factorization (\ref{FlatJFactoriz}).

\section{Flat space limit of AdS}
\label{sec:FSL}

The Euclidean conformal group in $d$ dimensions is isomorphic to $SO(d+1,1)$. The generators of the algebra $ {\mathscr{J}}_{A B}$ (with $A,B=0,1,\dots,d+1$) are antisymmetric and satisfy the usual commutation relations
\be
[ {\mathscr{J}}_{A B}, {\mathscr{J}}_{CD}]=i\left( \eta_{AD}  {\mathscr{J}}_{BC}+\eta_{BC}  {\mathscr{J}}_{A D}-\eta_{AC}  {\mathscr{J}}_{BD}-\eta_{BD}  {\mathscr{J}}_{A C} \right).
\ee
The quadratic and quartic Casimirs of the conformal group are  defined by
\begin{align}
\mathscr{C}^{(2)}=\frac{1}{2}  {\mathscr{J}}^{A B}  {\mathscr{J}}_{A B}\ , \qquad \qquad
\mathscr{C}^{(4)}=\frac{1}{2}  {\mathscr{J}}^{A B}  {\mathscr{J}}_{B C}  {\mathscr{J}}^{C D}  {\mathscr{J}}_{D A}\ .
\end{align}
A conformal primary operator   with scaling dimension $\D$ and spin $J$ is an eigenfunction of the Casimirs with eigenvalues
\begin{align}
c^{(2)}_{\D J}&=\D(\D - d)+J (J+d-2)\ ,\\
c^{(4)}_{\D J}&=\D^2(\D - d)^2+J^2 (J+d-2)^2+\frac{1}{2}(d-1) [d \D (\D-d)+(d-4) J (J+d-2)]\ .
\end{align}

\begin{figure}
\begin{centering}
\includegraphics[scale=0.35]{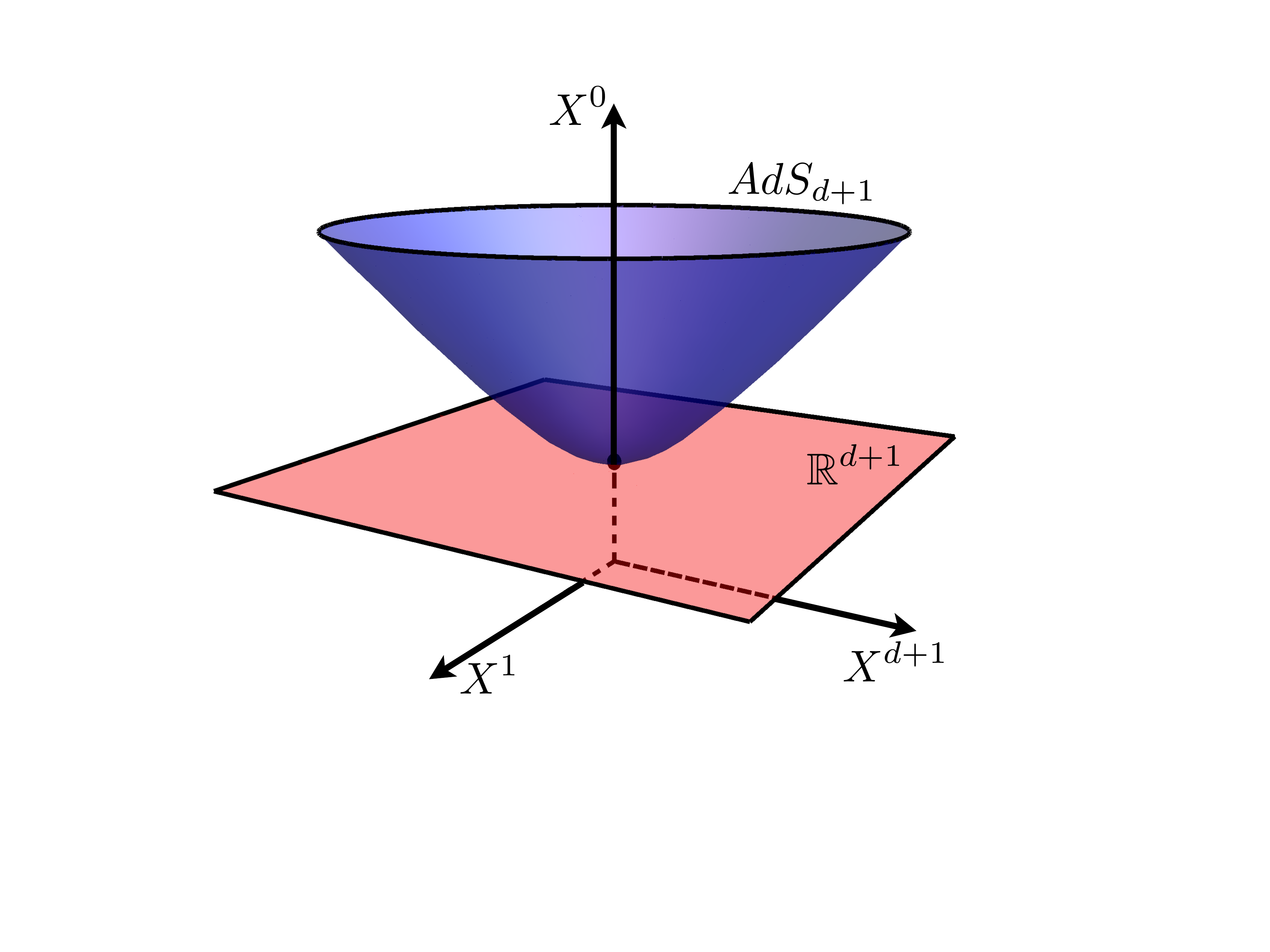}
\par\end{centering}
\caption{\label{fig:FSLAdS}
Euclidean AdS$_{d+1}$ embedded in Minkowski space
$\mathbb{M}^{d+2}$.
The tangent space $\mathbb{R}^{d+1}$ is a good local approximation to AdS in a region smaller than the AdS radius of curvature. 
}
\end{figure}

The group $SO(d+1,1)$ is also the isometry group of Euclidean AdS$_{d+1}$ (or hyperbolic space) defined by the hypersurface
\be
-(X^0)^2+(X^1)^2+\dots+(X^{d+1})^2 = -R^2\ ,
\ee
embedded in $(d+2)$-dimensional Minkowski space.
$R$ is the radius of curvature of AdS.
The flat space limit amounts to    approximating AdS by its tangent space $\mathbb{R}^{d+1}$ at the point $X^A=(R,0,\dots,0)$, as shown in figure \ref{fig:FSLAdS}. 
The Poincaré group of $\mathbb{R}^{d+1}$ is then obtained as the Inönü-Wigner contraction of the conformal group.
The translation generators are given by 
\begin{align}
 {\mathscr{P}}_{\mu}&=\lim_{R\to \infty}\frac{1}{R} {\mathscr{J}}_{\mu\,0},\qquad\qquad
\mu=1,\dots,d+1\ ,
\end{align}  
and the rotation generators are simply given by 
${\mathscr{J}}_{\mu\nu}$ with $\mu,\nu=1,\dots,d+1$.
This leads to the usual Poincaré algebra 
\begin{align}
\left[ {\mathscr{J}}_{\alpha \beta}, {\mathscr{J}}_{\gamma \d}\right]&=i\left( \eta_{\alpha \d}  {\mathscr{J}}_{\beta \gamma}+\eta_{\beta \gamma}  {\mathscr{J}}_{\alpha \d}-\eta_{\alpha \gamma}  {\mathscr{J}}_{\beta \delta}-\eta_{\beta \delta} {\mathscr{J}}_{\alpha \gamma} \right),\\
\left[ {\mathscr{J}}_{\alpha \beta}, {\mathscr{P}}_{\gamma}\right]&=i \left(\eta_{\beta \gamma}  {\mathscr{P}}_{\alpha}-\eta_{\alpha \gamma}  {\mathscr{P}}_{\beta}\right),
\qquad \ \ \ 
\left[ {\mathscr{P}}_{\mu}, {\mathscr{P}}_{\nu}\right]=0\ .
\end{align}
We can also write the Casimirs of the Poincaré group in terms of the  Casimirs of the conformal group,
\footnote{In 4 dimensional flat space, 
$\mathscr{W}_{\mu}=\frac{1}{2} 
\varepsilon_{\mu \nu \sigma \rho} 
\mathscr{P}^{\nu} 
\mathscr{J}^{\sigma \rho} $
is the Pauli-Lubanski pseudovector. }
\begin{align}
\mathscr{P}^2&=\lim_{R\to \infty}\frac{1}{R^2} \mathscr{C}^{(2)} \\
\mathscr{W}^2&\equiv
\frac{1}{2} \mathscr{P}^2 \mathscr{J}_{\mu\nu}^2 +  \mathscr{P}_\a  \mathscr{P}^\g \mathscr{J}^{\a \b} \mathscr{J}_{\b \g}
=\lim_{R\to \infty}\frac{1}{2R^2}\left[\left(\mathscr{C}^{(2)}\right)^2-\mathscr{C}^{(4)}+\frac{1}{2}d(d-1)\mathscr{C}^{(2)}\right].
\end{align}
From the physical point of view, the flat space limit requires  the radius of curvature of AdS to be much larger than any intrisic length $\ell_s$ of the bulk theory.
\footnote{In string theory, the intrinsic length $\ell_s$ can be the string length or the Planck length. }
In the dual CFT, the dimensionless ratio $R/\ell_s\equiv \theta$ is a coupling constant that parametrizes a family of theories.
However, not all observables of a CFT with $\theta \gg 1$ correspond to flat space observables of the dual bulk theory. For example, states of the CFT on the cylinder $S^{d-1}\times \mathbb{R}$ with energy $\D$ of order one, are dual to wavefunctions that spread over the scale $R$ and can not be described in flat space. 
On the other hand, 
states with large energy such that
$\lim_{\theta \to \infty} 
\D /\theta=  \a$   
correspond to states with mass $M=\a /\ell_s$ in flat space.
In fact, for such a state with energy $\D$ and spin $J$, we obtain the usual eigenvalues $M^2$ and $M^2 J(J+d-2)$ of the Casimirs $\mathscr{P}^2$ and $\mathscr{W}^2$, associated with massive particles of spin $J$ in flat space.

%

It is natural to ask what happens to other observables under this flat space limit.
In particular, in this section we will be interested in obtaining scattering amplitudes on $\mathbb{R}^{d+1}$ as a limit of the Mellin amplitudes of the same theory in AdS.
The Mellin amplitudes depend on the Mellin variables $\g_{ij}$ and on the CFT coupling constant $\theta$. As explained in \cite{JPMellin, JLAnalyticMellin}, the flat space limit corresponds to 
\be
\theta \to \infty\ , \qquad
\g_{ij} \to \infty\ , \qquad
{\rm with}\ \ 
\frac{\g_{ij}}{\theta^2}
\ \ {\rm fixed}\,. \label{LimitFSL}
\ee

\begin{figure}
\begin{centering}
\includegraphics[scale=0.4]{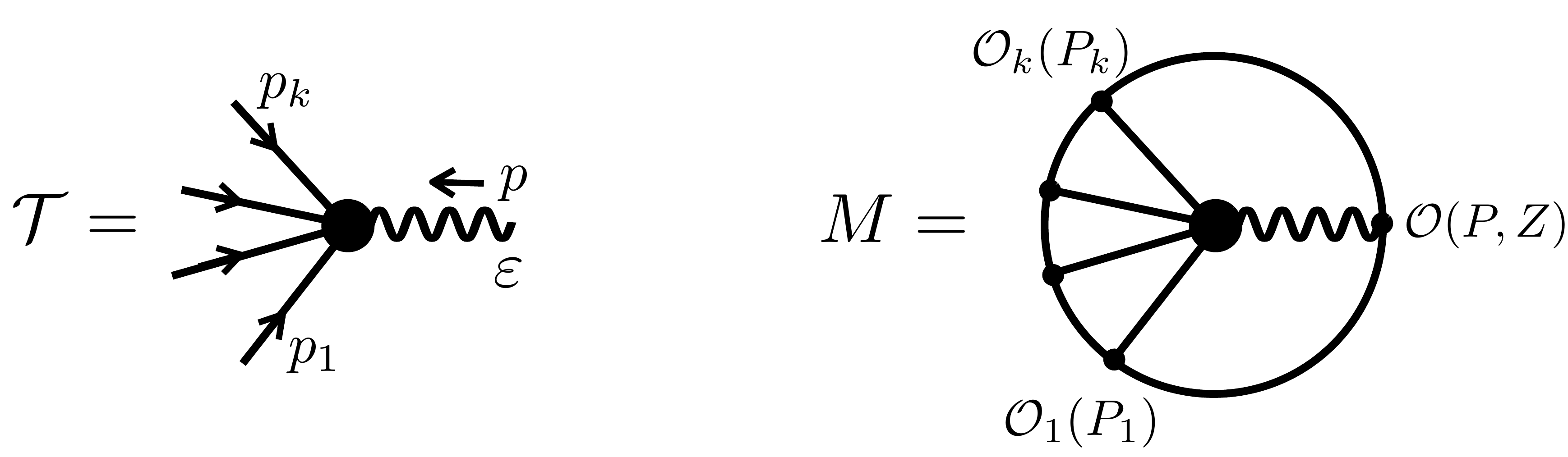}
\par\end{centering}
\caption{\label{fig:FSLidea}
(Left) Scattering amplitude associated with a tree level Feynman diagram with a single interaction vertex. 
(Right) Mellin amplitude associated with tree level Witten diagram using the same interaction vertex but now in AdS.
The flat space limit   of AdS leads to the general relation (\ref{eq:FSLRelationFinalSpinJ}) between the two amplitudes.
}
\end{figure}

We shall consider a local interaction vertex between a set of fields (scalar, vector, etc) and compare the associated tree level scattering amplitude in flat space with the corresponding Mellin amplitude of the CFT correlation function obtained by computing the tree-level Witten diagram using the same vertex in AdS (see figure
\ref{fig:FSLidea}). 
By considering an infinite class of local interactions we will be able to derive a general relation between the scattering amplitude and the Mellin amplitude, as it was done in \cite{JPMellin, JLAnalyticMellin} for scalar fields. To make the exposition pedagogical we will start with a simple  interaction vertex for a vector particle and $k$ scalar particles and then generalize to interactions involving a spin $J$ particle.

\subsection{Scattering amplitude with a vector particle}

\subsubsection{Simple local interaction}
Let us start with the simplest example involving a massive vector field $A_\mu$. Consider the local interaction vertex
\footnote{Generically, the coupling constant $g$ is dimensionful and therefore it defines an intrinsic length scale $\ell_s$ of the bulk theory, as discussed above.
}
\be
 g A_\mu (\nabla^\mu \phi_1 ) \phi_2 \dots \phi_k\label{eq:SimplestLocalInt}
\ee
with scalar fields $\phi_i$ and coupling constant $g$.
The associated scattering amplitude is
\be
\mathcal{T}= g \,\varepsilon \cdot p_1
\ee
where $\varepsilon$ is the polarization vector of the massive vector boson and $p_i$ is the momentum of the particle $\phi_i$. 
The polarization vector obeys
 $\varepsilon \cdot p=0$ where $p$ is the momentum of the vector boson.
Comparing with equation (\ref{ScatVector}), we conclude that all $\mathcal{T}^l$ vanish except $\mathcal{T}^1=g$.

The basic ingredients to evaluate the Mellin amplitude associated 
with the tree-level Witten diagram  using the same interaction vertex in AdS are the bulk to boundary propagators of a scalar field,
\footnote{We set the AdS curvature radius $R=1$.
We shall reintroduce $R$ at the end, in the final formulas.}
\be
\Pi_{\D,0}(X,P)=\frac{\mathcal{C}_{\D,0}}{(-2P\cdot X)^\Delta}\ ,
\ee
and of a vector field,
\be
\Pi_{\D,1}^{A}(X,P;Z)=
\frac{\mathcal{C}_{\D,1}}{\D}
\left[P^A (Z\cdot \partial_P)-Z^A(P\cdot \partial_P) \right]\frac{1}{(-2P\cdot X)^\Delta}\ ,
\ee
where   $X$ is a  point in AdS$_{d+1}$ embedded in $\mathbb{M}^{d+2}$ (\emph{i.e.} $X^2=-1$), $A$ is an embedding AdS index, and \cite{Costa:2014kfa}
\begin{align}
\mathcal{C}_{\D,J}=\frac{(J+\D-1)\G(\D)}{2
\pi^{\frac{d}{2}} (\D-1)\G\left(\D+1-\frac{d}{2}\right)}
\label{eq:SpinJBulkBoundaryPropagatorNormalization}\ .
\end{align}
The correlation function of $k$ scalars and one vector operator associated with the interaction (\ref{eq:SimplestLocalInt}) is 
\begin{align}
&\langle\Ocal(P,Z)\Ocal_1(P_1)\dots \Ocal_k(P_k)\rangle =
g \int_{\rm AdS}  dX\, 
\Pi_{\D ,1}^A(X,P;Z)
\nabla_A \Pi_{\D_1,0}(X,P_1)
\prod_{i=2}^k\Pi_{\D_i,0}(X,P_i)
\nonumber\\
&=
g   \frac{2\D_1\mathcal{C}_{\D,1}\mathcal{C}_{\D_1,0}}{\D}\,
D_1
\int_{\rm AdS} \frac{dX}{(-2P\cdot X)^\Delta
(-2P_1\cdot X)^{\Delta_1+1}}
\prod_{i=2}^k\Pi_{\D_i,0}(X,P_i)
\end{align}
where $D_1$ is precisely the differential operator  given in equation (\ref{eq:McheckSpinJ}) that was used to define the   Mellin amplitudes $\check{M}$.
The AdS covariant derivative $\nabla_A$ can be easily computed using the embedding formalism. As explained in \cite{Costa:2014kfa}, it amounts to projecting embedding space partial derivatives $\frac{\partial}{\partial X^A}$ with the projector 
\begin{align}
U_A^B=\delta_{A}^B+ X_AX^B \label{eq:ProjectorTransv}.
\end{align}
Performing the integral over the bulk position $X$, 
\footnote{This type of integral can be done using the (generalized) Symanzik formula \cite{Symanzik}, 
\begin{align}
\int_{AdS} dX \prod_{i=1}^n\frac{\G(\D_i)}{(-2P_i \cdot X)^{\D_i}} = \half \pi^{\frac{d}{2}}\G\left(\frac{\sum_{i=1}^n\D_i-d}{2}\right) \int [d\g] \prod_{i<j}^n  \frac{\G(\g_{ij})}{(-2P_i\cdot P_j)^{\g_{ij}}}\ ,\label{eq:SymanzikTypeIntegral}
\end{align}
where the integration variables satisfy $\sum_{i=1}^{n}\g_{ij}=0$ with $\g_{ii}=-\D_i$.
}
we conclude that the correlation function can be written as
\begin{align}
   g  \pi^{\frac{d}{2}}    
\frac{\mathcal{C}_{\D,1}
\G\left(\frac{\sum_{i}\D_i+\D+1-d}{2}\right)
}{\G(\D+1)}\prod_{i=1}^k
\frac{\mathcal{C}_{\D_i,0}}{\G(\Delta_i)}\,
D_1
\int[d\g] 
\prod_{i<j}^k \frac{\G(\g_{ij})}{(-2P_i\cdot P_j)^{\g_{ij}}}
\prod_{i=1}^k \frac{\G(\g_i +\d_i^1)}{(-2P\cdot P_i)^{\g_i +\d_i^1}}
\nonumber
\end{align}
where the Mellin variables $\g_{ij}$ obey the 
constraints (\ref{gammai}) and (\ref{constraintM1}).
Comparing with the Mellin representation (\ref{Mcheckrep}), we obtain
\be
\check{M}^1=  
 g  \pi^{\frac{d}{2}}  \G\!\left(\textstyle{\frac{\sum_{i=1}^k\D_i+\D+1-d}{2}}\right)  
\frac{\mathcal{C}_{\D,1}
}{\G(\D+1)}\prod_{i=1}^k
\frac{\mathcal{C}_{\D_i,0}}{\G(\Delta_i)}
\ee
and $\check{M}^l=0$ for $l=2,3,\dots,k$.
We conclude that for this simple interaction both the scattering amplitudes $\mathcal{T}^l$ and the Mellin amplitudes $\check{M}^l$ are constants.
In addition, they are proportional to each other $\check{M}^l \propto \mathcal{T}^l$.
This suggests, as already observed in \cite{PaulosMellin}, that the representation $\check{M}^l$ is more suitable to study the flat space limit of AdS than the representation $M^l$ also introduced in section \ref{sec:MellinSpin}.

\subsubsection{Generic local interaction}
The interaction (\ref{eq:SimplestLocalInt}) is the simplest local vertex for one vector and $k$ scalar operators. The generalization to other interactions follows essentially the same steps with minor modifications. Take the following local interaction vertex  
\be
g \nabla\dots \nabla A_\mu \nabla\dots \nabla(\nabla^\mu \phi_1 ) \nabla\dots \nabla\phi_2 \dots \nabla\dots \nabla\phi_k\label{eq:GeneralInteractionVertex}
\ee
where there are $\a_{ij}$ derivatives acting on the field $\phi_i$ contracted with derivatives acting on the field $\phi_j$ and $\a_i$ derivatives acting on $A_\mu$ contracted with derivatives acting on $\phi_i$. In total, the vertex contains 
$1+2\sum_{i=1}^k \a_{i}+2\sum_{i<j}^k \a_{ij}\equiv1+2N$ derivatives.
The  scattering amplitude associated to this interaction is given by
\be
\mathcal{T}(\varepsilon,p,p_i)= g \,\varepsilon \cdot p_1 
\prod_{i=1}^k (-p\cdot p_i)^{\a_{i}}
\prod_{i<j}^k  (-p_i\cdot p_j)^{\a_{ij}}=\sum_{l=1}^k \varepsilon\cdot p_l \,\mathcal{T}^{l}(p_i\cdot p_j)\label{eq:Spin1ScatteringAmplitudeFSL}.
\ee
where the last equality defines the partial amplitudes introduced in section \ref{subsec:SAspin}. Notice that, as in the previous example,  only $\mathcal{T}^1$ is non-zero.

Consider now the correlation function of $k$ scalars and a vector operator associated with the Witten diagram using the same interaction vertex but now in AdS. 
One should replace the fields in (\ref{eq:GeneralInteractionVertex}) by bulk-to-boundary propagators, compute their covariant derivatives, contract the indices and integrate over the interaction point $X$ in AdS.
As we shall argue below, in the flat space limit one can replace covariant derivatives by simple partial derivatives in the embedding space (\emph{i.e.}, we drop the second term in the projector (\ref{eq:ProjectorTransv})).
Using this simplifying assumption, the Witten diagram is given by 
\begin{align}
\label{genintespin1}
2g(-2)^N 
\frac{\mathcal{C}_{\D,1}}{\G(\D+1)}
\prod_{i=1}^k
&\frac{\mathcal{C}_{\D_i,0}}{\G(\D_i)}\,
D_1
\left[ \prod_{i<j}^k(-2P_i\cdot P_j)^{\alpha_{ij}} \prod_{i=1}^k(-2P_i\cdot P)^{\alpha_i}\right.
\\
&\left.\int dX
\frac{\G(\Delta+\sum_{i}\alpha_{i})}{(-2P\cdot X)^{\Delta+\sum_{i}\alpha_{i}}}
\prod_{i=1}^k\frac{\G(\Delta_i+\delta_{i}^1+\a_i+\sum_{j\neq i}^k\alpha_{ij})}{(-2P_i\cdot X)^{\Delta_i+\delta_{i}^1+\a_i+\sum_{j\neq i}^k\alpha_{ij}}}\right]\ .\nonumber
\end{align}
The integral over $X$ is again of Symanzik type and can be done using (\ref{eq:SymanzikTypeIntegral}).
After shifting the integration variables $\g_{ij}$ in  (\ref{eq:SymanzikTypeIntegral}) to bring the result to the standard form (\ref{Mcheckrep}), we obtain
\begin{align}
\!\!\check{M}^1= g(-2)^{N}\pi^\frac{d}{2} 
\G\!\left(\textstyle{\frac{\sum_{i=1}^k\D_i+\D+2N+1-d}{2}}\right) \frac{\mathcal{C}_{\D,1}}{\G(\D+1)}\prod_{i=1}^k\frac{\mathcal{C}_{\D_i,0}}{\G(\D_i)}\prod_{i<j}^k(\g_{ij})_{\alpha_{ij}}
\prod_{i=1}^k(\g_{i}+\d_i^1)_{\alpha_i}
\label{Mcheck1genSpin1}
\end{align} 
and $\check{M}^l=0$ for $l>1$.
We conclude that the Mellin amplitude is a polynomial of degree $N$. Moreover, its leading behaviour at large $\g_{ij}$ is 
$\check{M}^1  \propto  
  \prod_{i<j}^k \g_{ij}^{\alpha_{ij}}
\prod_{i=1}^k \g_{i}^{\alpha_i}
$.
Notice that this is exactly the form of the
 scattering amplitude (\ref{eq:Spin1ScatteringAmplitudeFSL})
  if we identify 
  $\g_{ij} \leftrightarrow p_i\cdot p_j$
  (which implies    $\g_i\leftrightarrow p\cdot p_i$).
In fact, we can write a general formula for the relation between the Mellin amplitude at large $\g_{ij}$ and dual scattering amplitude,
\begin{align} \label{McheckJ1FSL}
\check{M}^l(\g_{ij})\approx \pi^{\frac{d}{2}} \frac{\mathcal{C}_{\D,1}}{\G(\D+1)}\prod_{i=1}^k\frac{\mathcal{C}_{\D_i,0}}{\G(\D_i)}\int_{0}^{\infty}d\b 
\b^{\frac{\sum_{i}\D_i+\D-1-d }{2}}e^{-\b}\mathcal{T}^{l}(p_i\cdot p_j=2\b\g_{ij})\ ,
\end{align}
where the goal of the $\b$-integral is to create the first $\G$-function in (\ref{Mcheck1genSpin1}), which has information about the number of derivatives in the interaction vertex.

Let us return to the approximation used to compute AdS covariant derivatives. 
Notice that  the term we neglected in the projector (\ref{eq:ProjectorTransv})  would
contribute to (\ref{genintespin1}) with similar expressions but with
lower powers of $(-2P_i\cdot P_j)$.
In other words, the effect of the neglected terms can be thought of as an interaction with smaller number of derivatives.
Therefore, they give rise to subleading contributions in the large $\g_{ij}$ limit of the Mellin amplitude $\check{M}^l$.

\subsection{Scattering amplitude with a spin $J$ particle}
Consider a local  interaction of the form
\begin{align}
g (\nabla\dots \nabla h_{A_1\dots A_J})(\nabla\dots \nabla \phi_1)\dots (\nabla\dots \nabla \phi_k)\label{eq:MostGeneralInteractionVertexSpinJ}.
\end{align}
with a total of $2N+J$ derivatives distributed in the following way: there are $\alpha_{ij}$ derivatives acting on $\phi_i$ contracted with derivatives acting on $\phi_j$;
there are $\alpha_{i}$ derivatives acting on $h_{A_1\dots A_J}$ contracted with derivatives acting on $\phi_i$; there are $\b_i$ derivatives acting on $\phi_i$ contracted with   indices of the spin $J$ field. The  scattering amplitude associated to this interaction is given by
\begin{align}
\mathcal{T}(\varepsilon,p,p_i)=g \prod_{i=1}^k(\varepsilon\cdot p_i)^{\b_i}\prod_{i=1}^{k}(-p\cdot p_i)^{\a_i}\prod_{i <j}^{k}(-p_i\cdot p_j)^{\a_{ij}} \ .
\label{eq:SpinJGenericIntFSL}
\end{align}
Comparing with the representation (\ref{TRepres}), we conclude that the only non-zero components of $\mathcal{T}^{a_1 \dots a_J}$ are the ones with $\b_i$ indices equal to $i$,  \be
\mathcal{T}^{\tiny{\overset{\b_1}{\overbrace{1\dots 1}} \,\overset{\b_2}{\overbrace{2\dots 2}}\,\dots\, \overset{\b_k}{\overbrace{k\dots k}}}}
=g  \prod_{i=1}^{k}(-p\cdot p_i)^{\a_i}\prod_{i <j}^{k}(-p_i\cdot p_j)^{\a_{ij}} \ .
\ee

Consider now the correlation function of $k$ scalars and a tensor operator associated with the Witten diagram using the same interaction vertex but now in AdS. 
One should replace the fields in (\ref{eq:GeneralInteractionVertex}) by bulk-to-boundary propagators, compute their covariant derivatives, contract the indices and integrate over the interaction point $X$ in AdS. For the same reason that was explained in the last subsection, in the flat space limit, we can replace AdS covariant derivatives by the corresponding embedding partial derivatives. The main difference from the  spin one example is that the bulk-to-boundary propagator for a spin $J$ field has more indices. It is convenient to write the spin $J$  bulk-to-boundary propagator as 
\begin{align}
\Pi_{\D,J}(X,P;W,Z)&=\mathcal{C}_{\D,J}\frac{\big((-2P\cdot X) (W\cdot Z)+ 2(W\cdot P) (Z\cdot X) \big)^J}{(-2P \cdot X)^{\D+J}} 
\label{spinJpropagator}\\
&= \frac{\mathcal{C}_{\D,J}}{(\D)_J}
\big((P\cdot W) (Z\cdot \partial_P)-(Z\cdot W) (P\cdot \partial_P-Z\cdot \partial_Z) \big)^J
\frac{1}{(-2P \cdot X)^{\D }}
\nonumber
\end{align}
where the normalization constant $\mathcal{C}_{\D,J}$ is given by
(\ref{eq:SpinJBulkBoundaryPropagatorNormalization}) 
and the vector  $W$ is null, to encode the property that the field is symmetric and traceless. 
Notice that the vector  $W$ is just an artifact to hide bulk indices and for that reason it will not appear in the final formula.
 In fact these indices should be contracted with $J$ derivatives that act on the remaining fields. Let us focus on these contractions since they are the only difference compared to the previous case, 
\begin{align}
&h_{A_1\dots A_{\b_1} B_1\dots B_{\b_2}\dots }  
 (\nabla^{A_1}\dots \nabla^{A_{\b_1}}  \phi_1 )
  (\nabla^{B_1}\dots \nabla^{B_{\b_2}}  \phi_2 )
 \dots  
 \rightarrow \nonumber\\
&\Pi_{\D,J}(X,P;\vec{W},Z) \prod_{i=1}^k(W\cdot \partial_{P_i})^{\b_i}\Pi_{\D_i,0}(X,P_i)
\label{eq:DerivativesActingSpinJFieldFSL}
\end{align}
where the notation $ \vec{W}$  denotes that we should expand the expression and use 
\be
\vec{W}^{A_1}\dots \vec{W}^{A_J}
W^{B_1}\dots W^{B_J} = \mathcal{P}^{A_1 \dots A_J, B_1 \dots B_J}\ ,
\ee
where $\mathcal{P}$ is a projector onto   symmetric and traceless tensors. 
After taking the partial embedding derivatives 
and performing the index contractions encoded in $\vec{W}$, we obtain
\footnote{
Notice that the differential operator in equation (\ref{spinJpropagator}) can be written as $W^A \mathcal{D}_A$ where $\mathcal{D}_A$ is null, 
\emph{i.e.} $\mathcal{D}_A\mathcal{D}^A=0$ on the null cone ($P^2=P\cdot Z=Z^2=0$).
This implies that 
$\mathcal{D}_{A_1}\dots \mathcal{D}_{A_J}=\mathcal{P}^{A_1 \dots A_J, B_1 \dots B_J}\mathcal{D}_{B_1}\dots \mathcal{D}_{B_J}$,
which greatly simplifies 
the computation.
}
%
%
%
\begin{align}
\Pi_{\D,J}(X,P;\vec{W},Z) \prod_{i=1}^k(W\cdot \partial_{P_i})^{\b_i}\Pi_{\D_i,0}(X,P_i)=\frac{2^J}{(\D)_J}\bigg(\prod_{i=1}^k\frac{\mathcal{C}_{\D_i,0}(\D_i)_{\b_i}}{(-2P_i\cdot X)^{\D_i+\b_i}}D_{i}^{\b_i}\bigg)
\frac{\mathcal{C}_{\D,J}}{(-2P\cdot X)^{\D}}\ .
\nonumber
\end{align}
Acting with the remaining $2N$ derivatives, we conclude that the Witten diagram associated with (\ref
{eq:MostGeneralInteractionVertexSpinJ}), in the flat space limit, is given by
\begin{align}
g\,2^{J}(-2)^{N}\frac{\mathcal{C}_{\D,J}}{\G(\D+J)} &\bigg(\prod_{i=1}^{k}
\frac{\mathcal{C}_{\D_i,0}}{
\G(\D_i)}
D_i^{\b_i}\bigg)\left[ \prod_{i<j}^k(-2P_i\cdot P_j)^{\alpha_{ij}} \prod_{i=1}^k(-2P_i\cdot P)^{\alpha_i}\right.
\\
&\left.\int dX
\frac{\G(\Delta+\sum_{i}\alpha_{i})}{(-2P\cdot X)^{\Delta+\sum_{i}\alpha_{i}}}
\prod_{i=1}^k\frac{\G(\Delta_i+\b_i+\a_i+\sum_{j\neq i}^k\alpha_{ij})}{(-2P_i\cdot X)^{\Delta_i+\b_i+\a_i+\sum_{j\neq i}^k\alpha_{ij}}}\right].\nonumber
\end{align}
The integral over $X$ is again of  Symanzik type and can be done using (\ref{eq:SymanzikTypeIntegral}). After shifting the integration variables $\g_{ij}$ in (\ref{eq:SymanzikTypeIntegral}) to bring the result to the standard form (\ref{Mcheckrep}), we obtain
\begin{align}
\check{M}^{\tiny{\overset{\b_1}{\overbrace{1\dots 1}} \,\overset{\b_2}{\overbrace{2\dots 2}}\,\dots\, \overset{\b_k}{\overbrace{k\dots k}}}}
&= g(-2)^{N}2^{J-1}\pi^\frac{d}{2} 
\G\!\left(\textstyle{\frac{\sum_{i=1}^k\D_i+\D+2N+J-d}{2}}\right) \nonumber\\
&\times\frac{\mathcal{C}_{\D,J}}{\G(\D+J)}\prod_{i=1}^k\frac{\mathcal{C}_{\D_i,0}}{\G(\D_i)}\prod_{i<j}^k(\g_{ij})_{\alpha_{ij}}
\prod_{i=1}^k(\g_{i}+\b_i)_{\alpha_i}
\label{Mcheck1genSpinJ}
\end{align} 
and all other components of $\check{M}^{ a_1 \dots a_J}$ are zero. We conclude that the Mellin amplitude is a polynomial of degree $N$. Moreover, its leading behaviour at large $\g_{ij}$ is proportional to  the scattering amplitude (\ref{eq:SpinJGenericIntFSL}) if we identify $\g_{ij}\leftrightarrow p_i\cdot p_j$. In fact, we can write a general formula for the relation between the Mellin amplitude at large $\g_{ij}$ and dual scattering amplitude,  
 \begin{align}
\check{M}^{a_1\dots a_J} \approx
\mathcal{N} 
R^{\frac{(k+1)(1-d)}{2}+d+1-J}
 \int_{0}^{\infty}\frac{d\b }{\b}
\b^{\frac{\sum_{i }\D_i+\D-d+J}{2} 
}e^{-\b}\mathcal{T}^{a_1\dots a_J}\left(
p_i\cdot p_j=\frac{2\b}{R^2}\g_{ij}\right)\label{eq:FSLRelationFinalSpinJ}  ,
\end{align}
where we reintroduced the AdS radius $R$ and
\be \label{NormalizationN}
\mathcal{N} =
\pi^\frac{d}{2} 2^{J-1}
\frac{\sqrt{\mathcal{C}_{\D,J}}}
{  \G\big(\D+J\big)}\prod_{i=1}^k\frac{\sqrt{
\mathcal{C}_{\D_i,0}}}{\G(\D_i)}\ .
\ee
In the last equation, we have converted to the standard CFT normalization of operators, which corresponds to  
$ \langle \Ocal(x)\Ocal (0) \rangle = |x|^{-2\D}$
for scalar operators and 
(\ref{2ptNormalization}) for tensor operators.
This differs from the natural AdS normalization  by $\Ocal_{\rm AdS}(x)=\sqrt{\mathcal{C}_{\D,J}}\,\Ocal_{\rm CFT}(x)$.

The  inverted form of equation
(\ref{eq:FSLRelationFinalSpinJ}), 
 \begin{align}
\mathcal{T}^{a_1\dots a_J} (
p_i\cdot p_j )
=\lim_{R\to \infty}
\frac{1}{\mathcal{N} }
 \int_{-i \infty}^{i\infty}\frac{d\a }{2\pi i}
\a^{\frac{d-\sum_{i }\D_i-\D-J}{2} 
}e^{\a}\,
\frac{\check{M}^{a_1\dots a_J} 
\left(\g_{ij}
=\frac{R^2}{2\a}\,p_i\cdot p_j\right)
}{
R^{\frac{(k+1)(1-d)}{2}+d+1-J}
}
\label{eq:FSLRelationFinalSpinJinverse}  ,
\end{align} 
realizes the flat space limit intuition that the Mellin amplitude can be used to define the scattering amplitude.
The final formulas (\ref{eq:FSLRelationFinalSpinJ}) and
(\ref{eq:FSLRelationFinalSpinJinverse}) were derived based on the interaction vertex (\ref{eq:MostGeneralInteractionVertexSpinJ}). However, this  vertex is a basis for all possible interactions, so we expect the final formulas to be valid in general. As a consistency check,
we show in appendix \ref{app:FSLfactorization} that using (\ref{eq:FSLRelationFinalSpinJ}) in the factorization formulas for Mellin amplitudes derived in the previous sections, we recover the correct factorization properties of  flat space scattering amplitudes reviewed in section \ref{sec:reviewFS}.

\section{Conclusion}
\label{sec:Conclusion}

In the context of scattering amplitudes,
understanding their factorization properties is  the starting point for the construction of recursion relations (like BCFW \cite{BCFW}).
Such recursion relations determine $n$-particle scattering amplitudes in terms of scattering amplitudes with a smaller number of particles.
In some cases (gluons or gravitons), this can be iterated successively until all scattering amplitudes are fixed in terms of the 3-particle amplitudes.
Our long term goal is to generalize this type of recursion relations for Mellin amplitudes.
\footnote{See \cite{Raju:2010by} for a similar proposal for recursion relations for correlators associated with Witten diagrams.}

This work was the first step in this direction.
We derived factorization formulas for the residues of $n$-point Mellin amplitudes of scalar operators, associated with the exchange of  primary operators of spin $J=0,1,2$. 
For $J>2$, we only obtained partial results because formulas become rather complicated.

The next step is to understand the factorization of Mellin amplitudes of operators with spin.
In particular, the case of correlation functions of the stress-energy tensor $T_{\mu\nu}$ should be particularly interesting, in analogy to graviton scattering amplitudes.
This approach might eventually explain in what circumstances the 3-point function of $T_{\mu\nu}$ completely fixes all correlation functions of $T_{\mu\nu}$.
Notice that this is the missing link to prove the conjecture \cite{JP} that any CFT with a large $N$ expansion and with a parametrically large dimension of the lightest single-trace operator above the stress tensor, is well described by pure Einstein gravity in AdS.
In fact, an important part of this conjecture was recently proven in \cite{Camanho:2014apa}. In this work, the authors showed that the existence of such a large gap in the spectrum of single-trace operators implies that the 3-point function of $T_{\mu\nu}$ is given by Einstein gravity in AdS (with  higher curvature corrections parametrically small).  
 
Simpler but still very interesting objects to study  are correlation functions of conserved currents.
Here, one can explore  the analogy with gluon scattering amplitudes.
The first obstacle to surpass, is
to define a Mellin representation for the $n$-point function of conserved currents
\be
\langle
 \Ocal_1(P_1,Z_1)\dots \Ocal_k(P_n,Z_n) 
\rangle\ . 
\ee
A natural generalization of (\ref{Mrep}) is
\be 
\sum_{a_1,\dots,a_n=1 \atop a_i\neq i}^n   ( Z_1 \cdot P_{a_1})
\dots  ( Z_n \cdot P_{a_n})
  \int [d\g]\,
 M^{a_1 \dots a_n}
  \prod_{1\le i<j\le n} 
  \frac{\G(\g_{ij} +\d^i_{a_j} +\d^j_{a_i})}{(-2P_i\cdot P_j)^{\g_{ij}  +\d^i_{a_j} +\d^j_{a_i}}} \ , 
\ee 
where    the Mellin variables obey
\be 
\g_{ij}=\g_{ji}\ , \qquad
\g_{ii}=1-\D_i\ ,
\qquad
\sum_{ j=1}^n \g_{i j}=0 \ .
\ee
Unfortunately, this is incomplete because in general the correlation function also contains terms proportional to $Z_i\cdot Z_j$.
It is unclear what is the most convenient definition of the Mellin amplitudes in this case.
This is a question  for the future.


\section*{Acknowledgements}

J.P. is grateful to Perimeter Institute for the great hospitality in the summer of 2013 where this 
work was initiated.
V.G and E.T. also thank IPMU at Tokyo University for the great hospitality during the progress of this work.
The research leading to these results has received funding from the [European Union] Seventh Framework Programme under grant agreements No 269217 and No 317089.
This work was partially funded by the grant CERN/FP/123599/2011 and by the Matsumae International Foundation in Japan.
\emph{Centro de Fisica do Porto} is partially funded by the Foundation for 
Science and Technology of Portugal (FCT). The work of V.G. is supported  
by the FCT fellowship SFRH/BD/68313/2010. 
The work of E.T. is supported  
by the FCT fellowship SFRH/BD/51984/2012.

\pagebreak
\appendix
\section{Factorization from the shadow operator formalism}

In this appendix, we detail the calculations involved in the factorization method described in section \ref{sec:FacShadow}. We consider exchanged operators of spin 0, 1 and 2.
Auxiliary calculations are presented in the last three subsections: in \ref{ap:projector} we construct the projector for traceless symmetric tensors, in \ref{ap:conformalintegrals} we evaluate some useful conformal integrals and in \ref{ap:MellinIden}  we derive an identity involving Mellin integrals.

\subsection{Factorization on a scalar operator}
\label{ap:projectorScalar}
 
In this subsection we fill in the gaps of the derivation presented in section \ref{sec:FacShadow}.
We shall use the notation
\be
[x_{ab}]^{\g} =[x_a-x_b]^{\g} \equiv \frac{\G(\g)}{(x_a-x_b)^{2\g}}
\ee
to shorten the expressions that follow.
Using (\ref{MellinLeft}) and (\ref{MellinRight}), expression (\ref{CBplusShadow}) can  be written as follows
\begin{align}
\frac{1}{\Ncal_\D}&
\int [d\l] [d\r]\,
I M_L  
M_R 
\prod_{1\le a<b\le k} [x_{ab}]^{\l_{ab}}
\prod_{k< i<j\le n} [x_{ij}]^{\r_{ij}}\ ,
\label{eqscalarshadowfac1}
\end{align}
where $I$ is the scalar conformal integral
\begin{align}
I=\int dy dz \,\,[y-z]^{d-\D}\prod_{1\leq a\leq k}[x_a-y]^{\l_a}\prod_{k< i\leq n}[x_i-z]^{\r_i} 
\end{align}
which, in appendix (\ref{ap:ConfInt2p}), we show it can be written as
\begin{align}
&I=\pi^d 
\int [d\g] \; 
\frac{\G(B)\G(A-B)}{\G(A)} \prod_{1\leq \mu \leq \nu \leq n}[x_{\mu\nu}]^{\g_{\mu\nu}} 
\end{align}
with $B= \frac{2\D-d}{2}$ and $A=\sum_{1\leq a<b\leq k}\g_{ab}=\frac{\D-\g_{LR}}{2}$. 
Replacing this expression in (\ref{eqscalarshadowfac1}) and shifting
 the integration variables $\g_{\mu\nu}$   to absorb the factors $[x_{ab}]^{\l_{ab}}$ and $[x_{ij}]^{\r_{ij}}$
leads directly to  (\ref{MellinBlockPlusShadow}).
  

We shall now determine the residue of $F_L$ at $\g_{LR}=\D+2m$ by deforming the integration contours in (\ref{LeftPartB+S}).
Using the constraints (\ref{tildeconstraint}) we can solve for 
\be
\l_{12}=-\half \left[\D+2\sum_{a=3}^k\l_{1a}+\sum_{a,b=2}^k\l_{ab} \right]
\label{lambda12}
\ee
and use as independent integration variables 
$\l_{13},\l_{14},\dots ,\l_{1k}$ and
$\l_{ab}$ for $2 \le a<b\le k$.
Then, the measure reads
\be
  \int [d\l] = \int_{-i\infty}^{i\infty}\,
  \prod_{3\le a\le k}
\frac{d\l_{1a}}{2\pi i}
\prod_{2 \le a<b\le k} \frac{d\l_{ab}}{2\pi i}\ .
\ee
These $(k-2)(k+1)/2$ integrals   can be done by deforming the contour to the right and picking up poles of the integrand in (\ref{LeftPartB+S}). There are explicit poles of the $\G$-functions at $\l_{ab}=\g_{ab}+n_{ab}$ with $n_{ab}=0,1,2,\dots$, and possibly other poles of the Mellin amplitude $M_L$. This gives
\begin{align}
F_L=& 
\sum_{n_{ab}\ge 0} 
M_L(\g_{ab}+n_{ab})  
\frac{\G\left(\l_{12}\right)
\G\left(\g_{12}-\l_{12}\right)}{\G( \g_{12})}
\prod_{1\le a<b\le k}' \frac{(-1)^{n_{ab}}
\left(\g_{ab}\right)_{n_{ab}}}{n_{ab}!}
\label{Laux}
\\
&+{\rm contributions\ from\ poles\ of}\ M_L
\nonumber
\end{align}
where $\l_{12}$ is given by (\ref{lambda12}) with
$\l_{ab}=\g_{ab}+n_{ab}\,$, 
\be
\l_{12}=\g_{12}-\frac{1}{2}\left[
\D-\g_{LR}+2\sum_{1\le a<b\le k}'n_{ab}
\right] 
\ee
and the prime denotes that $ab=12$ is absent from the sum or product.
From (\ref{Laux}), it is clear that $F_L$ will have poles when $\g_{12}-\l_{12}=-n_{12}$ with $n_{12}=0,1,2,\dots$. This corresponds to a pole at
\be
\g_{LR}= 
\D+2m \ , \qquad
m=\sum_{1\le a<b\le k}n_{ab}\ ,
\ee
with residue
\be
 F_L\approx \frac{-2(-1)^m}{\g_{LR}-\D-2m} \sum_{n_{ab}\ge 0 \atop\sum n_{ab}=m} 
M_L(\g_{ab}+n_{ab})  
\prod_{1\le a<b\le k} \frac{ 
\left(\g_{ab}\right)_{n_{ab}}}{n_{ab}!}\ .
\ee

\subsection{Factorization on a vector operator }
\label{app:FacVector} 
This section will be very similar to the scalar case. The main difference is that we will use the embedding formalism to simplify the calculations.
The goal is to determine the poles and residues of the Mellin amplitude associated with 
\begin{align}
&\!\!\!\!\langle \Ocal_1(P_1)\dots \Ocal_k(P_k) |\Ocal| \Ocal_{k+1}(P_{k+1})\dots \Ocal_n(P_n)\rangle=\int dQ_1dQ_2
\langle \Ocal_1(P_1)\dots \Ocal_k(P_k) \Ocal(Q_1,Z_1) \rangle  
\nonumber
\\&\!\!\!\!\! \frac{\G(d-\D+1)}{\Ncal_{\D,1}}\frac{ (\cev{Z_1}\cdot \vec{Z_2})(Q_1\cdot Q_2)
-(\cev{Z_1}\cdot Q_2)(Q_1\cdot \vec{Z_2}) }{(-2Q_1\cdot Q_2)^{d-\D+1}}
\langle \Ocal(Q_2,Z_2) \Ocal_{k+1}(P_{k+1})\dots \Ocal_n(P_n)\rangle
\label{CBplusShadowVector}
\end{align} 
where we have used the projector for tensor operators described in appendix \ref{ap:projector}.

We shall use the notation
\be
[P,Q]^{a} =\frac{\G(a)}{(-2P\cdot Q)^{a}}\ ,\ \ \ \ \ \ \
[P_{ij}]^{a} =[P_i,P_j]^{a}=\frac{\G(a)}{(-2P_i\cdot P_j)^{a}} 
\ee
to shorten the expressions that follow.
We start by writing the correlation functions that appear in (\ref{CBplusShadowVector}) in the Mellin representation,
\be 
\label{SpinMellinRep}
\langle \Ocal_1(P_1)\dots \Ocal_k(P_k) \Ocal(Q_1,Z_1) \rangle = \int [d\l]
\sum_{l=1}^k (Z_1\cdot P_l)\,
 M^l_L  \prod_{1\le a<b\le k} [P_{ab}]^{\l_{ab}}
\prod_{1\le a\le k} [P_{a},Q_1]^{\l_{a}+\d_a^l} 
\nonumber
\ee 
where $\d_a^l$ is the Kronecker-delta and
\be
\l_{a}= -\sum_{ b=1}^k \l_{ab}\ ,\qquad
\l_{aa}=-\D_a\ ,\qquad
\l_{ab}=\l_{ba}\ ,\qquad
\sum_{a,b=1}^k \l_{ab}= 
1 - \D\ .
\ee 
Similarly,
\be
\langle \Ocal(Q_2,Z_2) \Ocal_{k+1}(P_{k+1})\dots \Ocal_n(P_n)  \rangle = \int [d\r]
\sum_{r=k+1}^n (  Z_2\cdot P_r) M^r_R \prod_{k< i<j\le n} [P_{ij}]^{\r_{ij}}
\prod_{k< i\le n} [P_{i},Q_2]^{\r_{i}+\d_{i}^r}
\nonumber
\ee
where 
\be
\r_{i}= -\sum_{ j=k+1}^n \r_{ij}\ ,\qquad
\r_{ii}=-\D_i\ ,\qquad
\r_{ij}=\r_{ji}\ ,\qquad
\sum_{i,j=k+1}^n \r_{ij}= 
1 - \D\ .
\ee 
Expression (\ref{CBplusShadowVector}) can then be written as follows
\begin{align}
 \frac{1}{\Ncal_{\D,1}} &
\int [d\l ] [d\r]\sum_{l=1}^k\sum_{r=k+1}^n
M^l_L
M^r_R
\prod_{1\le a<b\le k} [P_{ab}]^{\l_{ab}}
\prod_{k< i<j\le n} [P_{ij}]^{\r_{ij}}
\nonumber
\\ \times&
\int dQ_1dQ_2 
\prod_{1\le a\le k} [P_{a},Q_1]^{\l_{a}+\d_a^l}
\prod_{k< i\le n} [P_{i},Q_2]^{\r_{i}+\d_{i}^r}
\nonumber
\\
\times&\left(\frac{\D-d}{2}
 ( P_l\cdot P_r) [Q_{12}]^{d-\D}-
( P_l\cdot Q_2)( P_r\cdot Q_1)
 [Q_{12}]^{d-\D+1}
\right)
\label{2terms2}
\end{align}
where $\d_a^l$ and $\d_{i}^r$ are Kronecker-deltas.
Expanding the last line, we obtain two integrals with different structure. 
The integral over $Q_1$ and $Q_2$ of the first term in (\ref{2terms2}) can be done using the conformal integral for scalars (\ref{eq:ScalarIntegralBasicShadow}).
The integral over $Q_1$ and $Q_2$ in the second term of (\ref{2terms2}) can be done using the vector conformal integral (\ref{eq:StarterVectorConformalIntegrallr}). 
Putting all ingredients together we obtain that that the factorization for the vector case is given by
\begin{align}
M_{\Ocal}(\g_{\mu\nu})=\frac{\pi^d}{\mathcal{N}_{\D,1}}\frac{\G(B)\G(A-B)}{4\G(A+1)} \sum_{l=1}^k \sum_{r=k+1}^n\big(A(d-\D-1)\g_{lr}-B\g_{l}\g_{r} \big)F_{L}^{l}\times F_{R}^{r}
\end{align}
with   $B = \frac{2\D-d}{2}$, $A=\sum_{1\leq a<b\leq k}\g_{ab}=\frac{\D-1-\g_{LR}}{2}$ and 
\begin{align}
F_{L}^l&=\int [d\l]M_{L}^{l}(\l_{ab})\prod_{1\leq a<b\leq k}\frac{\G(\l_{ab})\G(\g_{ab}-\l_{ab})}{\G(\g_{ab})}\\
F_{R}^r&=\int [d\r]M_{R}^{r}(\r_{ij})\prod_{k< i<j\leq n}\frac{\G(\r_{ij})\G(\g_{ij}-\r_{ij})}{\G(\g_{ij})}
\end{align}
Looking for the poles of $\g_{LR}$ we obtain $\G(A+1)\approx \frac{(-2) }{(m-1)!}\frac{(-1)^{m-1}}{\g_{LR}-\D-2m+1}$ and $F_{L}^l\approx \frac{(-2) (-1)^m}{\g_{LR}-\D-2m+1} L^l_m$ where
\be
L^l_m= \sum_{\sum n_{ab}=m} 
M^l(\g_{ab}+n_{ab})  
\prod_{1\le a<b\le k} \frac{ 
\left(\g_{ab}\right)_{n_{ab}}}{n_{ab}!} \ 
\ee
and similarly for $R^r_m$. The contribution of the physical operator $\Ocal$ to the pole structure of $M_{O}$ can therefore be written as  
\begin{align}
\!\!\!\!\!\!\!\! M\approx
  \frac{    m!}{\left(1+\D-\frac{d}{2}\right)_m}
  \frac{\kappa_{\D,1}}{\g_{LR}-\D+1-2m}
  \sum_{l=1}^k \sum_{r=k+1}^n\left[\g_{l r}
  +\frac{d-2\D}{2 m(\D-d+1)} \g_{l}\g_{r}\right] L^l_m R^r_m \ ,
\end{align}
where we used $\G(A-B)=(-1)^m \frac{\G(d/2-\D)}{(1-d/2+\D)_m}$ and where we defined
\be
 \kappa_{\D,1}=\frac{\pi^d\G\left(\frac{2\D-d}{2}\right)\G\left(\frac{d-2\D }{2} \right)(\D-d-1)\G(\D-d+1)}
{2 \Ncal_{\D,1}  }=\Delta  \Gamma (\Delta -1) \ .
\ee

\subsection{Factorization on a spin 2 operator}
\label{ap:Spin2FacProj}
The derivation of factorization corresponding to the exchange of an operator of spin two follows the same steps as the scalar and vector cases.We first consider the projector for spin two operators defined in appendix \ref{ap:projector},
\be
\!\!\! |\Ocal|=\frac{\G(d-\D+2)}{\Ncal_{\D,2}} \int dQ_1 dQ_2 | \Ocal(Q_1,Z_1) \rangle
\frac{\left((\cev{Z_1}\cdot \vec{Z_2})(Q_1\cdot Q_2)
-(\cev{Z_1}\cdot Q_2)(Q_1\cdot \vec{Z_2}) \right)^2}{(-2Q_1\cdot Q_2)^{d-\D+2}}
\langle \Ocal(Q_2,Z_2) | \ ,
\nonumber
\ee
where spin two primary field $\mathcal{O}$ is inserted at the points $Q_1$ and $Q_2$.
We insert $|\Ocal|$ in the scalar $n-$point function in such a way to have $k+1$ operators on the left and $n-k+1$ on the right, namely
\begin{align}
&\langle \Ocal_1(P_1) \cdots \Ocal_k(P_k)|\Ocal|  \Ocal_{k+1}(P_{k+1}) \cdots \Ocal_n(P_n) \rangle \nonumber \\
& \qquad =\int\! \left[d\l_{ab}\right]\left[d\r_{ij}\right]\sum_{l_1,l_2=1}^k\sum_{r_1,r_2=k+1}^n \frac{M^{l_1l_2}(\l_{ab})M^{r_1r_2}(\r_{ij})}{16 \;\mathcal{N}_{\Delta,2}} \prod_{1\leq a<b\leq k}\left[P_{ab}\right]^{\l_{ab}}\prod_{k < i<j\leq n}\left[P_{ij}\right]^{\r_{ij}}\nonumber\\
 &\qquad \qquad  \times \int dQ_1dQ_2 \; \; W  \prod_{1\leq a \leq k} \left[P_a,Q_1\right]^{\l_a+\d_{a}^{l_1}+\d_{a}^{l_2}} \prod_{k< i \leq n}\left[P_i,Q_2\right]^{\r_i+\delta_{i}^{r_1}+\delta_{i}^{r_2}} \left[Q_{12}\right]^{d-\Delta+2}\label{eq:SpintwocomputationFactorizationStart}.
\end{align}
where the factor $W$ comes from the projector and is given by, 
\be
\!\!\! W=16 \, (P_{l_1}\cdot Z_1)(P_{l_2}\cdot Z_1) ((\overleftarrow{Z}_1\cdot \overrightarrow{Z}_2)(Q_2\cdot Q_1)-(Q_2\cdot \overleftarrow{Z}_1)(Q_1\cdot \overrightarrow{Z}_2))^2(P_{r_1}\cdot Z_2)(P_{r_2}\cdot Z_2) \ .
\ee
More concretely, $W$ is given by the sum of the following 6 terms
\be
\! \!\! \!\! \!\begin{array}{lll}
W_{I}= P_{l_1r_1}P_{l_2r_2}Q_{12}^2 \,, 
& W_{II}=-\dfrac{P_{l_1l_2}P_{r_1r_2} Q_{12}^2}{d} \,, 
&W_{III}=-2 P_{l_1r_1}\tilde{P}_{l_2,2}\tilde{P}_{r_2,1}Q_{12}\,, \\
W_{IV}=\dfrac{2 P_{r_1r_2} \tilde{P}_{l_1,1}\tilde{P}_{l_2,2} Q_{12}}{d}\,, \; \;
&W_V=\dfrac{2P_{l_1l_2}\tilde{P}_{r_1,1}\tilde{P}_{r_2,2} Q_{12}}{d}\,,\; \;
& W_{VI}=\tilde{P}_{l_1,2}\tilde{P}_{l_2,2} \tilde{P}_{r_1,1}\tilde{P}_{r_2,1} \,,
\end{array}
\ee
where we used the notation $\tilde{P}_{a,i}=-2P_a\cdot Q_i$. 
Notice that both $W_{IV}$ and $W_{V}$ have factors that can be   absorbed in (\ref{eq:SpintwocomputationFactorizationStart}) bringing down factors of $(\l_{l_1}+\d_{l_1}^{l_2})$ and $(\r_{r_1}+\d_{r_1}^{r_2})$ respectively.   These structures project to zero once the transversality condition (\ref{TransverseJ}) is used. The integrals involving the structures $W_{I}$ and $W_{II}$ can be done using the conformal integral for scalars (\ref{eq:ScalarIntegralBasicShadow}). 
Using (\ref{eq:SimpleScalarIntegralFinal}), they give the following contribution to the Mellin amplitude of (\ref{eq:SpintwocomputationFactorizationStart})
\begin{align}
\!\!\!\!\!\! M_{I}(\g_{\mu\nu})&=\frac{\pi^d \ (d-\D)_2}{16 \ \mathcal{N}_{\Delta,2}}\frac{\G(B)\G(A-B)}{\G(A)}\sum_{l_1,l_2=1}^k\sum_{r_1,r_2=k+1}^n\g_{l_1r_1}(\g_{l_2r_2}+\d_{l_1}^{l_2}\d_{r_1}^{r_2})F_{L}^{l_1l_2}F_{R}^{r_1r_2} ,
\\
\!\!\!\!\!\!M_{II}(\g_{\mu\nu})&=-\frac{\pi^d (d-\D)_2}{16 \ d \ \mathcal{N}_{\D,2}}\frac{\G(B)\G(A-B+1)}{\G(A+1)}\sum_{l_1, l_2=1 \atop l_1\neq l_2}^k \sum_{r_1,r_2=k+1 \atop r_1\neq r_2}^n \g_{l_1l_2}\g_{r_1r_2}[F_{L}^{l_1l_2}]^{l_1 l_2 } [F_{R}^{r_1r_2}]^{r_1 r_2}  , \label{MII}
\end{align}
where we have used $B=\frac{2\D-d}{2}$, $A=\frac{\D-2-\g_{LR}}{2}$ and
\begin{align}
F_{L}^{l_1l_2}&=\int [d\l]M_{L}^{l_1l_2}(\l_{ab})\prod_{1\leq a<b\leq k}\frac{\G(\l_{ab})\G(\g_{ab}-\l_{ab})}{\G(\g_{ab})} \ ,\\
F_{R}^{r_1r_2}&=\int [d\r]M_{R}^{r_1r_2}(\r_{ij})\prod_{k< i<j\leq n}\frac{\G(\r_{ij})\G(\g_{ij}-\r_{ij})}{\G(\g_{ij})}\ .
\end{align}
The square brackets $[\; \cdot \;]^{ab}$ shift the integration variable $\g_{ab}$ by one as defined in (\ref{def:SquareBrackets}) and they arise from the terms 
$\int [d\l](\g_{l_1 l_2}-\l_{l_1 l_2})M_{L}^{l_1l_2}(\l_{ab})\prod_{1\leq a<b\leq k}\frac{\G(\l_{ab})\G(\g_{ab}-\l_{ab})}{\G(\g_{ab})}=\g_{l_1l_2}[F_{L}^{l_1l_2}]^{l_1 l_2 }$. Notice that $M_{II}$ is written in terms of such square brackets while in $M_I$ they do not appear. This is because when we absorb the factors $[P_{ab}]^{\l_{ab}}$ (and $[P_{ij}]^{\r_{ij}}$) in the single term $[P_{\m \n}]^{\g_{\m \n}}$ we need to shift the variables $\g_{\m\nu}$ by $\l_{a b}$ (and by $\r_{i j}$) only if $\g_{\m\nu}$ is a variable with both indices $\m,\nu\le k $ (or $\m, \n >k$). Close to the pole in $\g_{LR}$ the following formulas hold
\be
F_{L}^{l_1l_2}\approx \frac{(-2) (-1)^m}{\g_{LR}-\D-2m+2} L^{l_1l_2}_m \ ,
\qquad\qquad [F_{L}^{l_1l_2}]^{l_1 l_2 }\approx \frac{(-2) (-1)^{m-1}}{\g_{LR}-\D-2m+2} [L^{l_1l_2}_{m-1}]^{l_1 l_2 } \ .
\ee
It is then easy to see that the structures coming from $M_{I}$ and $M_{II}$ have respectively the same form of $\mathcal{Q}_m^{(1)}$ and  $\mathcal{Q}_m^{(5)}$ in (\ref{QsJ=2}).  The contribution from the structure $W_{III}$ can be computed using the conformal integral for vectors (\ref{eq:StarterVectorConformalIntegrallr}),
\begin{align}
\!\!\!\! M_{III}=&-\frac{\pi^d(d-\D+1)}{8 \, \mathcal{N}_{\D,2}}\frac{\G(B)\G(A-B)}{\G(A+1)}\sum_{l_1,l_2=1}^k\sum_{r_1,r_2=k+1}^n\g_{l_1r_1}F_{L}^{l_1l_2}F_{R}^{r_1r_2} \nonumber\\
&\times \big[A(\g_{r_2l_2}+\d_{r_2}^{r_1}\d_{l_2}^{l_1})+B(\g_{l_2}+\d_{l_2}^{l_1})(\g_{r_2}+\d_{r_2}^{r_1})\big] \ .
\end{align}
The result of $M_{III}$ gives two terms that have the same form of $\mathcal{Q}_m^{(1)}$ and $\mathcal{Q}_m^{(2)}$.   
The structure $W_{VI}$ involves the spin two conformal integral (\ref{eq:SPinSpinTwoTwoIntegralStarter}). The contribution of $W_{VI}$ to the factorization is given by
\begin{align}
\!\!\!\!\!\!M_{VI}&=\frac{\pi^d}{16 \, \mathcal{N}_{\Delta,2}} \frac{\G(B)\G(A-B)}{\G(A+2)}  \sum_{l_1,l_2=1}^k\sum_{r_1,r_2=k+1}^n \biggl[ (\mathcal{Y}^{(1)}_{l_1l_2r_1r_2}+\mathcal{Y}^{(2)}_{l_1l_2r_1r_2}+\mathcal{Y}^{(3)}_{l_1l_2r_1r_2}) F_{L}^{l_1l_2}F_{R}^{r_1r_2} \nonumber \\
&+\left(\mathcal{Y}^{(4, L)}_{l_1l_2r_1r_2} [F_{L}^{l_1l_2}]^{l_1l_2} F_{R}^{r_1r_2}+\mathcal{Y}^{(4, R)}_{l_1l_2r_1r_2} F_{L}^{l_1l_2} [F_{R}^{r_1r_2}]^{r_1r_2}\right)+ \mathcal{Y}^{(5)}_{l_1l_2r_1r_2}[F_{L}^{l_1l_2}]^{l_1l_2}[F_{R}^{r_1r_2}]^{r_1r_2} \biggr]
\end{align}
where  $\mathcal{Y}^{(i)}_{l_1l_2r_1r_2}$ is defined in (\ref{eq:YconstantforSpinTwointegralFactorization}). Notice that every term $\mathcal{Y}^{(i)}_{l_1l_2r_1r_2}$ in $M_{VI}$ has the same form of the structure $\mathcal{Q}_{m}^{i}$. After gathering all structures from $M_{I},M_{II},M_{III}$ and $M_{VI}$, and looking for the poles in $\g_{LR}$, as was done for the scalar and vector cases, we obtain (\ref{finalresultJ2}).

\subsection{Projector for tensor operators}
\label{ap:projector}

In the embedding formalism, the projector for tensor operators takes the form
 \cite{SimmonsDuffin:2012uy} 
\be
|\Ocal|=\frac{\G(d-\D+J)}{\Ncal_{\D,J}} \int dP_1 dP_2 | \Ocal(P_1,Z_1) \rangle
\frac{\left((\cev{Z_1}\cdot \vec{Z_2})(P_1\cdot P_2)
-(\cev{Z_1}\cdot P_2)(P_1\cdot \vec{Z_2}) \right)^J}{(-2P_1\cdot P_2)^{d-\D+J}}
\langle \Ocal(P_2,Z_2) |
\nonumber
\ee
where the symbols $\cev{Z_1}$ and $\vec{Z_2}$ mean that we should  expand and contract using
\be
\vec{Z}^{B_1}\dots\vec{Z}^{B_J}\, Z^{A_1}\dots Z^{A_J} = \pi^{A_1\dots  A_J,B_1 \dots B_J}  \ ,
\ee
where $\pi^{A_1\dots  A_J,B_1 \dots B_J}$ is the projector onto traceless symmetric tensors with $J$ indices.
To determine the normalization constant $\Ncal_{\D,J}$ we impose that
\be
\langle \Ocal(P,Z) \dots \rangle =
\langle \Ocal(P,Z) |\Ocal| \dots \rangle
\label{eq:projector}
\ee
where the dots stand for any other operators.
We normalize the operator $\Ocal$ to have the following two point function 
\be
\langle \Ocal(P,Z) \Ocal(P_1,Z_1) \rangle= 
\frac{\left(
(Z \cdot Z_1)(-2 P\cdot P_1)- 
2(Z \cdot P_1)(P\cdot Z_1) 
\right)^J}{(-2 P\cdot P_1)^{\D+J}}\ .
\label{2ptNormalization}
\ee
In general, the correlation function $\langle \Ocal(P,Z) \dots \rangle$ of $\Ocal$ with any other operators  can be written as a linear combination (or integral) of
\be
\frac{\left((Z \cdot Y_1)(P\cdot Y_2)
-(Z \cdot Y_2)(P\cdot Y_1) \right)^J}{(-2 P\cdot X)^{\D+J}}
\ee
with different $X$, $Y_1$ and $Y_2$.
Therefore, equation (\ref{eq:projector}) is equivalent to 
\begin{align}
&\frac{\left((Z \cdot Y_1)(P\cdot Y_2)
-(Z \cdot Y_2)(P\cdot Y_1) \right)^J}{(-2 P\cdot X)^{\D+J}}\\
=&
\frac{2^J\G(d-\D+J)}{\Ncal_{\D,J}} \int   
\frac{dP_1 dP_2\ \Omega(Z,P,P_1,P_2,Y_1,Y_2)}{
(-2 P\cdot P_1)^{\D+J}
(-2P_1\cdot P_2)^{d-\D+J}
(-2 P_2\cdot X)^{\D+J}} 
\nonumber
\end{align}
where the numerator $\Omega(Z,P,P_1,P_2,Y_1,Y_2)$ is given by
\begin{align}
&\Big((Z \cdot P_1)(P\cdot Z_1) -(Z \cdot Z_1)(P\cdot P_1)
\Big)^J
\left((\cev{Z_1}\cdot \vec{Z_2})(P_1\cdot P_2)
-(\cev{Z_1}\cdot P_2)(P_1\cdot \vec{Z_2}) \right)^J\\&
\Big((Z_2 \cdot Y_1)(P_2\cdot Y_2)
-(Z_2 \cdot Y_2)(P_2\cdot Y_1) \Big)^J
\nonumber\\
=\big[&(P_1 \cdot Z)(P\cdot Y_1)(P_2 \cdot P_1)(P_2\cdot Y_2) -(P_1 \cdot P)(Z\cdot Y_1)(P_2 \cdot P_1)(P_2\cdot Y_2)  \nonumber\\
&(P_1 \cdot Z)(P\cdot P_2)(Y_2 \cdot P_1)(P_2\cdot Y_1) -(P_1 \cdot P)(Z\cdot P_2)(Y_2 \cdot P_1)(P_2\cdot Y_1) \\
&(P_1 \cdot P)(Z\cdot P_2)(Y_1 \cdot P_1)(P_2\cdot Y_2) -(P_1 \cdot Z)(P\cdot P_2)(Y_1 \cdot P_1)(P_2\cdot Y_2)
\nonumber\\
& (P_1 \cdot P)(Z\cdot Y_2)(P_1 \cdot P_2)(P_2\cdot Y_1) -(P_1 \cdot Z)(P\cdot Y_2)(P_1 \cdot P_2)(P_2\cdot Y_1) \big]^J
\nonumber
\end{align}
To perform the integrals we use the following trick
\begin{align}
&\int   
\frac{ dP_2\ \Omega(Z,P,P_1,P_2,Y_1,Y_2)}{
(-2P_1\cdot P_2)^{d-\D+J}
(-2 P_2\cdot X)^{\D+J}} \\
=& 
\frac{\G(\D-J)}{\G(\D+J)}
\Omega\left( P_2 \to \half \frac{\partial}{\partial X}\right)\int
\frac{ dP_2 }{
(-2P_1\cdot P_2)^{d-\D+J}
(-2 P_2\cdot X)^{\D-J}} \\
=& 
\frac{\G(\D-J)}{\G(\D+J)}
\Omega\left( P_2 \to \half \frac{\partial}{\partial X}\right) 
\frac{\pi^h \G(\D-J-h)}{\G(\D-J)}
\frac{(-X^2)^{h-\D+J} }{
(-2P_1\cdot X)^{d-\D+J} } 
\\
=& 
\frac{\pi^h \G(\D-J-h)}{\G(\D+J)}
(-X^2)^{h-\D+J}
\Omega\left( P_2 \to \half \frac{\partial}{\partial X}+(h-\D+J) \frac{X}{X^2}\right) 
\frac{ 1}{
(-2P_1\cdot X)^{d-\D+J} } \nonumber
\\
=& 
\frac{\pi^h \G(\D-J-h)}{(2J)!\G(\D+J)}
(-X^2)^{h-\D+J} 
\left( D_X \cdot \frac{\partial}{\partial P_2}  \right)^{2J} 
\frac{\Omega(Z,P,P_1,P_2,Y_1,Y_2) }{
(-2P_1\cdot X)^{d-\D+J} } 
\end{align}
where $h=d/2$
and 
\be
 D_X =\half \frac{\partial}{\partial X}+(h-\D+J) \frac{X}{X^2}\ .
\ee
Doing the integral over $P_1$ using the same technique we obtain
\begin{align}
&\int   
\frac{dP_1 dP_2\ \Omega(Z,P,P_1,P_2,Y_1,Y_2)}{
(-2 P\cdot P_1)^{\D+J}
(-2P_1\cdot P_2)^{d-\D+J}
(-2 P_2\cdot X)^{\D+J}} \\
=&
\frac{\pi^d \G(\D-J-h)\G(h-\D-J)}{(2J)!^2\G(\D+J)\G(d-\D+J)}
(-X^2)^{h-\D+J} 
\left( D_X \cdot \frac{\partial}{\partial P_2}  \right)^{2J}  \\&
\left( \half \frac{\partial}{\partial X}  \cdot \frac{\partial}{\partial P_1}  \right)^{2J}
\frac{\Omega(Z,P,P_1,P_2,Y_1,Y_2) }{
(-2P\cdot X)^{\D+J} } 
(-X^2)^{\D-h+J}
\nonumber 
\\
=&
\frac{\pi^d \G(\D-J-h)\G(h-\D-J)}{(2J)!^2\G(\D+J)\G(d-\D+J)}
\\&
\left(\half \frac{\partial}{\partial X}\cdot \frac{\partial}{\partial P_2}  \right)^{2J} (-X^2)^{h-\D+J} 
\left( \half \frac{\partial}{\partial X}  \cdot \frac{\partial}{\partial P_1}  \right)^{2J}
(-X^2)^{\D-h+J}
\frac{\Omega(Z,P,P_1,P_2,Y_1,Y_2) }{
(-2P\cdot X)^{\D+J} } 
\nonumber  \nonumber
\end{align}
It is not hard to see that expanding the derivatives in the last expression leads to  
\be
\sum_{n=0}^{4J} \frac{Q_n(X,Z,P,Y_1,Y_2)}{(-2 P\cdot X)^{\D+J+n}}
\ee
where $Q_n$ are homogeneous polynomials of degree $n$ in $X$, degree $(J+n)$ in $P$ and degree $J$ in $Z$, $Y_1$ and $Y_2$.
Moreover, the polynomials $Q_n$ inherit the following properties from the function $\Omega$,
\begin{align}
Q_n(X,Z,P , Y_1,Y_2)
&=Q_n(X,Z+\a P,P ,Y_1,Y_2)\\ 
&=Q_n(X, Z,P ,Y_1+\a Y_2,Y_2)\\
&=Q_n(X,Z,P ,Y_1,Y_2+\a Y_1)
\end{align}
This means that $Q_n$ can only depend on $Z$, $Y_1$ and $Y_2$ through the antisymmetric tensors
$Z^{[A}P^{B]}$ and  $Y_1^{[A}Y_2^{B]}$.
All these properties together, imply that
$Q_n$ must be proportional to 
\be
(-2 P\cdot X)^{n}\left((Z \cdot Y_1)(P\cdot Y_2)
-(Z \cdot Y_2)(P\cdot Y_1) \right)^J\ .
\ee
Therefore, we conclude that
\begin{align}
&\left(\half \frac{\partial}{\partial X}\cdot \frac{\partial}{\partial P_2}  \right)^{2J} (-X^2)^{h-\D+J} 
\left( \half \frac{\partial}{\partial X}  \cdot \frac{\partial}{\partial P_1}  \right)^{2J}
(-X^2)^{\D-h+J}
\frac{\Omega(Z,P,P_1,P_2,Y_1,Y_2) }{
(-2P\cdot X)^{\D+J} }  \nonumber\\=&
A_{\D,J} \frac{\left((Z \cdot Y_1)(P\cdot Y_2)
-(Z \cdot Y_2)(P\cdot Y_1) \right)^J}{(-2 P\cdot X)^{\D+J}}
\label{diffopsresult}
\end{align}
for some constant $A_{\D,J}$.
Putting everything together, the normalization constant is given by
\be
\Ncal_{\D,J}= 
\frac{2^J\pi^d 
\G\left(\D-\frac{d}{2}-J\right)
\G\left(\frac{d}{2}-\D-J\right)
}
{(2J)!^2\G(\D+J) }
A_{\D,J}\ .
\ee
Finally, we conjecture that
\be
A_{\D,J}=
 (-1 )^J 2^{-2J} (2 J)! ^2 (\Delta -1)_J
   \left(\Delta-\frac{d}{2} +1\right)_J 
   \left(\Delta-\frac{d}{2}-J  \right)_J
 (\Delta- d -J +2)_J\ .
\ee
Using \emph{Mathematica} we verified this formula up to $J=3$. Unfortunately, higher values of $J$ take too much time to compute all the derivatives in (\ref{diffopsresult}).

\subsection{Conformal integrals}
\label{ap:conformalintegrals}
\subsubsection{Integration over one point}
The basic integral we need is given by  Symanzik's formula
\be
I=\int dQ \prod_{\mu=1}^n [P_\mu,Q]^{\D_\mu}=
\pi^{\frac{d}{2}} \int [d\g_{\mu\nu}] \prod_{1\le \mu<\nu\le n} [P_{\mu\nu}]^{\g_{\mu\nu}} \label{eq:BasicSimanzik}
\ee
where $\sum_{\mu=1}^n\D_\mu=d$ and the measure $[d\g_{\mu\nu}]$ is the usual measure over the constraint surface $\sum_{\nu:\nu\neq \mu}^n \g_{\mu\nu} = \D_\mu$.
It will be useful to write this integral in alternative ways,
\begin{align}
I&=\int_0^\infty \prod_{\mu=1}^n dt_\mu t_\mu^{\D_\mu-1}
\int dQ e^{2Q\cdot \left(\sum t_\mu P_\mu \right)}\\
&=\int_0^\infty \prod_{\mu=1}^n dt_\mu t_\mu^{\D_\mu-1}
\int dQ e^{2Q\cdot T}
\int_0^\infty ds \,\d\left(s-\sum_{\m=1}^n t_\mu\right)
\\
&=\G(d) \int_0^\infty \prod_{\mu=1}^n dt_\mu t_\mu^{\D_\mu-1}\d\left(1-\sum_{\m=1}^n t_\mu\right)
\int \frac{dQ}{\left(-2Q\cdot T\right)^d} 
\\
&=\pi^{d/2} \int_0^\infty \prod_{\mu=1}^n dt_\mu t_\mu^{\D_\mu-1}\d\left(1-\sum_{\m=1}^n t_\mu\right)
  \frac{\G(d/2)}{\left(- T^2\right)^{d/2}}
\end{align}
where $t_\m$ are real variables, $T^A=\sum_{\mu=1}^n t_\mu P^A_\mu$ are vectors in the embedding space $\mathbb{M}^{d+2}$ and we have used results of \cite{SimmonsDuffin:2012uy} (for example, equation (2.21)).
Consider now the more general integral
\be
I^{A_1\dots A_l}=\int dQ \,Q^{A_1}\dots Q^{A_l}\prod_{\mu=1}^n [P_\mu,Q]^{\D_\mu}\label{eq:SpinLOnePointIntegrationStarter}
\ee
where  $\sum_{\mu=1}^n \D_\mu=d+l$.
We can write
\begin{align}
I^{A_1\dots A_l} 
&=\G(d+l) \int_0^\infty \prod_{\mu=1}^n dt_\mu t_\mu^{\D_\mu-1}\d\left(1-\sum t_\mu\right)
\int dQ\frac{Q^{A_1}\dots Q^{A_l}}{\left(-2Q\cdot T\right)^{d+l}} 
\\
&=\pi^{d/2} \G(d/2+l)\int_0^\infty \prod_{\mu=1}^n dt_\mu t_\mu^{\D_\mu-1}\d\left(1-\sum t_\mu\right)
  \frac{T^{A_1}\dots T^{A_l}-{\rm traces}}{\left(- T^2\right)^{d/2+l}}
  \\
&=\pi^{d/2} 
\sum_{\alpha_i=1}^n P_{\alpha_1}^{A_1} \dots P_{\alpha_l}^{A_l} \int [d\g_{\mu\nu}^{(\a)}] \prod_{1\le \mu<\nu\le n} [P_{\mu\nu}]^{\g_{\mu\nu}^{(\a)}}-{\rm traces} \label{eq:SpinLOnePointIntegration}
\end{align}
where the integration variables have to satisfy a constraint that depends on the set of $\a_1,\dots, \a_l$, namely $\sum_{\nu:\nu\neq\mu}\g_{\mu\nu}^{(\a)}=
\D_\mu+ 
\d_\mu^{\alpha_1}+\dots+\d_\mu^{\alpha_l}$.
In practice we will often need to compute only a piece of  (\ref{eq:SpinLOnePointIntegration}) because we will have a special point that we can now call $P_1$, and we will be interested only in the terms of $I^{A_1\dots A_l}$ that are not proportional to $P_1^{A_i}$ for any $i=1, \dots l$. With this simplification we can rewrite the integral $I^{A_1}$ in such a way to avoid the dependence on $\a$ of the constraint as follows
\begin{align}
I^{A_1}=\pi^{d/2} 
\sum_{\alpha_i=2}^n P_{\alpha_1}^{A_1} \int [d\g_{\mu\nu}] \prod_{2\le \mu<\nu\le n} [P_{\mu\nu}]^{\g_{\mu\nu}} \prod_{2\le \mu\le n}[P_{1 \mu}]^{\g_{\mu} +\d_\m^{\a_1}} + \dots \label{eq:SpinLOnePointIntegrationSpecialPoint}
\end{align}
where the dots stand by contributions proportional to $P_1^{A_1}$. We defined
\be
\g_{\m}=-\sum_{\n=2 \atop \m \neq \n}^n \g_{\m \n}+ \D_\m 
\ee
in such a way that  there is only one constraint to impose on the integration variables, namely 
\be 
\sum_{\m,\n=2  \atop \m \neq \n}^n \g_{\m \n}= \sum_{\m=2}^n \D_\m -\D_1+1=d+2-2 \D_1 \ .
\ee
We can similarly rewrite $I^{A_1 A_2}$ neglecting terms proportional to $P_1^{A_1}$ and $P_1^{A_2}$. In this case we have to be more careful since we also need to subtract the trace
\begin{align}
I^{A_1 A_2} 
=&  \pi^{d/2}  \sum_{\a_1,\a_2=1}^n \left( P_{\a_1}^{A_1} P_{\a_2}^{A_2}- \frac{\eta^{A_1 A_2}}{(d+2)} \frac{P_{\a_1 \a_2}}{-2} \right)\int [d\g_{\mu\nu}^{(\a)}] \prod_{1\le \mu<\nu\le n} [P_{\mu\nu}]^{\g_{\mu\nu}^{(\a)}} \\
=&\pi^{d/2}  \sum_{\a_1,\a_2=2}^n P_{\a_1}^{A_1} P_{\a_2}^{A_2} \int [d\g_{\mu\nu}^{(\a)}] \prod_{1\le \mu<\nu\le n} [P_{\mu\nu}]^{\g_{\mu\nu}^{(\a)}} + \dots  \nonumber \\
&-\pi^{d/2}  \frac{\eta^{A_1 A_2}}{-2(d+2)} \sum_{\a_1,\a_2=1 \atop \a_1 \neq \a_2 }^n   \int [d\g_{\mu\nu}^{(\a)}] (\g_{\a_1 \a_2}^{(\a)}-1) \prod_{1\le \mu<\nu\le n} 
[P_{\mu\nu}]^{\g_{\mu\nu}^{(\a)}-\d^{\a_1}_\m \d^{\a_2}_\n- \d^{\a_2}_\m \d^{\a_1}_\n }  
\end{align}
where the dots are the terms proportional to $P_1^{A_1}$ and $P_1^{A_2}$, and where in the second line we just absorbed $P_{\a_1 \a_2}$ in the integrand. Now we can simplify the first integral as we did in (\ref{eq:SpinLOnePointIntegrationSpecialPoint}). Moreover we can shift the integration variables of the second integral as follows $\g_{\mu\nu}^{(\a)}-\d^{\a_1}_\m \d^{\a_2}_\n- \d^{\a_2}_\m \d^{\a_1}_\n=\hat \g_{\m \n}$. In this way the sum over $\a_1$ and $\a_2$ only acts on the variable $\hat \g_{\a_1 \a_2}$ and can therefore be simplified using $\sum_{\a_1,\a_2=1 \atop \a_1 \neq \a_2}^n \hat \g_{\a_1 \a_2}=d+2$. The final result is 
\begin{align} 
I^{A_1 A_2}\;=&\;\; \pi^{d/2}  \sum_{\a_1,\a_2=2}^n P_{\a_1}^{A_1} P_{\a_2}^{A_2} \int [d\g_{\mu\nu}] \prod_{2\le \mu<\nu\le n} [P_{\mu\nu}]^{\g_{\mu\nu}} \prod_{2\le \mu \le n} [P_{1 \mu}]^{\g_{\mu}+\d_\m^{\a_1}+\d_\m^{\a_2}}  + \dots  \nonumber \\
& \; -\pi^{d/2}  \frac{\eta^{A_1 A_2}}{-2}    \int [d \hat \g_{\m \n}]  \prod_{2\le \mu<\nu\le n} 
[P_{\mu\nu}]^{\hat \g_{\mu\nu} }   \prod_{2\le \mu \le n} 
[P_{1 \mu}]^{\hat \g_{\mu} } \ , \label{eq:Spin2OnePointIntegrationSpecialPoint}
\end{align}
where the integration variables satisfy the constraint
\begin{eqnarray}
\g_{\m}= -\sum_{\myatop{\n=2 }{\m \neq \n}}^n \g_{\m \n} +\D_{\m} \, \qquad\qquad &&\sum_{\m ,\n=2 \atop \m \neq \n}^{n} \g_{\m \n}=\sum_{\m=2}^n \D_\m -\D_1 +2=d+4-2\D_1 \ ,\\
\hat \g_{\m}= -\sum_{\n=2 \atop \m \neq \n}^n \hat \g_{\m \n} +\D_{\m} \ ,\qquad\qquad& &\sum_{\m ,\n=2 \atop \m \neq \n}^{n} \hat \g_{\m \n}=\sum_{\m=2}^n \D_\m -\D_1=d+2-2\D_1 \ .
\end{eqnarray}
\subsubsection{Conformal integral - integrating over two points  }
\label{ap:ConfInt2p}
\subsubsection*{Scalar conformal integral}
The goal of this section is to compute the integral, 
\begin{align}
I=\int dQ_1 dQ_2 \,\,[Q_{12}]^{d-\D}\prod_{1\leq a\leq k}[P_a,Q_1]^{\l_a}\prod_{k< i\leq n}[P_i,Q_2]^{\r_i} \label{eq:ScalarIntegralBasicShadow}
\end{align} 
where the variables $\l_a$ and $\r_i$ satisfy, 
\begin{align}
\sum_{a=1}^k\l_a=\D\,, \ \ \ \ \ \ \ \sum_{i=k+1}^n\r_i=\D.
\end{align}

Let us compute first the $Q_1$ integral, 
\begin{align}
\int dQ_1 \,[Q_{12}]^{d-\D} \prod_{1\leq a\leq k}[P_a,Q_1]^{\l_a} =\pi^{\frac{d}{2}}\int [d\b] \prod_{1\leq a\leq b\leq k}[P_{ab}]^{\b_{ab}} \prod_{1\leq a\leq k} [P_a,Q_2]^{\b_a}
\end{align}
 where $\sum_{a=1}^k\b_a=d-\D$ and $\sum_{b:a\neq b}\b_{ab}=\l_a-\b_{a}$, in particular we have $\sum_{a,b:a\neq b}\b_{ab}=2\D-d$. The integration over $Q_2$ can also be done using Symanzik's rule, 
\begin{align}
\int dQ_2 \prod_{k< i\leq n}[P_i,Q_2]^{\r_i} \prod_{1\leq a\leq k} [P_a,Q_2]^{\b_a} =
\pi^{\frac{d}{2}}
\int [d\t] \prod_{1\leq \mu<\nu\leq n}[P_{\mu\nu}]^{\t_{\mu\nu}}
\end{align}
where the variables $\tau_{\mu\nu}$ satisfy $\sum_{\mu=1}^n\tau_{\mu\nu}=0$, $\tau_{aa}=-\b_a$ $\tau_{ii}=-\r_i$. The integral $I$ can be written as
\begin{align}
I=\pi^d \int [d\b][d\t] \prod_{1\leq a<b\leq k}[P_{ab}]^{\b_{ab}} \prod_{1\leq \mu <\nu\leq n}[P_{\mu\nu}]^{\tau_{\mu\nu}}.
\end{align}
We can change variables, 
\be
\t_{\mu\nu} = 
\left\{
\begin{array}{ l l }
  \g_{ab} -\b_{ab}\ \ \ \ \   
  & {\rm if}\ \  \mu=a\le k \ \  {\rm and} \ \ \nu=b\le k \\
\g_{\mu\nu} & {\rm otherwise}\\
\end{array}
 \right. \ .
\ee
The function $I$ can be rewritten as, 
\begin{align}
&I=\pi^d 
\int [d\g][d\b]\prod_{1\leq a< b\leq k}\frac{\G(\b_{ab})\G(\g_{ab}-\b_{ab})}{\G(\g_{ab})} \prod_{1\leq \mu \leq \nu \leq n}[P_{\mu\nu}]^{\g_{\mu\nu}} \ ,
\end{align}
 where the integration variables $\g_{\m \n}$ satisfy the following constraints
\be
\sum_{\m=1}^n \g_{\m \n} =0 \ , \qquad  \g_{aa}=-\l_a \ , \qquad \g_{ii}=-\r_i \ ,\qquad \mbox{ for } \left\{ \begin{array}{l} a=1,\dots k \\ i=k+1,\dots n \end{array} \right. \ .
\ee
The function $I$ is then given by, 
\begin{align}
&I=\pi^d \; 
\int [d\g]
\frac{\G(B)\G(A-B)}{\G(A)}
\prod_{1\leq \mu \leq \nu \leq n}[P_{\mu\nu}]^{\g_{\mu\nu}}\label{eq:SimpleScalarIntegralFinal}
\end{align}
with $B=\sum_{1\leq a<b\leq k}\b_{ab}=\frac{2\D-d}{2}$ and $A=\sum_{1\leq a<b\leq k}\g_{ab}=\frac{\D-\g_{LR}}{2}$. In the derivation of this result we have used the identity (\ref{eq:BarnesImportantIdentityConst}).

\subsubsection*{Vector integral}
The goal of this section is to compute the conformal integral,
\begin{align}
I_{l,\,r}=\int dQ_1 dQ_2 \,\,(P_l\cdot Q_2) (P_r\cdot Q_1)[Q_{12}]^{d-\D+1}\prod_{1\leq a\leq k}[P_a,Q_1]^{\l_a+\d_{a}^l}\prod_{k< i\leq n}[P_i,Q_2]^{\r_i+\d_{i}^r} \label{eq:StarterVectorConformalIntegrallr}
\end{align}
where the variables $\l_a$ and $\r_i$ satisfy,
\begin{align}
\sum_{a=1}^k\l_a=\D-1\,, \ \ \ \ \ \ \  \ \ \sum_{i=k+1}^n\r_i=\D-1.
\end{align}
This integral enters in the factorization of the vector and spin two, where we have the transversality condition (\ref{TransverseJ}).   A moment of thought shows that it is sufficient to compute  $I_{l,r}$ up to terms proportional to $\l_l$ or $\r_r$, so in the following we will drop these. Let us integrate first over $Q_1$, using (\ref{eq:SpinLOnePointIntegrationSpecialPoint}) we have
\begin{align}
&\int dQ_1 \; (Q_1\cdot P_r) \; [Q_{12}]^{d-\D+1}\prod_{1\leq a\leq k}[P_a,Q_1]^{\l_a+\d_{a}^l}\nonumber\\
& = \pi^{\frac{d}{2}}\sum_{c=1}^k  \frac{P_{c r}}{-2} \int [d\b] \prod_{1\leq a<b\leq k}[P_{ab}]^{\b_{ab}}\prod_{1\leq a\leq k}[P_{a},Q_2]^{\b_{a}+\d_{a}^l+\d_{a}^{c}}+\dots
\end{align}
where the dots stand by terms proportional to $P_{r}\cdot Q_2$ that we can drop since they give rise to a  contribution proportional to $\r_r$ and where the variables $\b_{ab}$ satisfy 
\begin{align}
\b_{a}=-\sum_{b=1}^k\b_{ab}\ , \qquad \qquad \b_{a a}=-\l_a \ , \qquad  \qquad \sum_{a,b=1 \atop a\neq b}^k\b_{ab}=
  2\D-d \ .
\end{align}
 Now we compute the integral over  $Q_2$ which is also of Symanzik type. Using (\ref{eq:BasicSimanzik}) and shifting the integration variables in order to absorb the factors $P_{c r}$ and $[P_{ab}]^{\b_{ab}}$ in a single term, we obtain
\begin{align}
&I_{l,r}=\frac{\pi^{d}}{4}\sum_{c=1}^{k}\int [d\b] [d\g]\;  \g_{cr} 
  (\b_l+\d_{l}^{c})
  \prod_{1\leq \mu<\nu\leq n} [P_{\mu\nu}]^{\g_{\mu\nu}}
  \prod_{1\leq a<b\leq k}\frac{\G(\b_{ab})\G(\g_{ab}-\b_{ab})}{\G(\g_{ab})} \ ,\nonumber
\end{align}
 where the integration variables $\g_{\m \n}$ have to satisfy the following constraints
\be
\sum_{\m=1}^n \g_{\m \n} =0 \ , \qquad  \g_{aa}=-\l_a \ , \qquad \g_{ii}=-\r_i \ ,\qquad \mbox{ for } \left\{ \begin{array}{l} a=1,\dots k \\ i=k+1,\dots n \end{array} \right. \ .
\ee
To integrate over $\b$ we use (\ref{eq:BarnesImportantIdentityConst}) and (\ref{eq:BarnesImportantIdentityConsAnother}). The function $I_{l,r}$ can be simplified to 
\begin{align}
&I_{l,r}=\frac{\pi^{d}}{4}   \sum_{c=1}^{k} \int [d\g]\;\frac{\G(B)\G(A-B)}{\G(A+1)}\;
\g_{cr} \left(A(\l_l+\d_{l}^{c})+B(\g_{l}-\l_{l}) \right)  \prod_{1\leq \mu<\nu\leq n} [P_{\mu\nu}]^{\g_{\mu\nu}}\nonumber
\end{align}
where $A=\frac{\D-1-\g_{LR}}{2}$ and $B = \frac{2\D-d}{2}$ and where we defined as usual $\g_l=\sum_{i=k+1}^n \g_{l i}$.
Simplifying the sum over $c$ and dropping terms proportional to $\l_l$ we finally obtain \footnote{The full integral, without dropping terms proportional to $\l_l$ or $\r_r$,  is obtained  by adding 
\begin{align}
\frac{1}{B-1}
\big[(A-B)(1+A-B)\l_l\r_r+(A-B)(B-1)(\g_{r}\l_{l}+\g_{l}\r_r)\big]\ .
\end{align}
to the bracket in the integrand of   (\ref{eq:FinalSpinOneVectorIntegralILR}).
}
\begin{align}
I_{l,r}=&\frac{\pi^d}{4} \int [d\g] \; \frac{\G(B)\G(A-B)}{\G(A+1)} \;  \big(A\g_{rl}+B\g_{l}\g_{r}\big)  \prod_{1\leq \mu<\nu\leq n} [P_{\mu\nu}]^{\g_{\mu\nu}}\label{eq:FinalSpinOneVectorIntegralILR} \ .
\end{align}

\subsubsection*{Spin two integral}
The goal of this section is to evaluate the integral,
\begin{align}
I_{l_1l_2,\,r_1 r_2}=&\int dQ_1 dQ_2 \,\,(P_{l_1}\cdot Q_2) (P_{l_2}\cdot Q_2) (P_{r_1}\cdot Q_1) ( P_{r_2}\cdot Q_1)\nonumber\\
&[Q_{12}]^{d-\D+2}\prod_{1\leq a\leq k}[P_a,Q_1]^{\l_a+\d_{a}^{l_1}+\d_{a}^{l_2}}\prod_{k\leq i\leq n}[P_i,Q_2]^{\r_i+\d_{i}^{r_1}+\d_{i}^{r_2}} \label{eq:SPinSpinTwoTwoIntegralStarter}
\end{align}
where
\be
\sum_{a=1}^k \l_{a}=\D-2 \qquad \qquad  \qquad \sum_{i=k+1}^n \r_{i}=\D-2 \ .
\ee
Such integral enters the factorization formula of the spin two operator where we can neglet terms proportional to $\r_{r_1}+\d_{r_1}^{r_2}$ or $\l_{l_1}+\d_{l_1}^{l_2}$ (and also for terms with $l_1\leftrightarrow l_2$ and $r_1\leftrightarrow r_2$) because they do not contribute in the final result due to the transversality condition (\ref{TransverseJ}).
As in the previous cases we start by  doing the integral over $Q_1$ 
\begin{align}
I_1=\int dQ_1 (P_{r_1}\cdot Q_1)  (P_{r_2}\cdot Q_1) [Q_{12}]^{d-\D+2}\prod_{1\leq a\leq k}[P_a,Q_1]^{\l_a+\d_{a}^{l_{1}}+\d_{a}^{l_{2}}} \ .
\end{align}
Using (\ref{eq:Spin2OnePointIntegrationSpecialPoint}) we obtain
\begin{align}
I_1 
=&\; \; \pi^{d/2} 
 \sum_{a_1,a_2=1}^k\frac{P_{r_1a_1}P_{r_2 a_2}}{4} \int [d\b_{ab}]\prod_{1\leq a\leq k}\big[P_{a},Q_2\big]^{\b_{a}+\d_{a}^{l_1}+\d_{a}^{l_2}+\d_{a}^{a_1}+\d_{a}^{a_2}} \prod_{1\le a<b\le k} [P_{ab}]^{\b_{ab}}+ \dots \nonumber\\
&- \pi^{d/2} (1-\d_{r_1}^{r_2})   \frac{P_{r_1r_2}}{4}\int [d \hat \b_{ab}]\prod_{1\leq a\leq k}\big[P_{a},Q_2\big]^{\hat \b_{a}+\d_{a}^{l_1}+\d_{a}^{l_2}} \prod_{1\le a<b\le k} [P_{ab}]^{\hat \b_{ab}}.
\end{align}
where the dots stand for contributions proportional to $Q_2 \cdot P_{r_1}$ and  $Q_2 \cdot P_{r_2}$ that do not contribute to the final result due to transversality. The integration variables satisfy the constraints
\begin{eqnarray}
\b_{a}=-\sum_{b=1 \atop a\neq b }^k \b_{a b} +\l_a \ ,&\qquad \qquad &\sum_{a,\,b=1 \atop a\neq b}^k \b_{ab}=  2\D-d  \ , \\
\hat \b_{a}=-\sum_{b=1\atop a\neq b}^k \hat \b_{a b} +\l_a  \ ,&\qquad \qquad &\sum_{a,\,b=1 \atop a\neq b }^k \hat \b_{ab}=2\D-d-2  \ .
\end {eqnarray}
Let us rewrite rewrite the two integrands absorbing the factor $(Q_2\cdot P_{l_1})  (Q_2 \cdot P_{l_2}) $,
\begin{align}
&(Q_2\cdot P_{l_1})   (Q_2\cdot P_{l_2}) \prod_{1\leq a \leq k}[P_{a},Q_2]^{\b_{a}+\d_{a}^{l_1}+\d_{a}^{l_2}+\d_{a}^{a_1}+\d_{a}^{a_2}} \nonumber \\
&\qquad =\frac{1}{4} (\b_{l_1}+\d_{l_1}^{a_1}+\d_{l_1}^{a_2}+\d_{l_1}^{l_2})(\b_{l_2}+\d_{l_2}^{a_1}+\d_{l_2}^{a_2})\prod_{1\leq a \leq k}[P_{a},Q_2]^{\b_{a}+\d_{a}^{a_1}+\d_{a}^{a_2}} \ ,\\
&(Q_2\cdot P_{l_1})  (Q_2 \cdot P_{l_2}) \prod_{1\leq a \leq k}[P_{a},Q_2]^{\hat \b_{a}+\d_{a}^{l_1}+\d_{a}^{l_2}}=\frac{1}{4}  (\hat \b_{l_1}+\d_{l_1}^{l_2})\hat \b_{l_2} \prod_{1\leq a \leq k}[P_{a},Q_2]^{\hat \b_{a}} \ .
\end{align}
The integral over $Q_2$ can be done using Symanzik formula
%
\begin{align}
&\!\!\! I_{l_1l_2r_1r_2} = \pi^d \int [d\g]\prod_{1\leq \mu<\nu\leq n} [P_{\mu\nu}]^{\g_{\mu\nu}} 
\biggl[ -\g_{r_1r_2}(1-\d_{r_1}^{r_2}) \int [d\hat{\b}](\hat{\b}_{l_1}+\d_{l_1}^{l_2})\hat{\b}_{l_2}\prod_{1\leq a <b\leq k} {\textstyle \frac{\G(\hat{\b}_{ab})\G(\g_{ab}- \hat{\b}_{ab})}{16 \; \G(\g_{ab})} }\nonumber \\
& \!\!\!\!\!  +\sum_{a_1, a_2=1}^k \g_{a_2r_2}(\g_{r_1 a_1}+\d_{a_1}^{a_2}\d_{r_1}^{r_2}) \int [d\b]   (\b_{l_1}+\d_{l_1}^{l_2}+\d_{a_1}^{l_1}+\d_{a_2}^{l_1})(\b_{l_2}+\d_{a_1}^{l_2}+\d_{a_2}^{l_2})\prod_{1\leq a <b\leq k}  {\textstyle \frac{\G(\b_{ab})\G(\g_{ab}-\b_{ab})}{16 \; \G(\g_{ab})} }
\biggr]\nonumber
\end{align} 
where we have shifted the integration variables to repack all the factors $[P_{ab}]^{\b_{ab}}$, $(P_{l_1}\cdot Q_2)$, $(P_{l_2}\cdot Q_2)$, $P_{r_1r_2}$ in the single term $[P_{\mu\nu}]^{\g_{\mu\nu}} $. The new integration variables $\g_{\m \n}$ are constrained as follows
\be
\sum_{\n=1 \atop \n \neq a}^n \g_{a \n} = \l_a \qquad \qquad  \sum_{\n=1 \atop \n \neq i}^n \g_{i \n} = \r_i \qquad \qquad \mbox{ for } \left\{ \begin{array}{l} a=1,\dots k \\ i=k+1,\dots n \end{array} \right. \ .
\ee
%
We then integrate over $\b_{ab}$ and $\hat{\b}_{ab}$ using  (\ref{eq:BarnesImportantIdentityConst}),(\ref{eq:BarnesImportantIdentityConsAnother}) and (\ref{eq:BarnesImportantIdentityConsAnother2}).  Thus we conclude that
\begin{align}
&I_{l_1l_2r_1r_2}
=\pi^d \int [d\g] \; \frac{\G(A-B)\G(B)}{\G(A+2)}  \; \mathcal{Y}_{l_1l_2r_1r_2}\prod_{1\leq \mu<\nu\leq n}[P_{\mu\nu}]^{\g_{\mu\nu}} \ ,
\end{align}
where $\mathcal{Y}_{l_1l_2r_1r_2}$ is given by the sum of the following structures
\begin{align}
\mathcal{Y}^{(1)}_{l_1l_2r_1r_2}=& 2 A (A+1)\g_{l_1r_1}(\g_{l_2r_2}+\d_{l_1}^{l_2}\d_{r_1}^{r_2}) \\ 
\mathcal{Y}^{(2)}_{l_1l_2r_1r_2}=& 4 B(A+1)\g_{l_1r_1}(\g_{l_2}+\d_{l_1}^{l_2})(\g_{r_2}+\d_{r_1}^{r_2}) \\
\mathcal{Y}^{(3)}_{l_1l_2r_1r_2}=& B(B+1)\g_{l_2}\g_{r_2}(\g_{l_1}+\d_{l_1}^{l_2})(\g_{r_1}+\d_{r_1}^{r_2})\\
\mathcal{Y}^{(4,L)}_{l_1l_2r_1r_2}=& B(A-B)\g_{l_1l_2}\g_{r_2}(\g_{r_1}+\d_{r_1}^{r_2})(1-\d_{l_1}^{l_2})\\
\mathcal{Y}^{(4,R)}_{l_1l_2r_1r_2}=& B(A-B)\g_{r_1r_2}\g_{l_1}(\g_{l_2}+\d_{l_1}^{l_2})(1-\d_{r_1}^{r_2})\\
\mathcal{Y}^{(5)}_{l_1l_2r_1r_2}=& (A-B+1)(A-B)(1-\d_{l_1}^{l_2})(1-\d_{r_1}^{r_2})\g_{l_1l_2}\g_{r_1r_2} 
\, , \label{eq:YconstantforSpinTwointegralFactorization}
\end{align}
with 
\begin{eqnarray}
\! A=\sum_{1 \leq a < b \leq k} \g_{a b}=\frac{\D -2-\g_{LR}}{2} \ , \quad  && B=\sum_{1 \leq a < b \leq k} \b_{a b}=1+\sum_{1 \leq a < b \leq k} \hat\b_{a b}=\frac{2\D -d}{2} \ , \qquad
\end{eqnarray}
and where the variables $\g_\m$ are defined as usual by 
\begin{eqnarray}
\g_a=\sum_{i=k+1}^n \g_{a i} \ , \qquad \qquad &&\g_i=\sum_{a=1}^k \g_{ai} \ , \qquad \qquad  \mbox{ for } \left\{ \begin{array}{l} a=1,\dots k \\ i=k+1,\dots n \end{array} \right. \ .
\end{eqnarray}
\subsection{Constrained Mellin integral identity}
\label{ap:MellinIden}
The goal of this section is to analyze an integral over Mellin variables $\beta_{ab}$ constrained by $\sum_{1\leq a< b\leq k }\b_{ab}=B$. Using recursively the first Barnes lemma
 we can prove the following identity 
 \footnote{
Imposing the constraint $\sum_{1 \leq a < b \leq k} \b_{a b}=B$, one can solve for $\b_{12}$ in terms of the other $(k+1)(k-2)/2$ variables  $\b_{1a}$ for $a=3,\dots,k$ and $\b_{ab}$ for $2\le a<b\le k$.
Then, the integrals over $$\int [d\b]=
\int_{-i\infty}^{i\infty}
\prod_{a=3}^k\frac{d\b_{1a}}{2\pi i} 
\prod_{2\le a<b\le k} \frac{d\b_{ab}}{2\pi i} $$ can be done by successive use of the first Barnes lemma.}
\begin{align}
\int [d\b] \prod_{1\leq a< b\leq k }\frac{\G(\b_{ab})\G(\a_{ab}-\b_{ab})}{\G(\a_{ab})}=\frac{\G(B)\G(A-B)}{\G(A)}\label{eq:BarnesImportantIdentityConst}
\end{align}
where $A=\sum_{1\leq a< b\leq k}\a_{ab}$. This type of integral can be easily generalized to the case where we have also a linear or quadratic dependence in $\b_{ab}$. Let us consider first the linear case
\begin{align}
\int [d\b] \b_{f_1p_1}\prod_{1\leq a<b\leq k}\frac{\G(\b_{ab})\G(\a_{ab}-\b_{ab})}{\G(\a_{ab})}=\a_{f_1p_1} \frac{\G(B+1)\G(A-B)}{\G(A+1)}\label{eq:BarnesImportantIdentityConsAnother}.
\end{align}
where we have shifted the integration variables to reduce this case to the previous one. Given a function defined by
\be
\mathcal{F}_{\{f_i, p_i\}}(\b_{ab})\equiv  \prod_{j=1}^J\left(\b_{f_j p_j}+\sum_{\ell=1}^{j-1} \d_{f_j}^{f_\ell}\d_{p_j}^{p_\ell}+\d_{f_j}^{p_\ell}\d_{p_j}^{f_\ell}\right),
\ee
it easy to check that, shifting the integration variables, formula (\ref{eq:BarnesImportantIdentityConst}) can be generalized as follows 
\begin{align}
\int [d\b]\mathcal{F}_{\{f_i, p_i\}}(\b_{ab})\prod_{1\leq a<b\leq k} \frac{\G(\b_{ab})\G(\a_{ab}-\b_{ab})}{\G(\a_{ab})}=\mathcal{F}_{\{f_i, p_i\}}(\a_{ab}) \frac{\G(B+J)\G(A-B)}{\G(A+J)}\label{eq:BarnesImportantIdentityGeneric}.
\end{align}
We can now generalize this type of integrals to the case of any polynomial dependence in $\b_{ab}$ just taking linear combination of (\ref{eq:BarnesImportantIdentityGeneric}), since $\mathcal{F}_{\{f_i, p_i\}}(\b_{ab})$ can be used as a basis for the polynomials in $\b_{ab}$. A useful example is given by a quadratic term in $\b_{ab}$. In fact, using  $\b_{f_1p_1}\b_{f_2p_2}=(\b_{f_1p_1}+\d_{f_1}^{f_2}\d_{p_1}^{p_2} +\d_{f_1}^{p_2}\d^{f_2}_{p_1} -\d_{f_1}^{f_2}\d_{p_1}^{p_2}-\d_{f_1}^{p_2}\d^{f_2}_{p_1})\b_{f_2p_2}$ and (\ref{eq:BarnesImportantIdentityGeneric}), we find
\begin{align}
&\!\!\int [d\b] \b_{f_1p_1}\b_{f_2p_2}\prod_{1\leq a<b\leq k}\frac{\G(\b_{ab})\G(\a_{ab}-\b_{ab})}{\G(\a_{ab})} \nonumber \\
&\quad =\a_{f_1p_1}\a_{f_2 p_2}\frac{\G(B+2)\G(A-B)}{\G(A+2)}- \a_{f_2 p_2}(\d_{f_1}^{f_2}\d_{p_1}^{p_2}+\d_{f_1}^{p_2}\d^{f_2}_{p_1})\frac{\G(B+1)\G(A-B+1)}{\G(A+2)}
\label{eq:BarnesImportantIdentityConsAnother2}\ .
\end{align}

\section{Factorization from the Casimir equation}

The goal of this appendix is to fill in the gaps in the derivation of the factorization formulas in section \ref{sec:FacCasimir} of the main text. In the next subsection, we detail the vector case and in subsection \ref{App. SpinJ=2 Fact Formula}, we discuss    the spin $J=2$ case. In subsection \ref{ap:CasimirFormulasJ1}, we prove some recurrence formulas that are useful  in the cases  $J=0,1,2$. In the last subsection \ref{App:Q0DownIndex}, we provide more details about the residue of the first pole of the Mellin amplitude associated with the exchange of an operator of general spin $J$. In particular, we will match with the results for the four point function found in \cite{Mack, Costa:2012cb}.

\subsection{Factorization for spin $J=1$} \label{App:J=1CasimirFac}
In this subsection we prove formulas (\ref{C(Q1)J=1}) and (\ref{C(Q2)J=1}).
First we consider that in the action of the casimir operator $\hat{C}$ defined in  (\ref{CasimirOperatorJm}) there is a term of the kind $[\,\cdot\,]^{a i, b j}_{a j, b i}$, which only shifts the mixed variable and it does not act on $L_m$ and $R_m$. Moreover the action of $[\,\cdot\,]^{a i, b j}_{a j, b i}$ on a mixed variable $\g_{cl}$ can be written in a simple way, namely
\be
[\g_{cl}]^{a i, b j}_{a j, b i}=\g_{cl}+ \d_{c}^{a} \d_{l}^{i}+\d_{c}^{b} \d_{l}^{j}-\d_{c}^{a} \d_{l}^{j}-\d_{c}^{b} \d_{l}^{i} \ .
\ee
So that we easily obtain
\begin{align}
&\sum_{a, b=1 \atop a\neq b}^k
\sum_{i ,j=k+1  \atop i\neq j}^n \g_{a i}\g_{bj}\left( \mathcal{Q}^{(1)}_{m}- [\mathcal{Q}^{(1)}_{m}]^{a i, b j}_{a j, b i}\right)=-2 (\D-1+2m)\mathcal{Q}^{(1)}_{m}+2 \mathcal{Q}^{(2)}_{m}   \ ,\label{rel1RES1}\\
&\sum_{a, b=1 \atop a\neq b}^k
\sum_{i ,j=k+1  \atop i\neq j}^n  \g_{a i}\g_{bj}\left( \mathcal{Q}^{(2)}_{m}- [\mathcal{Q}^{(2)}_{m}]^{a i, b j}_{a j, b i}\right)=0 \ .\label{rel1RES2}
\end{align}
The second part of the computation is more subtle since $\hat{C}$ also contains a term $[\mathcal{Q}_{m-1}]^{a b, i j}_{a i, b j}$ that has a shift both in the Mellin variables and in $m$. To simplify this term we find (see appendix \ref{ap:CasimirFormulasJ1}) the following recurrence relations to connect structures defined at $m-1$ to structures defined at $m$
\begin{align}
&L^c_m=\frac{1}{2 m}\sum_{a, b=1 \atop a\neq b}^k
 \g_{a b} \left[ L^c_{m-1} \right]^{a b} \label{Lcm1} \ ,\\
&\dot{L}_m=\sum_{a, b=1 \atop a\neq b}^k \g_{a b} \left[ L^a_{m-1} \right]^{a b} \ , \label{Lhatm2}\\
&\dot{L}_m=\frac{1}{2 (m-1)} \sum_{a, b=1 \atop a\neq b}^k \g_{a b} \left[ \dot{L}_{m-1} \right]^{a b}\ . \label{Lhatm1}
\end{align}
Using  (\ref{Lcm1}), (\ref{Lhatm2}), (\ref{Lhatm1}) we easily find
\begin{align}
&\sum_{a, b=1 \atop a\neq b}^k
\sum_{i ,j=k+1  \atop i\neq j}^n  \g_{ab}\g_{ij} [\mathcal{Q}^{(1)}_{m-1}]^{a b, i j}_{a i, b j}=-2 \mathcal{Q}^{(2)}_{m}+(2 m)^2 \mathcal{Q}^{(1)}_{m} \ , \label{rel2RES1}\\
&\sum_{a, b=1 \atop a\neq b}^k
\sum_{i ,j=k+1  \atop i\neq j}^n  \g_{ab}\g_{ij} [\mathcal{Q}^{(2)}_{m-1}]^{a b, i j}_{a i, b j}=4 (m-1)^2 \mathcal{Q}^{(2)}_{m}\label{rel2RES2}.
\end{align}
Formulas (\ref{C(Q1)J=1}) and (\ref{C(Q2)J=1}) descend respectively from  (\ref{rel1RES1}) and  (\ref{rel2RES1}) and from (\ref{rel1RES2}) and  (\ref{rel2RES2}).

\subsection{Factorization for spin $J=2$} \label{App. SpinJ=2 Fact Formula}
\subsubsection{Solving the $m=0$ case} 
In the case of spin $J=2$ we consider the following ansatz for the first residue
\be \label{res3J2}
\mathcal{Q}^{(1)}_{0}= \sum_{a, b=1 }^k
\sum_{i ,j=k+1 }^n  \g_{a i}\g_{bj} M^{a b}_L M^{i j}_R \ .
\ee
Acting with the Casimir equation operator (\ref{CasimirOperatorJm}) on (\ref{res3J2}) and using the transverse relation (\ref{TransverseJ})
we obtain
\begin{align}
& \hat{C}(\mathcal{Q}^{(1)}_{0})=- 2 \D \mathcal{Q}^{(2)}_{0}- 2  \mathcal{Q}^{(3)}_{0} \ , \qquad \mbox{with} \\ 
&\mathcal{Q}^{(2)}_{0}=\sum_{a=1 }^k
\sum_{i=k+1 }^n \g_{a i} M_L^{a a} M_R^{i i} \ ,\qquad
\mathcal{Q}^{(3)}_{0}= \left(\sum_{a=1 }^k
 \g_{a} M_L^{a a}\right) \left(\sum_{b=1 }^k
 \g_{j} M_R^{j j}\right).
\end{align}
The action of $ \hat{C}$ on $\mathcal{Q}^{(2)}_{0}$ is
\begin{align}
 \hat{C}(\mathcal{Q}^{(2)}_{0})&=2\D \mathcal{Q}^{(2)}_{0}  +2  \mathcal{Q}^{(3)}_{0}.\end{align}
It is trivial to see that $\mathcal{Q}^{(1)}_{0}+\mathcal{Q}^{(2)}_{0}$ solves the Casimir equation. Moreover we can rewrite the final result in a compact form as shown in formula (\ref{ResJ2m0}).

\subsubsection{Solving for a general $m$}
As a new ansatz we consider the natural generalization of (\ref{ResJ2m0}) to any $m$
\be \label{Res3J2m}
\mathcal{Q}^{(1)}_{m} = \sum_{a, b=1 }^k
\sum_{i ,j=k+1 }^n  \g_{a i}(\g_{bj}+\d^a_b \d^i_j) L^{a b}_m R^{i j}_m.
\ee
We start applying the Casimir operator (\ref{CasimirOperatorJm}) and the transverse relation (\ref{TransverseJ}) to (\ref{Res3J2m}). This action generates some structures. We then act with $\hat{C}$ on the new structures until its action closes. We find
\begin{align}
\hat{C}(f^{(1)}_m \mathcal{Q}^{(1)}_{m})=&\left[(\eta-4(\bar \g_{LR}+1))f^{(1)}_m+(2m)^2 f^{(1)}_{m-1}\right]\mathcal{Q}^{(1)}_{m} +4[f^{(1)}_m -f^{(1)}_{m-1}]\mathcal{Q}^{(2)}_{m}+2 f^{(1)}_{m-1} \mathcal{Q}^{(5)}_{m} \nonumber\\
\hat{C}(f^{(2)}_{m} \mathcal{Q}^{(2)}_{m})=&\left[(\eta-2 \bar \g_{LR})f^{(2)}_{m} +(2(m-1))^2 f^{(2)}_{m-1} \right] \mathcal{Q}^{(2)}_{m}+ 2 [f^{(2)}_{m}- f^{(2)}_{m-1}] \mathcal{Q}^{(3)}_{m}+2 f^{(2)}_{m-1} \mathcal{Q}^{(4)}_{m}  \nonumber \\
&-2 f^{(2)}_{m-1} \mathcal{Q}^{(5)}_{m}\nonumber\\
\hat{C}(f^{(3)}_{m}\mathcal{Q}^{(3)}_{m})=&\left[\eta f^{(3)}_{m}+(2(m-2))^2 f^{(3)}_{m-1}\right] \mathcal{Q}^{(3)}_{m}
+ 4(m-2)f^{(3)}_{m-1} \mathcal{Q}^{(4)}_{m}
+4f^{(3)}_{m-1}\mathcal{Q}^{(5)}_{m} 
\nonumber \\
\hat{C}(f^{(4)}_{m} \mathcal{Q}^{(4)}_{m})=&\left[\eta f^{(4)}_{m}+4(m-1)(m-2)f^{(4)}_{m-1}\right] \mathcal{Q}^{(4)}_{m} +8(m-1) f^{(4)}_{m-1} \mathcal{Q}^{(5)}_{m}\ ,\nonumber \\
 \hat{C}(f^{(5)}_{m} \mathcal{Q}^{(5)}_{m})=&\left[\eta f^{(5)}_{m} - (2 (m - 1))^2 f^{(5)}_{m-1}\right] \mathcal{Q}^{(5)}_{m}\nonumber
 \end{align}
 where the definitions of $ \mathcal{Q}^{(s)}_{m}$ are given in (\ref{QsJ=2}) and where $\bar \g_{LR}=\D +2m -2$. The computations are similar to the $J=1$ case and we had to make use of some recurrence relations that are written in appendix \ref{LmJ2Formulas}. One more time to fix the $f^{(s)}_m$  we need to set $\hat{C}(\sum_{s=1}^5 f^{(s)}_m \mathcal{Q}^{(s)}_m)=0$
 and require that $f^{(s)}_m=f^{(1)}_m h^{(s)}_m$ with $h^{(s)}_m$ rational functions of $m$. 
 In this way we get formula (\ref{finalresultJ2}).

\subsection{Recurrence relations}
\label{ap:CasimirFormulasJ1}

In this appendix we demonstrate formula (\ref{LmRecScalar}) for the scalar case and  (\ref{Lcm1}), (\ref{Lhatm2}), (\ref{Lhatm1}) for the vector case. We also show some similar formulas useful in the spin two case.\\

First we write some formulas important to demonstrate the following results. 
We often deal with the set of integer compositions 
\be
\mathcal{A}^{m}=\left\{ \{\l_1,\dots \l_n\} : \sum_{i=1}^n \l_i = m \,, \, \l_i\in \mathbb{N}_0  \right\} \ , \nonumber
\ee
where from now on we will denote $\{\l_1,\dots \l_n\} \equiv \{\l_i\} $.
We shall now show a simple property of the integer composition that will be very useful in the rest of this section. 
It is a trivial fact that for any $j\in \{1,\dots,n\}$ the set 
\be
\mathcal{A}^m_j \equiv \left\{\{\l_i+\d_i^j\} : \sum_{i=1}^n \l_i =m \,,\  \l_i \in \mathbb{N}_0 \right\} \nonumber
\ee 
is contained in the set $\mathcal{A}^{m+1}$. Moreover, it is clear that $\mathcal{A}_j^{m}$ contains all the elements of $\mathcal{A}^{m+1}$ except the ones that  have $\l_j=0$, namely
\be
\mathcal{A}^{m+1} \setminus  \mathcal{A}^{m}_j = \left\{\{\l_i\} : \sum_{i=1}^n  \l_i =m+1\,,\  \l_j=0 \,, \, \l_i \in \mathbb{N}_0  \right\} \ .\nonumber
\ee 
Using this fact it is easy to show the following formula.
Given a function of $n$ variables $F(\{ \l_1, \dots \l_n\}) \equiv F(\{ \l_i\})$. If $ F(\{ \l_i\}) \big|_{\l_j=0}=0$ for a fixed $j \in \{1,\dots,n\}$ then \footnote{
Property (\ref{partition1}) holds because
\begin{align}
\sum_{\{ \l_i\} \in \mathcal{A}^{m}_j} F(\{  \l_i \}) = \sum_{\{ \l_i\} \in \mathcal{A}^{m+1}} F(\{  \l_i \}) - \sum_{ \{ \l_i\} \in \mathcal{A}^{m+1}  \setminus  \mathcal{A}_j^{m} } F(\{  \l_i \}) 
= \sum_{\{ \l_i\} \in \mathcal{A}^{m+1}} F(\{  \l_i \}) \ , \nonumber 
\end{align}
where we used the fact that $F(\{\l_i \})=0$ for any $\{ \l_i\} \in  \mathcal{A}^{m+1}  \setminus  \mathcal{A}_j^m$.
}
\begin{align}
\label{partition1}
  \sum_{\sum_{i=1}^n \l_i =m}F(\{ \l_i + \d_{i}^j\}) &= \sum_{\sum_{i=1}^n \l_i =m+1}  F(\{ \l_i\}) \ .
\end{align}
We can use (\ref{partition1}) to find 
\begin{align}
 \sum_{\sum_{i=1}^n \l_i =m} (\l_j+ 1) G(\{ \l_i + \d_{i}^j\}) &= \sum_{\sum_{i=1}^n \l_i =m+1}  \l_j  G(\{ \l_i\}) \ , \label{formulapart2}\\
\sum_{\sum_{i=1}^n \l_i =m}\sum_{j=1}^n (\l_j +1) G(\{ \l_i + \d_{i}^j\}) &= (m+1)  \sum_{\sum_{i=1}^n \l_i =m+1}G(\{ \l_i\}) \ , \label{formulapart1} 
\end{align}
for any function $G(\{ \l_i\})$.

\subsubsection{Demonstration of (\ref{LmRecScalar}) and (\ref{Lcm1})} \label{App:DemRec1}
A proof of (\ref{LmRecScalar}) can be given as follows
\begin{align}
\sum_{e,f=1 \atop e<f}^k \g_{e f} [L_{m-1}]^{e f}
&=\sum_{e,f=1 \atop e<f}^k \g_{e f} \sum_{\sum n_{ab}=m-1} \left[M(\g_{ab}+n_{ab})\right]^{e f} \prod_{a,b=1 \atop a< b}^k\frac{\left[(\g_{ab})_{n_{ab}}\right]^{ef}}{n_{ab}!} \nonumber \\ 
&=\sum_{e,f=1 \atop e<f}^k \sum_{\sum n_{ab}=m-1} (n_{e f}+1) M(\g_{ab}+n_{ab}+\d_a^e\d_b^f) \prod_{a,b=1 \atop a< b}^k\frac{(\g_{ab})_{n_{ab}+\d_a^e\d_b^f}}{(n_{ab}+\d_a^e\d_b^f)!}\nonumber  \\
&=m \sum_{\sum n_{ab}=m} M(\g_{ab}+n_{ab}) \prod_{a,b=1 \atop a<b}^k\frac{(\g_{ab})_{n_{ab}}}{n_{ab}!}\nonumber \\
&=m L_m \ , \nonumber
\end{align}
where  we used
$
\prod_{a<b}\left[(\g_{ab})_{n_{ab}}\right]^{ef}=\frac{1}{\g_{ef} } \prod_{a< b} (\g_{ab})_{n_{ab}+\d_a^e\d_b^f} 
$ 
and (\ref{formulapart1}). Clearly the same demonstration holds for a Mellin amplitude with one or more indices that are not summed, namely
\be
\sum_{e,f=1 \atop e<f}^k \g_{e f} [L^{a_1 \dots a_J}_{m-1}]^{e f}=m L ^{a_1 \dots a_J}_m \ , \qquad \mbox{with } L_m^{a_1 \dots a_J }=\sum_{\sum n_{e f}=m}  M^{a_1 \dots a_J}(\g_{ef}+n_{ef}) \prod_{e,f=1 \atop e< f}^k\frac{(\g_{ef})_{n_{ef}}}{n_{ef}!} \ . \nonumber
\ee
In particular (\ref{Lcm1}) holds.
\subsubsection{Demonstration of  (\ref{Lhatm2})}
First we note that $\dot{L}_m\equiv \sum_{e=1}^k \g_e L^{e}_m$ can be also defined as follows
\be \label{LdotmTransversality}
\dot{L}_m=\sum_{\sum n_{ab}=m} \sum_{e,f=1 \atop e\neq f}^k n_{e f} M^{e}(\g_{ab}+n_{ab}) \prod_{a,b=1 \atop a<b}^k\frac{(\g_{ab})_{n_{ab}}}{n_{ab}!}\ , 
\ee
where we consider $n_{ef}$ symmetric in its indeces, so that $n_{fe}\equiv n_{ef}$ when $f>e$. 
Formula (\ref{LdotmTransversality}) is true because of the transversality condition (\ref{constraintMa}), in fact 
\begin{align}
\sum_{e=1}^k \g_e M^{e}(\g_{ab}+n_{ab})&=-\sum_{e,f=1 \atop e \neq f}^k (\g_{ef}+n_{ef}-n_{ef}) M^{e}(\g_{ab}+n_{ab})=\sum_{e,f=1 \atop e \neq f}^k n_{ef} M^{e}(\g_{ab}+n_{ab}) \ .\nonumber 
\end{align}
Using (\ref{LdotmTransversality}) we can now prove (\ref{Lhatm2}) 
\begin{align}
\sum_{e,f=1 \atop e\neq f}^k \g_{e f} [L^e_{m-1}]^{e f}
&=\sum_{e,f=1 \atop e < f}^k \g_{e f} [L^e_{m-1}+L^f_{m-1}]^{e f} \nonumber \\
&=\sum_{e,f=1 \atop e<f}^k \g_{e f} \sum_{\sum n_{ab}=m-1} \left[(M^e+M^f)(\g_{ab}+n_{ab})\right]^{e f} \prod_{a,b=1 \atop a< b}^k\frac{\left[(\g_{ab})_{n_{ab}}\right]^{ef}}{n_{ab}!} \nonumber \\
&= \sum_{\sum n_{ab}=m} \sum_{e,f=1 \atop e<f}^k n_{e f} (M^e+M^f)(\g_{ab}+n_{ab}) \prod_{a,b=1 \atop a< b}^k\frac{(\g_{ab})_{n_{ab}}}{(n_{ab})!}\nonumber  \\
&=\dot{L}_m \ , \nonumber
\end{align}
where followed the same steps of demonstration in the previous subsection except that we used formula (\ref{formulapart1}) instead of (\ref{formulapart2}).
\subsubsection{ Demonstration of (\ref{Lhatm1})}
\begin{align}
\sum_{a,b=1 \atop a \neq b}^k\g_{a b} \biggl[ \dot{L}_{m-1} \biggr]^{a b}
&=\sum_{a,b=1 \atop a \neq b}^k \g_{a b} \biggl[\sum_{c=1}^k \g_{c} L^c_{m-1} \biggr]^{a b} \nonumber \\ 
&=\sum_{a,b=1 \atop a \neq b}^k \g_{a b}\biggl( \sum_{c=1}^k \g_{c}  \left[L^c_{m-1} \right]^{a b} - \left[L^a_{m-1} \right]^{a b}- \left[L^b_{m-1} \right]^{a b}) \biggr) \nonumber \\ 
&= \sum_{c=1}^k \g_{c}\sum_{a,b=1 \atop a \neq b}^k \g_{a b}  \left[L^c_{m-1} \right]^{a b} -2 \sum_{a,b=1 \atop a \neq b}^k \g_{a b} \left[L^a_{m-1} \right]^{a b} \nonumber \\ 
&=2m \sum_{c=1}^k \g_{c} L^c_{m}  -2 \dot{L}_{m} \nonumber \\ 
&=2(m-1) \dot{L}_{m} \nonumber 
\end{align}
where in we have used the relations  (\ref{Lcm1}) and (\ref{Lhatm2}).

\subsubsection{Recurrence relations for the $J=2$ factorization formula}
\label{LmJ2Formulas}
In this section we present the $J=2$ analog of the recurrence relations (\ref{Lcm1}), (\ref{Lhatm2}), (\ref{Lhatm1}). Since the technology we needed was similar to the $J=1$ case we will not write here any demonstration. The formulas are
\begin{align}
\dot{L}^c_m&=\sum_{a=1}^k (\g_{a}+\d^a_{c} )L^{a c}_m \ ,\\ 
\dot{L}^c_m&=\sum_{\sum n_{ab}=m} \sum_{e,f=1 \atop e \neq f}^k n_{e f} M^{e c}(\g_{ab}+n_{ab}) \prod_{a<b}\frac{(\g_{ab})_{n_{ab}}}{n_{ab}!}\ , \\
\dot{L}^c_m&=\frac{1}{2(m-1)}\sum_{a,b=1 \atop a \neq b}^k \g_{a b} \left[ \dot{L}_{m-1}^c \right]^{a b} \label{App:RecRelLHatc}\ ,\\
\dot{L}^c_m&=\sum_{a,b=1 \atop a \neq b}^k \g_{a b} \left[L^{a c}_{m-1}\right]^{a b} \ ,
\end{align}
\begin{align}
\ddot{L}_m&=\sum_{a=1}^k \g_{a} \dot{L}^{a}_m \ , \\
\ddot{L}_m&=\tilde{L}_m+\sum_{a,b=1 \atop a \neq b}^k \g_{a b} \left[\dot{L}^{a}_{m-1}\right]^{a b} \ ,  \qquad \qquad \qquad \qquad
\end{align}
\begin{align}
\tilde{L}_m&=\sum_{\sum n_{ab}=m} \sum_{e,f=1 \atop e \neq f}^k n_{e f} M^{e f}(\g_{ab}+n_{ab}) \prod_{a<b}\frac{(\g_{ab})_{n_{ab}}}{n_{ab}!} \ ,\\
\tilde{L}_m&=\frac{1}{2(m-1)}\sum_{a,b=1 \atop a \neq b}^k \g_{a b} \left[ \tilde{L}_{m-1} \right]^{a b} \label{App:RecRelLtilde}\ , \\
\tilde{L}_m&=\sum_{a,b=1 \atop a \neq b}^k  \g_{a b} \left[L^{a b}_{m-1}\right]^{a b}\ .  
\end{align}


\subsection{The first residue} \label{App:Q0DownIndex}
In this subsection we study the first residue 
$\mathcal{Q}_0$
of the factorization formula for a generic spin $J$ exchange. We first rewrite formula (\ref{genericm0}) in a nicer way and we then match it with the known result for the first residue of the four point Mellin amplitude.

Any object $\mathcal{S}^{\m_1 \dots \m_J}$ symmetric in its indices $\m_1, \dots, \m_J$  (where $\m_i=1,\dots n$) can be written in terms of the occurrence of the values $1,\dots, n$ in the indices $\m_1, \dots , \m_J$. In particular given $\a_\ell$ occurrences of $\ell \in\{1,\dots, n\}$  we define a new object $\mathcal{S}_{\a_1 \dots \a_n }$ (with down indices) as follows
\be \label{Def:downindices}
\mathcal{S}_{\a_1 \dots \a_n }\equiv \mathcal{S}^{\scriptsize \overbrace{ \mathsmaller{ 1}  \dots  \mathsmaller{ 1} }^{\mathlarger{\a_1}}\overbrace{\tiny 2 \dots 2}^{\mathlarger{\a_2}} \mathlarger{\dots} \overbrace{  \mathlarger{ n} \dots   \mathlarger{ n}}^{ \mathlarger{\a_n}}} \ .
\ee
Rewriting (\ref{genericm0}) in the down indices notation (\ref{Def:downindices}) we find a remarkably simple formula 
\be
\mathcal{Q}_0= \k_{\D J} \,J! \sum_{\sum_{ai} j_{ai}=J \atop j_{ai} \ge 0 } \left( \prod_{1 \le a \le k \atop k < i \le n} \frac{(\g_{a i})_{j_{a i}}}{j_{ai}!} \right) M_{j_1 \dots j_k} M_{j_{k+1}\dots j_n} \label{genericm0A} \ ,
\ee
where $j_{a}=\sum_{i= k+1}^n j_{ai}$ and $j_{i}=\sum_{a=1}^k j_{ai}$ (where $a=1,\dots,k$ and $i=k+1,\dots n$).

\subsubsection{The first residue for $k=2$ and $n=4$} \label{Qm=0anyJ}


We want to show  that  for $k=2$ and $n=4$  (\ref{genericm0}) reduces to the known formula for the first residue of the four point function computed   in \cite{Mack, Costa:2012cb} once we fix 
\be \label{AppkDeltaJ}
\k_{\D J} =(-2)^{1-J} (\D+J-1) \G(\D-1) \ . 
\ee
The first part of the computation consists in finding the Mellin amplitude associated to a three point function. Then we will plug such a formula in (\ref{genericm0A}) and we will compare the result with \cite{Costa:2012cb}.

A three point function of two scalars with and a spin $J$ operator is given by
\begin{align}
\! \! \! \left\langle \mathcal{O}_1(P_1)\mathcal{O}_2(P_2) \mathcal{O}(Z_3,P_3) \right\rangle 
&=\frac{ c_{1 2 \Ocal}\big((Z_3\cdot P_1)(-2 P_2\cdot P_3)-(Z_3\cdot P_2)(- 2 P_1\cdot P_3)\big)^{J}}
{(-2P_{1}\cdot P_{2})^{\frac{\Delta_1+\Delta_2-\Delta_3+J}{2}}
(-2P_{1}\cdot P_{3})^{\frac{\Delta_1+\Delta_3-\Delta_2+J}{2}}
(-2P_{2}\cdot P_{3})^{\frac{\Delta_2+\Delta_3-\Delta_1+J}{2}}
} \nonumber \\
& =\frac{ c_{1 2 \Ocal}}{(-2P_{1}\cdot P_{2})^{\g_{12}}} \sum_{j=0}^J  \binom{J}{j}  \frac{(Z_3\cdot P_1)^j (-Z_3\cdot P_2)^{J-j}}{(-2P_{1} \cdot P_{3})^{\g_1+j} (-2P_{2} \cdot P_{3})^{\g_2+J-j}}
\end{align}
with 
\begin{align}
\g_{12}=\frac{\Delta_1+\Delta_2-\Delta_3+J}{2}\,, \ \ \g_{1}=\frac{\Delta_1+\Delta_3-\Delta_2-J}{2}\,, \ \ \g_{2}=\frac{\Delta_2+\Delta_3-\Delta_1-J}{2} \ .
\end{align}
According to (\ref{SpinJRep1}) we have,
\begin{align} 
&\left\langle \mathcal{O}_1(P_1)\mathcal{O}_2(P_2) \mathcal{O}(Z_3,P_3) \right\rangle= \\
&= \frac{\G(\g_{12})}{(-2P_{1} \cdot P_{2})^{\g_{12}}} \sum_{j=0}^J  \binom{J}{j} (Z_3\cdot P_1)^j (Z_3\cdot P_2)^{J-j}\frac{\G(\g_{1}+j)}{(-2P_{1}\cdot P_{3})^{\g_{1}+j}} \frac{\G(\g_{2}+J-j)}{(-2P_{2}\cdot P_{3})^{\g_{2}+J-j}}M^{\overset{j}{\scriptsize  \overbrace{ 1\dots 1}}{\overset{J-j}{\scriptsize  \overbrace{ 2\dots 2}}}} \nonumber
\end{align}
so that the Mellin amplitudes 
can be written in the down indices notation as
\begin{align} \label{Mellin3ptSpinJ}
M_{j \; J-j} =c_{1 2 \Ocal} \frac{ (-1)^{J-j} }{ \Gamma(\g_{1 2}) \Gamma(\g_1+j) \Gamma(\g_2+J-j)}.
\end{align}
Replacing (\ref{Mellin3ptSpinJ}) in (\ref{genericm0A})
and using the identity $\frac{1}{\G(\g+n)}=\frac{(-1)^{n+J}}{\G(\g+J)} (1-\g-J)_{J-n}$ (that holds for any $n,J \in \mathbb{Z}$) we find
\be
\mathcal{Q}_0= \frac{c_{1 2 \Ocal} c_{3 4 \Ocal} \k_{\D J}\ (-1)^J J! }{\G(\g_{12}) \G(\g_{34}) \prod_{\ell=1}^4 \G(\g_{\ell}+J)} \sum_{\sum_{a i} j_{a i}=J \atop j_{ai} \ge 0} (-1)^{j_{1 3}+j_{2 4}} \left( \prod_{a=1,2 \atop i=3,4} \frac{(\g_{a i})_{j_{ai}}}{j_{ai}!} \right)  \prod_{\ell=1}^4 (1-\g_{\ell}-J)_{J-j_{\ell}} \label{4ptm0}
\ee
where $j_a=j_{a 3}+j_{a 4}$ and  $j_i=j_{1 i}+j_{2 i}$ (with $a=1,2$ and $i=3,4$).


We can match (\ref{4ptm0}) with formula (166) in \cite{Costa:2012cb}. The formulas look similar but the sum in (\ref{4ptm0}) contains only mixed variables $\g_{a i}$ while the one in (166) also depends on the spacetime dimension $d$. To factorize such a dependence from (166)  we need to use the identity (\ref{firstconjecture}) with $x_1=\D-\frac{d}{2}=x_2$. In this way, once we fix (\ref{AppkDeltaJ}),  we find an agreement between the two formulas.

We can also match equation (127) in \cite{Costa:2012cb}.
In fact we can simplify (\ref{4ptm0}) applying the conjectured identity (\ref{SecondConjecture})
\begin{align} \label{M0Simple}
\mathcal{Q}_0&=c_{1 2 \Ocal} c_{3 4 \Ocal} \k_{\D J} \frac{ (-1)^J (\g_{14})_J (\g_{23})_J  (\D-1)_J }{\G(\g_{12}) \G(\g_{34}) \prod_{t=1}^{4}\G(\g_t+J )} \; {_{3}F_2} (-J,\g_{13},\g_{24};1-J-\g_{14},1-J-\g_{23};1) \ , \nonumber
\end{align}
where we used the constraint $\g_{13}+\g_{23}+\g_{14}+\g_{24}=\D-J$. We can further use the following hypergeometric relation 
\be
 {_{3}F_2} (-J,b,c;d,e;1)=\frac{(d-b)_J (e-b)_J}{(d)_J (e)_J}  {_{3}F_2} (-J,b+c-d-e-J+1,b;b-d-J+1,b-e-J+1;1) \nonumber
\ee
valid for any integer $J$ to get
\begin{align} \label{M0Simple1App}
\mathcal{Q}_0&=c_{1 2 \Ocal} c_{3 4 \Ocal}   \k_{\D J}  \frac{(-1)^J(\D-1)_J \, (\g_{1})_J (\g_{3})_J}{\G(\g_{12}) \G(\g_{34}) \prod_{t=1}^{4}\G(\g_t+J ) } \, {_{3}F_2} (-J,\D-1,\g_{13};\g_1,\g_3;1)
\end{align}
where we could have also exchanged $1 \rightarrow 2$ and $3 \rightarrow 4$ adding a factor $(-1)^J$ in the formula.
Formula (\ref{M0Simple1App}) matches exactly (127) in \cite{Costa:2012cb} if we fix (\ref{AppkDeltaJ}).

\subsubsection{Conjectured Identity}

Consider the following polynomial in the variables
$\g_{13},\g_{14},\g_{23},\g_{24},x_1,x_2$,
\be
f^J(x_1,x_2;\g_{a i})= \!\!\sum_{\sum_{a i} j_{a i}=J} (-1)^{j_{1 3}+j_{2 4}} \left( \prod_{a=1,2 \atop i=3,4} \frac{(\g_{a i})_{j_{ai}}}{j_{ai}!} \right) \prod_{a=1}^2 (1-\g_a+x_a-J)_{J-j_a}  \prod_{i=3}^4 (1-\g_{i}-J)_{J-j_i}
\nonumber
\ee
 where the sum is over $j_{ai}\ge 0$ with $j_{13}+j_{23}+j_{14}+j_{24}=J$. We also defined $\g_a=\g_{a 3}+\g_{a 4}$ and $\g_{i}=\g_{1 i}+\g_{2 i}$ and the same for $j_a$ and $j_i$.
It is clear from the definition that $f^J$ is a polynomial of degree $3J$ in the variables $\g_{ai}$ and degree $J$ in the variables $x_a$.
We first conjecture that the following ratio is independent of $x_1$ and $x_2$,
 \be
\frac{f^J(x_1,x_2;\g_{a i})}
{(\g_{LR}-x_1-x_2 +J -1)_J}
= \frac{f^J(0,0;\g_{a i})}{(\g_{LR}+J-1)_J} 
\label{firstconjecture} \ ,
\ee
where $\g_{LR}= \g_{13}+\g_{14}+\g_{23}+\g_{24}$.
We conclude that (\ref{firstconjecture}) is a polynomial of degree $2J$ in the variables $\g_{ai}$.
In fact, we also conjecture that
\begin{align}
\frac{f^J(0,0;\g_{a i})}{(\g_{LR}+J-1)_J}&=
\frac{1}{J!}\sum _{j=0}^J (-1)^j \binom{J}{j}  (\g_{14})_{J-j}   (\g_{23})_{J-j}   (\g_{13})_{j}    (\g_{24})_{j} 
\label{SecondConjecture}\\
&= \frac{1}{J!} (\g_{14})_J (\g_{2 3})_J \; {_{3}F_2} (-J,\g_{13},\g_{24};1-J-\g_{14},1-J-\g_{23};1) \ .
\end{align}
Where in the last line we can also exchange $1\rightarrow 2$ or $3\rightarrow 4$ adding a factor $(-1)^J$ in the formula. 
We verified both conjectures in general up to $J=8$. We also checked their consistency up to $J=80$ setting all the entries of the functions to random integers between $1$ and $100$.



\section{Flat space limit of the factorization formulas}
\label{app:FSLfactorization}
The mass of a spin $J$ field in $AdS_{d+1}$  is $M^2=\frac{\D(\D-d)-J}{R^2}$ where $\D$ is the conformal dimension of the operator in the dual CFT and $R$ is the radius of $AdS_{d+1}$. In the flat space limit $R$ goes to infinity and to keep $M$ finite we need to assure that $\D$ scales like $R$. In terms of the Mellin variables the flat space limit amounts to (\ref{LimitFSL}), so that we can be more precise and ask for $\d_{ij},\D \rightarrow \infty$ with fixed ratio $\frac{\d_{ij}}{\D^2}$. 

The factorization formulas often contain functions of the kind
\be
f_m(\g_{ij})=\sum_{\myatop{n_{i j}\geq 0}{\sum n_{i j}= m}} g(\g_{ij}+n_{i j}) \prod_{i<j} \frac{(\g_{ij})_{n_{i j}}}{n_{i j}!} 
\ee
where $g(\g_{ij})$ is a function of $\g_{ij}$. We want to know what is the leading behavior of $f_m(\g_{ij})$ in the flat space limit. 
It is easy to to show \cite{NaturalMellin} that if $g(\g_{ij})$ is a polynomial of degree $\k$ in the variables $\g_{ij}$, then, for any finite $m$,  $f_m(\g_{ij})$ is also a polynomial of degree $\k+m$ and its leading term (the highest degree part of the polynomial) is the same as that  of 
\be
f_m(\g_{ij}) \simeq  \mathcal{D}^{(m,\sum_{i<j}\g_{ij}+m)}_t   g(t \g_{ij}) \big{|}_{t=1}  \ ,  \qquad \mbox{where } \mathcal{D}^{(m,\g)}_t\equiv \frac{1}{m!} \partial_t^m t^{\g-1}\ .\label{FSL:fm}
\ee


With this in mind, we can now review the scalar case. To keep the formulas more compact we rewrite (\ref{MellinFacPoles}--\ref{LmSum}) as
\begin{align} \label{App:FacFormulaScalar}
M&\approx  \sum_{m=0}^\infty \frac{g(m)}{\g_{LR}-\bar{\g}_{LR}} L_m R_m  \ , \qquad  \mbox{where }g(m)=\k_{\D 0} \frac{m!}{(1-\frac{d}{2}+\D)_m}
\end{align}
and $\bar{\g}_{LR}=\D+2m$ is the position of the poles. The leading behavior of  $ L_m$ and $R_m$ is found using (\ref{FSL:fm}) and moreover we can apply formula (\ref{eq:FSLRelationFinalSpinJ}). Using units where $R=1$ the result for $L_m$ (and similarly for $R_m$) is
\begin{align}
L_m 
&\simeq \mathcal{D}^{(m,\g_L+m)}_{t_L} M_L(t_L \g_{ab})\big{|}_{t_L=1} \nonumber \\
&\simeq \mathcal{D}^{(m,\g_L+m)}_{t_L} \mathcal{N}_{L}\int_{0}^{\infty}\frac{d\b_L }{\b_L}\b_L^{\frac{\sum_{a }\D_a+\D-d}{2} }e^{-\b_L}\mathcal{T}_L\left(2\b_L t_L \g_{ab}\right) \big{|}_{t_L=1} \nonumber \\
&= \mathcal{N}_{L} \int_{0}^{\infty}\frac{d\b_L }{\b_L} 
\b_L^{\frac{\sum_{a }\D_a+\D-d}{2}}\mathcal{T}_L\left(2\b_L\g_{ab}\right) \mathcal{D}^{(m,\frac{d}{2}-\D)}_{t_L}
e^{-\b_L/t_L} \big{|}_{t_L=1}  \label{FSL:Lm}
\end{align}
where $\g_L=\sum_{a<b\leq k} \g_{a b}=\sum_{a=1}^k \frac{\D_a}{2} -\frac{\bar{\g}_{LR}}{2}$ and $\mathcal{N}_{L}$ is the normalization of the left Mellin amplitude defined in (\ref{NormalizationN}). Notice that to get to (\ref{FSL:Lm}) we just performed the replacement $\b_L \rightarrow \b_L/t_L$ in the integration variable. Replacing in formula (\ref{App:FacFormulaScalar}) the flat space limit of $L_m$ found in (\ref{FSL:Lm}) and a similar formula for $R_m$,  we obtain
\begin{align} 
 M
\simeq \mathcal{N} \int_{0}^{\infty}\frac{d\b_L }{\b_L}
\b_L^{\frac{\sum_{a=1 }^k \D_a-d}{2} 
} \int_{0}^{\infty}\frac{d\b_R }{\b_R}
\b_R^{\frac{\sum_{i=k+1 }^n\D_i-d}{2} 
}   \mathcal{T}_L\left(2 \b_L\g_{ab}\right)\mathcal{T}_R\left(2 \b_R\g_{ij}\right) S(\b_L,\b_R) \ ,
\end{align}
where  $\mathcal{N}$ is the normalization of (\ref{eq:FSLRelationFinalSpinJ}) associated to the full scalar $n$-point function and where we defined
\begin{align}
S(\b_L,\b_R)&= \tilde{\mathcal{N}}  \sum_{m=0}^\infty  \frac{g(m)}{\g_{LR}-\bar{\g}_{LR}}  \left[ \mathcal{D}^{(m,-\D+\frac{d}{2})}_{t_L}e^{-\frac{\b_L}{t_L}}\b_L^{\frac{\D}{2}} \right]_{t_L=1} \left[\mathcal{D}^{(m,-\D+\frac{d}{2})}_{t_R} e^{-\frac{\b_R}{t_R}}\b_R^{\frac{\D}{2}} \right]_{t_R=1}
\end{align}
with $\tilde{\mathcal{N}}=\frac{\mathcal{N}_{L}\mathcal{N}_{R}}{\mathcal{N} }=\frac{\, \, \pi^\frac{d}{2}}{2}\frac{\mathcal{C}_{\D,0}}{ \G(\D)^2}$. As we will see later we can drastically simplify $S(\b_L,\b_R)$ to get
\be
S(\b_L,\b_R)\simeq  \d(\b_L-\b_R)  e^{-\b_L} \b_L^{\frac{d}{2}+1}\frac{1}{-2 \b_L \g_{LR}+\D^2}. \label{App:FSLofS}
\ee
With this simplification we finally match the flat space factorization of scalar scattering amplitudes (\ref{FSLfacJ0})
\be \label{FSL:FacFormulaScalar}
M \simeq \mathcal{N} \int_{0}^{\infty}\frac{d\b }{\b}
\b^{\frac{\sum_{\a=1}^n \D_\a-d}{2} 
}   e^{-\b}  \biggl[  \frac{ \mathcal{T}_L\left(p_a \cdot p_b\right)\mathcal{T}_R\left(p_i \cdot p_j\right) }{p^2+M^2}  \biggr]_{p_\m \cdot p_\n=2 \b \g_{\m \n}}
\ee 
where we defined $p=\sum_{a=1}^k p_a=-\sum_{i=k+1}^n p_i$ (so that $p^2\big{|}_{p_i\cdot p_j=2 \b \g_{i j}}= -2 \b \g_{LR}$) and we identified $M^2=\D^2$. This shows that the poles of $M$, in the flat space limit, give rise to a cut that can be obtained by the integral over $\b$ in (\ref{FSL:FacFormulaScalar}) of the unique pole of the scattering amplitude $ \mathcal{T}$.

We now explain  how to find formula (\ref{App:FSLofS}).
We need  the following Mellin transform
\be
f(\b)=\left. \mathcal{D}^{(m,\g)}_{t}e^{-\frac{\b}{t}} \b^y
\right|_{t=1}\longrightarrow \hat{f}(x)
\equiv \int_0^\infty \frac{d\b}{\b} \b^{x} f(\b)
=\frac{\G(x+y)}{m! (\g+x+y)_{-m}} \label{App:Dproperty}.
\ee
Using (\ref{App:Dproperty}) we can Mellin transform $S(\b_L,\b_R)$ in both the variables $\b_L$ and $\b_R$,
\be
\hat{S}(x_L,x_R)=\tilde{\mathcal{N}} \sum_{m=0}^\infty \frac{g(m)}{\g_{LR} -\bar{\g}_{LR}}  \; \frac{\G(x_L+\frac{\D}{2})}{m! (x_L+\frac{d-\D}{2})_{-m}}\frac{\G(x_R+\frac{\D}{2})}{m! (x_R+\frac{d-\D}{2})_{-m}}.
\ee
When $m$ is of order one, the summand gives a negligible contribution. In fact the terms that are going to contribute are the ones with $m$ of order $\D^2$. We can then define a new variable $s=\frac{4m}{\D^2}$ and, since $s$ will slowly vary when $m$ increases by a step of one, we can take a continuum limit and turn the sum into an integral
\be
\sum_{m=0}^\infty \frac{4}{\D^2} \dots \longrightarrow \int_0^\infty d s \dots \; .
\ee
Using the Stirling approximation in the integrand we then find
\begin{align}
\hat{S}(x_L,x_R)
&\simeq \int_0^\infty \frac{d s}{s} \, \frac{e^{-1/s}}{s^{x_L+x_R+d/2}}  \; \frac{1}{-2 s^{-1} \g_{LR}+\D^2}	\nonumber \\
&=\int_0^\infty \frac{d\b_L}{\b_L}  \b_L^{x_L} \int_0^\infty \frac{d\b_R}{\b_R}  \b_R^{x_R} \biggl[ \d(\b_L-\b_R)  e^{-\b_L} \b_L^{\frac{d}{2}+1} \;  \frac{1}{-2 \b_L \g_{LR}+\D^2} \biggr] \label{App:FSLShat}
\end{align}
where we performed the change of variable $s=1/\b_L$. Formula (\ref{App:FSLShat}) is the definition of a double Mellin transform, therefore we can identify the term in the square brackets with $S(\b_L,\b_R)$ and finally recover formula (\ref{App:FSLofS}).
\subsection{Vector operator}
The position of the poles of the Mellin amplitude at $\g_{LR}=\D+2m$, turned into poles at $\g_{LR}= \D^2/(2\b)$ in the integral in formula  (\ref{FSL:FacFormulaScalar}). This means that in the flat space limit the contribution of the infinite sum over $m$ will be dominated by the poles which have a value of $m$ that scales like $\D^2$. This simple observation is proven to be very effective to simplify the factorization formulas for the exchange of operators with non zero spin. In fact, in these cases, one has to deal with formulas of the kind
\begin{align}
M\approx  \sum_{m=0}^\infty \frac{\mathcal{Q}_m}{\g_{LR}-\bar{\g}_{LR}} \ ,   \nonumber 
\end{align}
with
\begin{align}
\mathcal{Q}_m=g(m) \sum_s h_s(m) \mathcal{Q}^{(s)}_m\ ,   \qquad 
\qquad g(m)=\k_{\D J} \frac{m!}{(1-\frac{d}{2}+\D)_m} \ ,\label{FactorizationFormulaGeneric}
\end{align}
 where the position of the poles is at $\bar{\g}_{LR}=\D+2m-J$. The residue $\mathcal{Q}_m$ is expressed as a sum over structures labeled by $s$ and we will now see that it is easy to drop some of  such structures checking that they give rise to a subdominant contribution in the flat space limit (once we consider that $m$ scales like $\D^2$).

The factorization formula for the vector case is (\ref{FactorizationFormulaGeneric}) where $s$ runs from $1$ to $2$ and the two structures are defined as follows
\begin{align}
h_1(m)=&1 \ ,   &
\mathcal{Q}^{(1)}_m&= \sum_{a=1}^k \sum_{i=k+1 }^n\underset{\sim \D^2}{\underbrace{\g_{ai}}} L^a_m R^i_m  \ , \label{h1Q1}\\
h_2(m)=&\underset{\sim \D^{-2}}{\underbrace{\frac{d-2\D}{2m (\D-d+1)}}} \ ,
 &\mathcal{Q}^{(2)}_m&\equiv \sum_{a=1}^k \sum_{i=k+1 }^n\underset{\sim \D^4}{\underbrace{ \g_{a} \g_{i}}} L^a_m R^i_m \label{h2Q2} \ ,
\end{align}
where we put in evidence the scaling behavior in $\D$.
Naively one would think that the two structures (\ref{h1Q1}) and  (\ref{h2Q2}) have the same leading behavior in $\D$ since $\mathcal{Q}^{(1)}_m$ contains only one Mellin variable that scales like $\D^2$ and $\mathcal{Q}^{(2)}_m$ has two Mellin variables which scale as $\D^4$ but it is suppressed by $\D^{-2}$ because of the contribution of $h_2(m)$. This analysis is too simplistic because we did not consider that the Mellin amplitudes satisfy the transversality condition (\ref{constraintMa}), namely $\sum_{a=1}^k\g_aM^a=0$. Using this condition is enough to ensure that the second structure (\ref{h2Q2}) is subdominant because its naive leading term in $\D$ is actually zero. To see this we recall that the flat space limit of $L^a_m$ and $R^i_m$ can be computed using (\ref{FSL:fm}), then
\begin{align}
\sum_{a=1}^k \g_{a} L^a_m &= \sum_{a=1}^k \g_{a} \mathcal{D}^{(m,\g_L+m)}_{t_L} M^a_L(t_L \g_{l_1 l_2})\big{|}_{t_L=1} + \dots \nonumber \\
&= \mathcal{D}^{(m,\g_L+m-1)}_{t_L} \sum_{a=1}^k (t_L\g_{a}) M^a_L(t_L \g_{l_1 l_2})\big{|}_{t_L=1} + \dots \nonumber \\
&=0 + \dots \ ,
\end{align}
where we used  (\ref{constraintMa}) with $\g_L=\sum_{a=1}^k \frac{\D_a}{2} -\frac{\bar{\g}_{LR}}{2}$ and where the dots stand for subleading contributions in the flat space limit ($\g_{ij} \sim \D^2 \to \infty$). This analysis allows us to drop (\ref{h2Q2}) and to just compute the flat space limit of (\ref{h1Q1}). The computation is similar to the scalar case but now, in order to use formula  (\ref{eq:FSLRelationFinalSpinJ}), we need to express the Mellin amplitudes $M^a$ in terms of their {\it check} representation  $\check M^a$.\footnote{Notice that in the scalar case $M=\check{M}$.} In particular the following equation holds 
\begin{align}
\sum_{a=1}^k \sum_{i=k+1 }^n\g_{ai} M^a_L M^i_R
&=\sum_{a,b=1}^k \sum_{i,j=k+1 }^n\g_{ai}\g_b\g_j (\check{M}_L^b- \check{M}_L^a)(\check{M}_R^j- \check{M}_R^i)\\
&=\sum_{a,b=1}^k \sum_{i,j=k+1 }^n \g_{ai}\g_b\g_j (\check{M}_L^b \check{M}_R^j-\check{M}_L^b \check{M}_R^i-\check{M}_L^a \check{M}_R^j+ \check{M}_L^a \check{M}_R^i)\\
&=\g_{LR}\sum_{a=1}^k \sum_{i=k+1 }^n (-\g_a\g_i +\g_{ai} \g_{LR}) \check{M}_L^a \check{M}_R^i \label{eq:MaMi->McheckaMchecki}
\end{align}
in which we used $\sum_{a=1}^k \g_a=\g_{LR}=\sum_{i=k+1}^n \g_i$. When the on shell condition $\g_{LR}=\bar{\g}_{LR}$ holds, we can rewrite (\ref{eq:MaMi->McheckaMchecki}) as follows
\begin{align}
\sum_{a=1}^k \sum_{i=k+1 }^n\g_{ai} M^a_L M^i_R&=\bar{\g}_{LR}^2 \sum_{a=1}^k \sum_{i=k+1 }^n\check{\Omega}_{a i} \check{M}^a_L \check{M}^i_R \ ,\\
\check{\Omega}_{a i}&= (\g_{a i }-\frac{1}{\bar{\g}_{LR}}\g_a \g_i ) \qquad 
{ \scriptsize \left\{
\begin{array}{lcl}
a&=& 1,\dots, k\\
i&=&k+1,\dots, n
\end{array}
\right.
}
\label{CheckOmega} \ .
\end{align}
In this way we can compute the flat space limit of the structure $\mathcal{Q}^{(1)}_m$
\begin{align}
\mathcal{Q}^{(1)}_m &=\sum_{a=1}^k \sum_{i=k+1 }^n\g_{ai} L_m^a R_m^i \simeq \mathcal{D}^{(m,\g_L)}_{t_L} \mathcal{D}^{(m,\g_R)}_{t_R}\sum_{a=1}^k \sum_{i=k+1 }^n\g_{ai}  M_L^a(t_L \g_{l_1 l_2}) M_R^i(t_R \g_{r_1 r_2}) \nonumber  \\
&=\mathcal{D}^{(m,\g_L+1+m)}_{t_L} \mathcal{D}^{(m,\g_R+1+m)}_{t_R} \bar{\g}^2_{LR}\sum_{a=1}^k \sum_{i=k+1 }^n\check{\Omega}_{ai} \check{M}_L^a(t_L \g_{l_1 l_2})  \check{M}_R^i(t_R \g_{r_1 r_2}) \ . \label{FSL:MJ1Q1}
\end{align}
Replacing this result in (\ref{FactorizationFormulaGeneric}), dropping the contribution of $\mathcal{Q}^{(2)}_m$ and using (\ref{eq:FSLRelationFinalSpinJ}) we find
\begin{eqnarray}
\! M&\simeq& \sum_{m=0}^\infty  \frac{ g(m)}{\g_{LR}-\bar{\g}_{LR}}h_1(m) \mathcal{Q}^{(1)}_{m}\\
&\simeq&\mathcal{N} \int_{0}^{\infty}\frac{d\b_L}{\b_L} \b_L^{\frac{\sum_{a}\D_a+1-d}{2}}
\int_{0}^{\infty}\frac{d\b_R}{\b_R} \b_R^{\frac{\sum_{i}\D_i+1-d}{2}} \nonumber  \sum_{a=1}^k \sum_{i=k+1 }^n   \mathcal{T}_L^{a}(2\b_L \g_{l_1 l_2}) \mathcal{T}_R^{i}(2\b_R \g_{r_1 r_2})  S_{ai}^{(1)}(\b_L,\b_R) \label{StructureFSLJ1} \ ,
\end{eqnarray}
where we rescaled the integration variables $\b_L \to \b_L/t_L$ and $\b_R \to \b_R /t_R$ and we defined $\mathcal{N}$ to be the normalization of (\ref{eq:FSLRelationFinalSpinJ}) associated to the full scalar $n$-point function and 
\begin{align}
S_{a i}^{(1)}(\b_L,\b_R)&=\tilde{\Ncal}\sum_{m=0}^\infty \frac{g(m)}{\g_{LR}-\bar{\g}_{LR}} \bar \g_{LR}^{2}\;  \check{\Omega}_{a i}  \left[\mathcal{D}^{(m,1-\D+h)}_{t_L} e^{-\frac{\b_L}{t_L}} \b_L^{\frac{\D}{2}}\right]_{t_L=1} \left[\mathcal{D}^{(m,1-\D+h)}_{t_R} e^{-\frac{\b_R}{t_R}}\b_R^{\frac{\D}{2}} \right]_{t_R=1} \ ,\nonumber \\
\end{align}
where $\tilde{\Ncal}=\frac{\Ncal_L \Ncal_R}{\Ncal}=\pi^{\frac{d}{2}} 2 \frac{\mathcal{C}_{\D,1}}{\G(\D+1)^2}$.
Using the same kind of computation as in the scalar case and identifying $\D=M$,  we find
\be \label{SJ1Final}
S_{a i}^{(1)}(\b_L,\b_R)\simeq  \d(\b_L-\b_R)  e^{-\b_L}  \b_L^{\frac{d}{2}}  \left[\frac{\Omega_{ai}}{ p^2+M^2}\right]_{p_i \cdot p_j= 2 \b_L\g_{i j }} 
\ ,
\ee
where $\Omega_{ai}$ is defined in (\ref{Omegaij}).
Replacing (\ref{SJ1Final}) into (\ref{FSL:MJ1Q1}) we get to the final formula
\begin{eqnarray}
\!\!\!\!\!\!\! M&\simeq&\mathcal{N} \int_{0}^{\infty}\frac{d\b}{\b} \b^{\frac{\sum_{\a=1}^n\D_\a-d}{2}}  e^{-\b}   \left[\sum_{a=1}^k \sum_{i=k+1 }^n \frac{ \mathcal{T}^{a}_L(p_{l_1} \cdot p_{l_2}) \mathcal{T}^{i}_R(p_{r_1} \cdot p_{r_2}) }{ p^2+M^2} \; \Omega_{ai} \right]_{p_\m \cdot p_\n= 2 \b_L\g_{\m \n }} .
\end{eqnarray}
\subsection{Spin two operator}
The flat space limit for the factorization formula for spin $J=2$ can be found in a similar way as in the previous cases. First we find that in  (\ref{finalresultJ2}) there are only two structures that contribute, namely the first and the last one. The procedure to see this is analogous to the vector case. Once we consider $\g_{ij}\sim m \sim \D^2$ all the structures naively seem to have the same leading term in $\D$ (except the fourth one that is obviously subdominant). It is then easy to show that the naive leading term   of $\dot L^{a}_{m}$ and of $\ddot L^{a}_{m}$ is actually zero because of the tranversality condition $\sum_{i} (\g_i+\d^{j}_i) M^{i j}=0$. This fact allows us to drop all the structures which contain $\dot L^{a}_{m}$ or $\ddot L^{a}_{m}$, therefore we can restrict our computation to just $\mathcal{Q}^{(1)}_{m}$ and $\mathcal{Q}^{(5)}_{m}$.

We then write the flat space limit of $L^{a b}_{m}$ and $\tilde L_{m}$
\begin{align}
L^{a b}_{m}&\simeq \mathcal{D}^{(m,\g_L+m)}_t  M^{a b}(t \g_{ef})|_{t=1} \ ,\label{FSL:Lmab}\\ 
\tilde L_{m}&\simeq -\sum_{a,b=1}^k \g_{a b} \mathcal{D}^{(m,\g_L+m)}_t  M^{a b}(t \g_{ef})|_{t=1} \ ,\label{FSL:Ltilde}
\end{align}
where $\g_L=\sum_{a=1}^k \frac{\D_a}{2} -\frac{\bar{\g}_{LR}}{2}$. 
Formula (\ref{FSL:Lmab}) is obtained using (\ref{FSL:fm}) and to find the flat space limit of $\tilde L_{m}=\sum_{a b} \g_{a b} \left[L^{a b}_{m-1}\right]^{a b}$ we first use  (\ref{FSL:fm}) to get \footnote{We assume that $L_m^{ab}$ is polynamially bounded.}
\be \label{eqShiftLtilde}
\left[L^{a b}_{m-1}\right]^{a b} \simeq
L^{a b}_{m-1}\simeq \mathcal{D}^{(m-1,\d_L+m)}_t  M^{a b}(t \d_{ef}) |_{t=1} \ .
\ee
Then, since $m \sim \g_{ij}$, we can further simplify (\ref{eqShiftLtilde}) considering the following relation   
 \be
\mathcal{D}^{(m-1,\g )}_t M^{a b}(t \g_{ef})|_{t=1}\simeq-\mathcal{D}^{(m,\g )}_t M^{a b}(t \g_{ef})|_{t=1}\ , \qquad m\gg \g, 1\ . \label{FSL:DerivativesOfM}
\ee  
Combining  (\ref{eqShiftLtilde})and (\ref{FSL:DerivativesOfM}) we find that (\ref{FSL:Ltilde}) holds.\\
Using formulas (\ref{FSL:Lmab}) and (\ref{FSL:Ltilde}) we can write the flat space limit of $\mathcal{Q}^{(1)}_{m} $ and $\mathcal{Q}^{(5)}_{m} $ as follows
\begin{align}
\mathcal{Q}^{(1)}_{m} \simeq \mathcal{D}^{(m,\g_L+m)}_{t_L}\mathcal{D}^{(m,\g_R+m)}_{t_R}  \sum_{a,b=1}^k \sum_{i,j=k+1 }^n  \g_{a i}\g_{b j} M_L^{a b}(t_L \g_{l_1 l_2})  M_R^{i j}(t_R \g_{r_1 r_2})\big{|}_{t_L,t_R=1} \ ,\\
\mathcal{Q}^{(5)}_{m} \simeq \mathcal{D}^{(m,\g_L+m)}_{t_L}\mathcal{D}^{(m,\g_R+m)}_{t_R}  \sum_{a,b=1}^k \sum_{i,j=k+1 }^n  \g_{a b}\g_{i j} M_L^{a b}(t_L \g_{l_1 l_2})  M_R^{i j}(t_R \g_{r_1 r_2})\big{|}_{t_L,t_R=1} \ .\label{formulasRes1Res3}
\end{align}
Now we need to express $M$ in terms of $\check{M}$. To do so we use the following identities\footnote{A proof of such identities is given in appendix \ref{ProofOfGammaToOmegaJ}. Notice that the difference of signature in the definition of $\check \Omega$ comes from the fact that $\sum_{a,b=1}^k \g_{ab}=-\g_{LR}$ while $\sum_{a=1}^k\sum_{i=k+1}^n \g_{ai}=\g_{LR}$.} valid in the flat space limit and where $\g_{LR}=\bar\g_{LR}$
\begin{align}
\sum_{a,b=1}^k \sum_{i,j=k+1 }^n  \g_{a b}\g_{i j}  M_L^{a b} M_R^{i j} \simeq  \bar\g_{LR}^4 \sum_{a,b=1}^k \sum_{i,j=k+1 }^n \check{\Omega}_{a b} \check{\Omega}_{i j} \check{M}_L^{a b} \check{M}_R^{i j} \ , \nonumber \\
\sum_{a,b=1}^k \sum_{i,j=k+1 }^n  \g_{a i}\g_{b j}  M_L^{a b} M_R^{i j} \simeq \bar \g_{LR}^4 \sum_{a,b=1}^k \sum_{i,j=k+1 }^n \check{\Omega}_{a i} \check{\Omega}_{b j} \check{M}_L^{a b} \check{M}_R^{i j} \ , \label{FormulasSumMtoMcheckJ2}
\end{align}
where $\check{\Omega}_{a i}$ was defined in (\ref{CheckOmega}) and 
\be
\check{\Omega}_{a b}= (\g_{a b }+\frac{1}{\bar{\g}_{LR}}\g_a \g_b ) \qquad { \scriptsize \left\{ 
\begin{array}{l}
a,b = 1,\dots, k\\
\qquad \quad \mbox{or}\\
a,b=k+1,\dots, n
\end{array} \right.
}
.
\ee
Putting everything together and using (\ref{eq:FSLRelationFinalSpinJ}) we find
\begin{eqnarray}
M&\simeq& \sum_{m=0}^\infty  \frac{ g(m)}{\g_{LR}-\bar \g_{LR}}\left[h_1(m) \mathcal{Q}^{(1)}_{m}+h_5(m) \mathcal{Q}^{(5)}_{m} \right] \\
&\simeq&\Ncal \int_{0}^{\infty} \frac{d\b_L}{\b_L} \, \b_L^{\frac{\sum_{a}\D_a+2-d}{2}}
\int_{0}^{\infty} \frac{d\b_R}{\b_R} \, \b_R^{\frac{\sum_{i}\D_i+2-d}{2}} \nonumber \\
&& \sum_{a,b=1}^k \sum_{i,j=k+1 }^n   \mathcal{T}_L^{a b}(2\b_L \g_{l_1 l_2}) \mathcal{T}_R^{i j}(2\b_R \g_{r_1 r_2})  \left[ S_{a b i j}^{(1)}(\b_L,\b_R)+ S_{a b i j}^{(5)}(\b_L,\b_R) \right] \label{FSL:MJ2S1S5} \ ,
\end{eqnarray}
where $\Ncal$ is the normalization (\ref{NormalizationN}) for a scalar $n-$point function, 
\begin{align}
 \! \! \!S_{a b i j}^{(1)}(\b_L,\b_R)&=\tilde{\Ncal}\sum_{m=0}^\infty\frac{  g(m)}{\g_{LR}-\bar{\g}_{LR}} \bar \g_{LR}^{4} \;  \check{\Omega}_{a i} \check{\Omega}_{b j} \left[\mathcal{D}^{(m,2-\D+h)}_{t_L} e^{-\frac{\b_L}{t_L}} \b_L^{\frac{\D}{2}}\right]_{t_L=1} \left[\mathcal{D}^{(m,2-\D+h)}_{t_R} e^{-\frac{\b_R}{t_R}}\b_R^{\frac{\D}{2}} \right]_{t_R=1} \ ,\nonumber \\
\! \! \!S_{a b i j}^{(5)}(\b_L,\b_R)&=-\frac{1}{d} S_{a i b j}^{(1)}(\b_L,\b_R) \ , \label{FSL:S3}
\end{align}
and $\tilde{\Ncal}=\frac{\Ncal_L\Ncal_R}{\Ncal}=
\pi^\frac{d}{2} 2^{3} \frac{\mathcal{C}_{\D,2}}
{  \G\left(\D+2 \right)^2}$, where $\Ncal_L$ and $\Ncal_R$ are   normalization constants that arise from using  (\ref{eq:FSLRelationFinalSpinJ}) respectively on the left and on the right Mellin amplitudes.
Notice that the factor $-\frac{1}{d}$ in (\ref{FSL:S3}) comes from the leading term in $\D$ in the expansion of $h_5(m)$ (when we consider $m\sim \D^2$). Once again following the same steps that we wrote in the scalar case we can simplify the form of $S_{a b i j}^{(1)}$ and $S_{a b i j}^{(5)}$, 
\begin{align} \label{FSL:S1}
S^{(1)}_{a b i j}(\b_L,\b_R) \simeq  \d(\b_L-\b_R) e^{-\b_L}  \b_L^{\frac{d}{2}-1} \left[\frac{ \Omega_{ai}\Omega_{b j}}{p^2 + M^2}\right]_{p_i \cdot p_j =2 \b_L \g_{i j}}  .
\end{align}
Replacing  (\ref{FSL:S1}) in (\ref{FSL:MJ2S1S5}) one gets the final result 
\begin{align} \nonumber
\! M& \simeq \Ncal \int_{0}^{\infty}\frac{d\b}{\b} \b^{\frac{\sum_{\a=1}^n \D_\a-d}{2}} e^{-\b} \left[ \sum_{a,b=1}^k \sum_{i,j=k+1 }^n \frac{ \mathcal{T}^{a b}_L(p_{l_1}\cdot p_{l_2}) \mathcal{T}_R^{i j}(p_{r_1}\cdot p_{r_2}) }{p^2+M^2} \left(  \Omega_{ai} \Omega_{bj}  - \frac{1}{d} \Omega_{ab} \Omega_{ij}\right) \right]_{p_\m \cdot p_\n=2 \b \g_{\m \n}} ,
\end{align} 
 that is exactly what we expect from the flat space factorization formula (\ref{FS:J2Factoriz}).
\subsection{General spin $J$ }
\label{App: FSL general spin J factorization}
In (\ref{genericm0}) we wrote the factorization formula for any $J$ but just for $m=0$. We now want to check that this contribution reduces in the flat space limit to one piece of the result (\ref{FlatJFactoriz}).
If we use the notation (\ref{FactorizationFormulaGeneric}) we now have only one structure (we denote it with $s=1$) and the other ones are unknown. From our recipe we can uplift the result of (\ref{genericm0}) to a general $m$ just putting the prefactor $g(m)$ in front of it and replacing $M_L \rightarrow L_m$ and $M_R \rightarrow R_m$. In this way the contribution that we want to analyze (that we will call $M^{(1)}$) becomes
\begin{align}
 M^{(1)}&=\sum_{m=0}^\infty g(m) \frac{\mathcal{Q}^{(1)}_{m}}{\g_{LR}-\bar\g_{LR}}\\
\mathcal{Q}^{(1)}_{m}&=\prod_{j=1}^J (\g_{a_j i_j}+\sum_{q=j+1}^J \d_{a_j}^{a_q} \d_{i_j}^{i_q}) L_m^{a_1 \dots a_J} R_m^{i_1 \dots i_J}  
\end{align}
with $\bar{\g}_{LR}=\D+2m-J$. In particular in the flat space limit we have
\begin{align}
\mathcal{Q}^{(1)}_{m} \simeq \mathcal{D}^{(m,\g_L+m)}_{t_L}\mathcal{D}^{(m,\g_R+m)}_{t_R}  \sum_{\{a\}=1}^k \sum_{\{i\}=k+1}^n \left( \prod_{\ell=1}^J\g_{a_\ell i_\ell} \right)  M_L^{\{a\}}(t_L \g_{ab})  M_R^{\{i\}}(t_R \g_{ij})\big{|}_{t_L,t_R=1}.
\end{align}
where $\g_L=\sum_{a=1}^k \frac{\D_a}{2} -\frac{\bar{\g}_{LR}}{2}$ (and similarly for $\g_R$) and   $\{a\}=a_1,\dots,a_J$ (and the same for $\{i\}$). 
We need to express the result in terms of $\check{M}$. We will use the following identity  that can be proved in the flat space limit when $\g_{LR}=\bar \g_{LR}$
\be \label{FSL:GammaToOmegaJ}
\sum_{\{a\}=1}^k \sum_{\{i\}=k+1}^n \left( \prod_{\ell=1}^J\g_{a_\ell i_\ell} \right)  M_L^{\{a\}}(\g_{ab})  M_R^{\{i\}}(\g_{ij}) \simeq  \bar\g_{LR}^{2J} \sum_{\{a\}=1}^k \sum_{\{i\}=k+1}^n \left( \prod_{\ell=1}^J\check{\Omega}_{a_\ell i_\ell} \right)  \check{M}_L^{\{a\}}(\g_{ab})  \check{M}_R^{\{i\}}(\g_{ij})
\ee
where $\check{\Omega}_{ai}$ is  defined in (\ref{CheckOmega}). 
A proof of (\ref{FSL:GammaToOmegaJ}) is given in appendix \ref{ProofOfGammaToOmegaJ}. 
Putting everything together and using formula (\ref{eq:FSLRelationFinalSpinJ}) we find
\begin{align}
M^{(1)}\simeq& \; \Ncal \int_{0}^{\infty}\frac{d\b_L}{\b_L} \b_L^{\frac{\sum_{a}\D_a+J-d}{2}}
\int_{0}^{\infty} \frac{d\b_R}{\b_R} \b_R^{\frac{\sum_{i}\D_i+J-d}{2}} \nonumber \\
&\sum_{\{a\}=1}^k \sum_{\{i\}=k+1}^n  \mathcal{T}_L^{\{a\}}(2\b_L \g_{a b}) \mathcal{T}_R^{\{i\}}(2\b_R \g_{i j}) S^{(1)}_{\{a\} \{i\}}(\b_L,\b_R). \label{FSL:JM1}
\end{align}
where
\begin{align}
\!\!\!\!\!\!S^{(1)}_{\{a\} \{i\}}(\b_L,\b_R)=& \tilde{\Ncal} \sum_{m=0}^\infty \frac{g(m)}{\g_{LR}-\bar\g_{LR}}  \bar\g_{LR}^{2J}   \left( \prod_{\ell=1}^J\check{\Omega}_{a_\ell i_\ell} \right) \left[ \mathcal{D}^{(m,J-\D+\frac{d}{2})}_{t_L} \b_L^{\frac{\D}{2}}e^{-\frac{\b_L}{t_L}}\right]_{t_L=1} \left[ \mathcal{D}^{(m,J-\D+\frac{d}{2})}_{t_R} \b_R^{\frac{\D}{2}}e^{-\frac{\b_R}{t_R}}\right]_{t_R=1} \nonumber
\end{align} 
with $\tilde{\Ncal}=\frac{\Ncal_L \Ncal_R}{\Ncal}=\pi^{\frac{d}{2}} 2^{2J-1} \frac{\mathcal{C}_{\D,J}}{\G(\D+J)^2}$.
In the flat space limit we can simplify $S^{(1)}_{\{a\} \{i\}}$ as explained in the scalar case. The result is
\begin{align}
S^{(1)}_{\{a\} \{i\}}(\b_L,\b_R)& \simeq \d(\b_L-\b_R) e^{-\b_L}  \b_L^{\frac{d}{2}-J+1} \left( \frac{\prod_{\ell=1}^J \Omega_{a_\ell i_\ell}}{p^2+M^2}\right)_{p_i \cdot p_j =2 \b_L \g_{i j}} \ . \label{FSLJS1}
\end{align}
Replacing (\ref{FSLJS1}) in (\ref{FSL:JM1}) we get
\begin{align} 
M^{(1)}&\simeq \Ncal \int_{0}^{\infty}\frac{d\b}{\b} \b^{\frac{\sum_{\a=1}^n \D_\a-d}{2}} e^{-\b} \left[  \sum_{\{a\}=1}^k \sum_{\{i\}=k+1}^n \frac{ \mathcal{T}_L^{a_1 \dots a_J}(p_a\cdot p_b) \mathcal{T}_R^{i_1 \dots i_J}(p_i\cdot p_j) }{p^2+M^2}  \prod_{\ell=1}^J \Omega_{a_\ell i_\ell} \right]_{p_\m \cdot p_\n=2 \b \g_{\m \n}} \label{FSLm0anyJ} \nonumber \ ,
\end{align}
that corresponds to the $r=0$ contribution of (\ref{FlatJFactoriz}).
 
 \subsubsection{Proof of (\ref{FSL:GammaToOmegaJ})} \label{ProofOfGammaToOmegaJ}
We
 start considering that the relation (\ref{MofMcheck}) to express the representation $M$ in terms of $\check M$ simplifies drastically in the flat space limit, since we can drop all the Kronecker-deltas. Thus we can define the leading behavior of $M$ in the flat space limit as  follows
\be
  M^{a_1 \dots a_J}_{fsl}= \sum_{b_1, \dots b_J=1}^k  \g_{b_1} \cdots  \g_{b_J}  \sum_{q=0}^J (-1)^q \check{M}^{\{b\}\{a\}}_{q} \ ,
  \label{FSL:MofMcheck}
\ee
where $\check{M}^{\{b\}\{a\}}_{q} $ is defined in (\ref{Mpermutations}). In this notation, the leading contribution (in the flat space limit) of the left hand side of (\ref{FSL:GammaToOmegaJ}) is given by 
\be
\mathcal{L}=\sum_{\{a\}=1}^k \sum_{\{i\}=k+1}^n \g_{a_1 i_1} \cdots \g_{a_J i_J}  M_{L \, fsl}^{\{a\}}  M_{R \, fsl}^{\{i\}} \ .
\ee
 Equation (\ref{FSL:MofMcheck}) also makes clear that $M_{fsl}$ satisfies a simpler transversality condition, namely
\be \label{FSL:transverse}
\sum_{a_1=1}^k   \g_{a_1} M_{fsl}^{a_1 \dots a_J}= 0 \ .
\ee
For convenience we define $\mathcal{S}_{i_1 \dots i_J}\equiv\sum_{\{a\}=1}^{k}\g_{a_1 i_1} \cdots \g_{a_J i_J} M_{L \,fsl}^{\{a \}}$, a totally symmetric object in the indices $\{ i \}$. Using property  (\ref{FSL:transverse}) we immediately find $\sum_{i_1=k+1}^n \mathcal{S}_{i_1 \dots i_J}= 0$, that can be used to show  
\begin{align}
\mathcal{L}
&=\sum_{\{i\},\{j\}=k+1}^n \mathcal{S}_{i_1 \dots i_J} \g_{j_1} \cdots  \g_{j_J}  \sum_{q=0}^J (-1)^q \check{M}^{\{j\}\{i\}}_{q} \nonumber \\ 
&= \sum_{\{i\},\{j\}=k+1}^n \mathcal{S}_{i_1 \dots i_J} \g_{j_1} \cdots  \g_{j_J}   (\check M_R^{i_1 \dots i_J}- \check M_R^{i_1 \dots i_{J-1} j_{J}}- \dots) \nonumber \\
&=\sum_{\{i\},\{j\}=k+1}^n \mathcal{S}_{i_1 \dots i_J}  \g_{j_1} \cdots  \g_{j_J}   \check M_R^{i_1 \dots i_J} = \g_{LR}^J \sum_{\{i\}=k+1}^n \mathcal{S}_{i_1 \dots i_J}   \check M_R^{i_1 \dots i_J} \ . \label{FSL:eqSMR}
\end{align}
We can further simplify  $\mathcal{L}$ replacing the definition of $\mathcal{S}_{i_1 \dots i_J}$, using (\ref{FSL:MofMcheck}) and considering that for symmetry reasons each $\check{M}^{\{b\}\{a\}}_{q} $ just contributes as $\binom{J}{q} \check M^{b_{1} \dots b_q a_{q+1} \dots a_J}$. The result is
\begin{align}
\mathcal{L}&=\g_{LR}^J \sum_{\{i\}=k+1}^n \sum_{\{a\},\{b\}=1}^{k}\g_{a_1 i_1} \cdots \g_{a_J i_J}   \g_{b_1} \cdots  \g_{b_J}  \sum_{q=0}^J (-1)^q \binom{J}{q} \check M_{L}^{b_{1} \dots b_q a_{q+1} \dots a_n}\check M_R^{i_1 \dots i_J} \nonumber \\
&=\g_{LR}^{2J}  \sum_{\{ i\} =k+1}^n \sum_{\{a\}=1}^{k}\sum_{q=0}^J  \binom{J}{q} \left(-\frac{1}{\g_{LR}} \g_{a_1}\g_{i_1}\right)\cdots  \left(-\frac{1}{\g_{LR}}\g_{a_q}  \g_{i_q}\right)  \g_{a_{q+1} i_{q+1}}  \cdots \g_{a_J i_J}    \check M_L^{\{a\}}\check M_R^{\{i\}} \nonumber \\
&=\g_{LR}^{2J}  \sum_{\{ i\} =k+1}^n \sum_{\{a\}=1}^{k}  \Omega_{a_1 i_1}\cdots  \Omega_{a_J i_J}  \check M_L^{\{a\}}\check M_R^{\{i\}} \nonumber
\end{align}
where $\Omega_{ai}$ is defined in (\ref{CheckOmega}). Once we set $\g_{LR}=\bar \g_{LR}$ we recover formula (\ref{FSL:GammaToOmegaJ}).

\bibliographystyle{./utphys}
\bibliography{./mybib}

\end{document}